\documentclass[aps,prx,twocolumn,superscriptaddress,floatfix,showpacs,longbibliography]{revtex4-2}

\usepackage{graphicx}
\usepackage{amsmath}
\usepackage{dsfont}
\usepackage[utf8]{inputenc}
\usepackage[english]{babel}
\usepackage{color}

\newcommand{\beq}{\begin{equation}}
\newcommand{\eeq}{\end{equation}}
\newcommand{\beqa}{\begin{eqnarray}}
\newcommand{\eeqa}{\end{eqnarray}}
\newcommand{\bsub}{\begin{subequations}}
\newcommand{\esub}{\end{subequations}}

\newcommand{\rem}[1]{}
\newcommand{\refE}[1]{Eq.~(\ref{#1})}
\newcommand{\refe}[1]{(\ref{#1})}
\newcommand{\refF}[1]{Fig.~\ref{#1}}

\newcommand{\notes}[1]{}
\newcommand{\omegam}{\omega_{\rm m}}

\newcommand{\xzp}{{x_{\rm z}}}

\newcommand{\hx}{{\hat x}}
\newcommand{\hp}{{\hat p}}

\begin{document}
\title{Proposal for a nanomechanical qubit} 
\author{F.~Pistolesi}
\affiliation{
Université de Bordeaux, CNRS, LOMA, UMR 5798, F-33400 Talence, France 
}
\email{Fabio.Pistolesi@u-bordeaux.fr}
\author{A.N.~Cleland}
\affiliation{
Pritzker School of Molecular Engineering, University of Chicago, Chicago IL 60637, USA
}
\author{A.~Bachtold}
\affiliation{
ICFO - Institut de Ciencies Fotoniques, The Barcelona Institute of Science and Technology, 08860 Castelldefels, Barcelona, Spain
}

\begin{abstract}
Mechanical oscillators have been demonstrated with very high quality factors over a wide range of frequencies. These also couple to a wide variety of fields and forces, making them ideal as sensors.   
The realization of a mechanically-based quantum bit could therefore provide an important new platform for quantum computation and sensing. 
Here we show that by coupling one of the flexural modes of a suspended carbon nanotube to the charge states of a double quantum dot defined in the nanotube, it is possible to induce sufficient anharmonicity in the mechanical oscillator so that the coupled system can be used as a mechanical quantum bit. This can however only be achieved when the device enters the ultrastrong coupling regime.
We discuss the conditions for the anharmonicity to  appear, and we show that the Hamiltonian can be mapped onto an anharmonic oscillator, allowing us to work out the energy level structure and how decoherence from the quantum dot and the mechanical oscillator are inherited by the qubit.
Remarkably, the dephasing due to the quantum dot is expected to be reduced by several orders of magnitude in the coupled system.
We outline qubit control, readout protocols, the realization of a CNOT gate
by coupling two qubits to microwave cavity, and finally how the qubit can 
be used as a static force quantum sensor. 

\end{abstract}

\date{5th of May 2021}

\maketitle

\section{Introduction}

Mechanical systems have important applications in quantum information and quantum sensing, with for example significant recent interest in their use for frequency 
conversion between optical and microwave signals
 \cite{barzanjeh_entangling_2011,palomaki_coherent_2013,andrews_bidirectional_2014,lecocq_mechanically_2016,vainsencher_bi-directional_2016, bochmann_nanomechanical_2013},
the sensing of weak forces using position detection at or beyond the standard quantum limit  \cite{ockeloen-korppi_quantum_2016},
and demonstrations of mechanically-based quantum buses and memory elements \cite{rabl_quantum_2010,stannigel_optomechanical_2010,satzinger_quantum_2018,
bienfait_phonon-mediated_2019,bienfait_quantum_2020}.
Realizing a quantum bit (qubit) based on a mechanical oscillator is thus a highly desirable goal, providing the quantum information community with a new platform for quantum information processing and storage with a number of unique features. 
A hallmark of mechanical resonators is their ability to couple to a variety of external perturbations, as any force leads to a mechanical displacement; a mechanical qubit could thus enable quantum sensing \cite{degen_quantum_2017} of a wide range of force-generating fields.
Another outstanding aspect is that mechanical oscillators can be designed to exhibit
very large quality factors \cite{urgell_cooling_2020,maccabe_nano-acoustic_2020}, thus well-isolated from their environment, with correspondingly long coherence times. 
Mechanical devices may offer the possibility to develop quantum circuits with both a large number of qubits and a long qubit decoherence time. 
This is of considerable relevance to quantum computing, since the decoherence times of superconducting qubits integrated in large scale circuits \cite{arute_quantum_2019} are reduced to of order 10~$\mu s$, which is much lower than what can be achieved when operating single superconducting qubits \cite{rigetti_superconducting_2012}.

A mechanical oscillator can be made into a qubit by introducing a controlled
anharmonicity, thereby introducing energy-dependent spacing in the oscillator's quantized energy spectrum  \cite{rips_quantum_2013,rips_nonlinear_2014}. 
The anharmonicity then enables the controlled and selective excitation of energy states of the system, 
for example the ground and first excited state, without populating other states, breaking the strong correspondence principle that otherwise limits the quantum control of harmonic systems. 
\begin{figure}[h]
\begin{center}
\includegraphics[width=8cm]{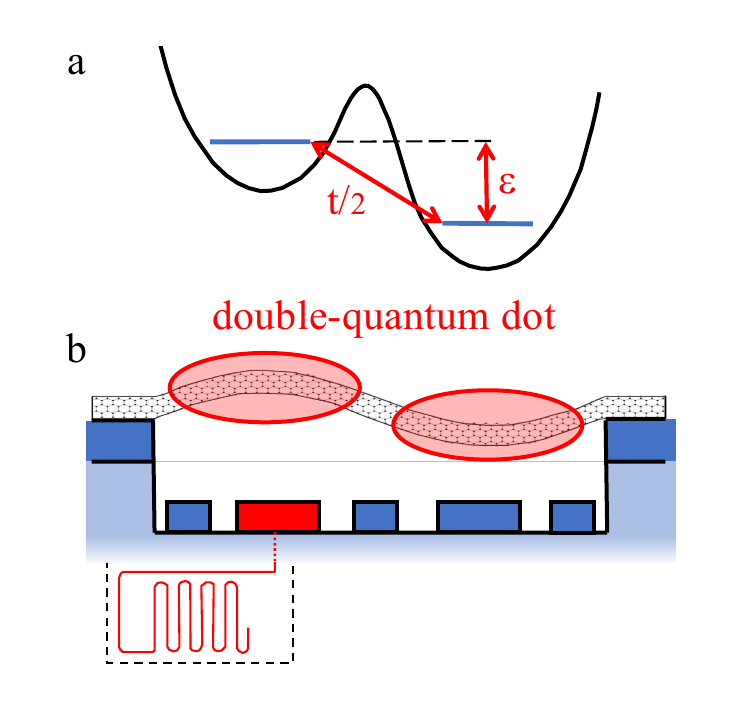}
\end{center}
\caption{Schematic of the proposed setup. A suspended carbon nanotube hosting a double quantum dot, whose one-electron charged state is coupled to the second flexural mode. 
(a) Sketch of the electronic confinement potential and of the two main parameters, the hopping amplitude $t$ and the energy 
difference $\epsilon$ between the two single-charge states.
(b) Physical realization. 
One of the gate electrodes is connected to a microwave cavity for dispersive qubit readout.
}
\label{fig1}
\end{figure}

Notwithstanding the apparent simplicity of this idea, finding mechanical oscillators with  sufficiently strong and controllable anharmonicity is not trivial. 
In Ref.~\onlinecite{rips_quantum_2013,rips_nonlinear_2014}, anharmonicity induced by proximity to a buckling instability has been proposed.
However, such a scheme is difficult to achieve experimentally.
Here we consider the possibility of coupling one of the flexural modes of a carbon nanotube to an integrated double quantum dot, the dot itself defined in the nanotube (cf.~Fig.~\ref{fig1}). 
By tuning independently the gate voltages for the two quantum dots, it is possible to select the 
low-energy electronic states so that only those with a single (additional) electron on 
the double quantum dot are energetically accessible.
The excess electron can sit either on the left or the right dot.
This charged two-level system is electrostatically coupled to the displacement of the oscillator, in particular to the second flexural mode, as illustrated in \refF{fig1}.

We show in the following that for sufficiently strong electro-mechanical coupling, the double quantum dot induces a bistability in the mechanical mode, 
by reducing and then changing the sign of the quadratic term of the effective mechanical potential.  
We find that for strong, but nonetheless reachable coupling constants, it is possible in this way to generate an anharmonicity sufficient to transform the 
mechanical oscillator into a qubit; this does however require entering the so-called ultrastrong coupling regime, where the coupling strength is larger than the mechanical  energy level spacing.

Remarkably, we also find that in the dispersive limit of large detuning  of the oscillator frequency and the electronic two-level system energy splitting, the problem can be mapped onto the Hamiltonian of the quantum-anharmonic oscillator, allowing use of results from that system in this work.
Following a description of the anharmonically-coupled system, we investigate the decoherence induced by the charged two-level system on the mechanical qubit, as well as how standard protocols for quantum manipulation can be implemented. 
The reduction of the pure-dephasing rate of the mechanical qubit with respect to that of the charged two-level system can be made larger than $10^3$ with parameters accessible experimentally.
We show how qubit readout and manipulation can be achieved as 
well as how a CNOT gate for two nanomechanical qubits could be realized by 
coupling them to the same microwave cavity. 
We also show that the mechanical qubit can be used as a quantum sensor for 
any static force that could displace the oscillator.
The static force sensitivity can reach values as good as $10^{-21} {\rm N}/{\rm Hz}^{1/2}$.  

\section{Model}
We consider a nanomechanical system \cite{armour_classical_2004, blanter_single-electron_2004,
chtchelkatchev_charge_2004,clerk_quantum_2005,koch_franck-condon_2005,mozyrsky_intermittent_2006,doiron_electrical_2006,pistolesi_current_2007} 
based on a 
suspended carbon nanotube (cf.~\refF{fig1}) similar to those demonstrated by a number of groups \cite{de_bonis_ultrasensitive_2018,khivrich_nanomechanical_2019,blien_quantum_2020,wen_coherent_2020}. 
It has been shown that it is possible to use multiples gates to fine-tune the electrostatic potential along 
the suspended part of the nanotube \cite{benyamini_real-space_2014,hamo_electron_2016,khivrich_nanomechanical_2019}. 
It is thus possible to form a double-well 
potential to engineer a double quantum dot.
We consider the case when 
only two states, each with one excess electron, are energetically accessible \cite{van_der_wiel_electron_2002}, the other states being at higher energy due to the Coulomb interaction. 
The two single-charge states, corresponding to an electron on the left or right dot, are coupled by a hopping term $t/2$.
Their relative energy 
difference, $\epsilon$, can be controlled by varying the two gate voltages.
The two states couple to the nanotube flexural modes.
By placing the double dot in the center of the nanotube, the coupling 
of the two charge states with the second (anti-symmetric) mechanical mode is maximized
(cf.~\refF{fig1}).

A model Hamiltonian capturing the basic physics of this system can be written down:
\beq
	H={p^2\over 2m}+\frac{m \omegam^2 x^2}{2} 
	+ \frac{\epsilon}{2} \sigma_z+\frac{t}{2} 
	\sigma_x    -    \hbar g \frac{x}{\xzp} \sigma_z \,,
	\label{hamiltonian}
\eeq
where the first two terms describe the relevant mechanical mode of frequency 
$\omegam/2\pi$ with
effective mass $m$, displacement $x$, momentum $p$, and we have introduced 
the zero-point quantum fluctuation $\xzp=(\hbar/2m\omegam)^{1/2}$ with
$\hbar$ the reduced Planck constant. 
The electronic response has been reduced to a two-level system, where the two Pauli matrices $\sigma_z$ and $\sigma_x$ represent the dot charge energy splitting and inter-dot charge hopping, respectively.
Finally $\hbar g/\xzp$ is the variation of the force acting on the mechanical mode 
when the charge switches from one dot to the other. 
The value and sign of $g$ can be tuned over a large range by adjusting the gate voltages \cite{urgell_cooling_2020}. 
In Appendix \ref{Electrostatics} we give a microscopic derivation of the Hamiltonian with the explicit form of the 
coupling terms.

\section{Born-Oppenheimer picture} 
To gain an insight into the physics of the problem, it is instructive to first consider a  semi-classical Born-Oppenheimer picture valid for $\hbar \omegam \ll \sqrt{t^2+\epsilon^2}$. 
We diagonalize $H$ given by \refE{hamiltonian}, neglecting the $p^2$ term and regarding 
$x$ as a classical variable.
The two eigenvalues read:
\beq
	\varepsilon_\pm(x)={m\omegam^2  x^2/2} \pm \sqrt{(\epsilon-2\hbar g x/\xzp)^2+t^2}/2 .
	\label{epsilonPM} 
\eeq
In the spirit of Born-Oppenheimer approximation, 
the energy profile $\varepsilon_\pm$ can be regarded as an 
effective potential for the 
oscillator, which depends on which charge quantum level is occupied. 
Taylor-expanding $\varepsilon_\pm(x)$ for small $x$ and $\epsilon=0$ one finds:
\beq
	\varepsilon_\pm= \pm{t\over 2} + {m\omegam^2 \over2}
	\left(1\pm{4\hbar g^2 \over\omegam t}\right) x^2
	\mp{4m^2 \omegam^2 \hbar^2 g^4 \over t^3} x^4
	+ \dots
\, .
\eeq
The coupling to the double dot leads to a renormalization of the 
quadratic coefficient and the appearance of a quartic and higher terms.
The interaction stiffens the resonating 
frequency of the upper branch while 
softening the lower one.
In particular, for $g>g^{\rm sc}_c=(\omegam t/4\hbar)^{1/2}$, 
the quadratic coefficient of the lower branch becomes negative.
This leads to a double-well potential and a bistability
similar to that predicted for a single quantum dot coupled to a mechanical oscillator \cite{galperin_hysteresis_2005,mozyrsky_intermittent_2006,pistolesi_current_2007,micchi_mechanical_2015,avriller_bistability_2018}.

\begin{figure}[h]
\begin{center}
\includegraphics[width=8cm]{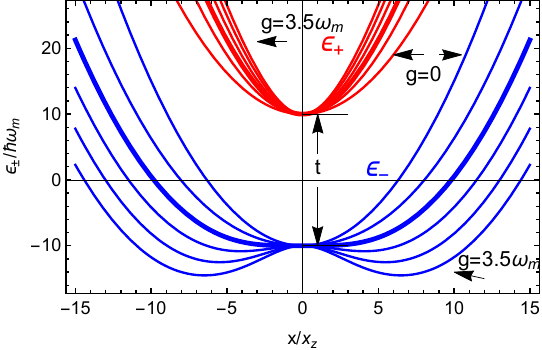}
\end{center}
\caption{Effective potentials $\varepsilon_+(x)$ (red) and  
$\varepsilon_-(x)$ (blue) 
from Eq.~\refe{epsilonPM} for $t/\hbar \omegam=20$ and 
the values of $(4g/\omegam)^2=0$, 10, 20, 30, 40, 50, with 
the first and last line explicitly indicated in the figure.
The potential for $g=g_c^{\rm sc}=\omegam \sqrt{5}$ is shown with a 
thicker line. 
}
\label{figSC}
\end{figure}

Figure~\ref{figSC} shows the evolution of the two branches of the potential as a function 
of the coupling constant $g$, for an experimentally-accessible value of $t=20\hbar \omegam$. 
One clearly sees the formation of the double well-potential for $g>g^{\rm sc}_c$. 
For $g=g^{\rm sc}_c$ the potential of the lower branch is purely quartic (thick line). 
Thus one expects that tuning $g$ close to this critical value, it should be possible 
to modify, over a large range, the ratio between the quadratic and quartic terms and consequently tune the degree of anharmonicity of the system at will. 
%


\section{Full Quantum description}

\subsection{Conditions for anharmonicity}
\begin{figure}[h]
\begin{center}
	\includegraphics[width=8cm]{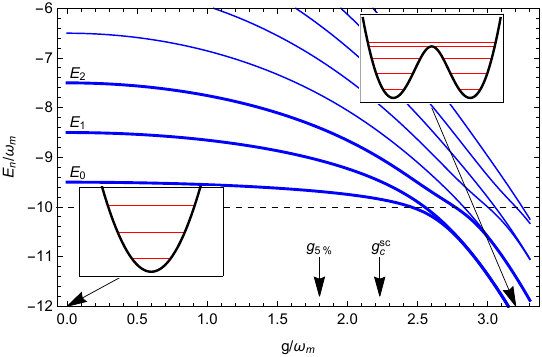}
\end{center}
\caption{
Lowest-lying energy eigenvalues $E_n$ of the Hamiltonian \refe{hamiltonian}  
for $\epsilon=0$ and $t=20\hbar\omegam$ as a function of $g/\omegam$.
The Born-Oppenheimer potential given by \refE{epsilonPM} and the energy levels are shown in the
insets for $g=0$ and $g=3.2\omegam$.
The dashed line indicates the lowest non-interacting electronic level $-t/2$.
The semiclassical critical value for the bistability is $g^{\rm sc}_c/\omegam=\sqrt5\approx 2.23$.
The value of $g=g_{5\%}\approx 1.8 \omegam$ for which the anharmonicity is $5\%$ is also shown.
}
\label{fig2}
\end{figure}

The validity of the qualitative description of the previous section can be confirmed in the general case by 
numerical diagonalization of the Hamiltonian 
given by \refE{hamiltonian} in a truncated Hilbert space. 
Using a basis comprising the $10^2$ lowest harmonic oscillator states largely suffices to reach convergence 
and we find the Hamiltonian 
eigenvectors $|n\rangle$ and eigenstates $E_n$ for the problem.
The result for the lowest set of energy levels is shown in Fig.~\ref{fig2}.

We first notice that for $g \sim g_{\rm c}^{\rm sc}$, the ground state  crosses the  lowest non-interacting electronic level, indicated by the dashed line $-t/2$, preceding the formation of two bound states
in the double-well.
Note that one expects that this crossing should occur for a coupling 
larger than $g_{\rm c}^{\rm sc}$, since for this value the problem reduces to a quartic oscillator, for which the ground state has a positive value \cite{hioe_quantum_1975}  similarly to the harmonic oscillator zero-point motion $\hbar \omegam/2$.
For $g\gg g_{\rm c}^{\rm sc} $, the above mentioned bound states have the same energy (cf.~the upper-right inset in \refF{fig2}) and are sufficiently far from each other that their overlap is negligible.
In \refF{fig2}, the third level remains well separated from the first two, and merges with the fourth level for large $g$.
We introduce the transition frequencies 
$\omega_{nm}=(E_n-E_m)/\hbar$.
The anharmonicity, defined as
\beq
    a= {\omega_{21}-\omega_{10} \over \omega_{10} } \, ,
    \label{anarmonicity} 
\eeq
thus diverges as we increase $g$ from 0 to a value of the order of $g_{\rm c}^{\rm sc}$.
%
%
As discussed in the introduction, this anharmonicity is crucial to enabling quantum control of the qubit formed by the first two levels, $|0\rangle$ and $|1\rangle$.
A minimum requirement is that 
the transition frequency $\omega_{10}$ between $|0\rangle$ and $|1\rangle$ needs 
to differ from $\omega_{12}$ between 
$|1\rangle$ and $|2\rangle$ by much more than the spectral linewidth of the states.  
As a practical example, in the superconducting transmon qubit \cite{schreier_suppressing_2008}, an anharmonicity 
of the order of 5\% suffices to afford full quantum control of the qubit states. 
In the following we will thus consider 5\% anharmonicity as a (somewhat arbitrary) requirement; this is sufficient to find the relevant coupling scale required to implement the mechanical qubit.
\begin{figure}[h]
\begin{center}
	\includegraphics[width=8cm]{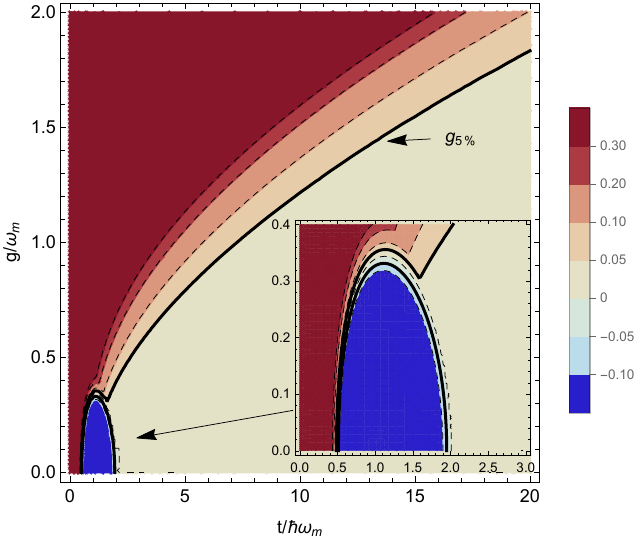}
\end{center}
\caption{
Contour plot of the anharmonicity $a$ in the $(t,g)$ plane.
The contour line for $a=0.05$ is thicker, and defines 
the function $g_{5\%}(t)$. 
The kink at $t\approx1.54\hbar \omegam$ of this function,
better seen in the inset, 
is due to the avoided crossing between the 
charge and oscillator eigenstates that occurs at that value of $t$. It indicates the region
where the eigenstate begins to have a predominantly charge 
nature. 
}
\label{fig3}
\end{figure}
%
%
%
\begin{figure*}
    \centering
    \includegraphics[width=\textwidth]{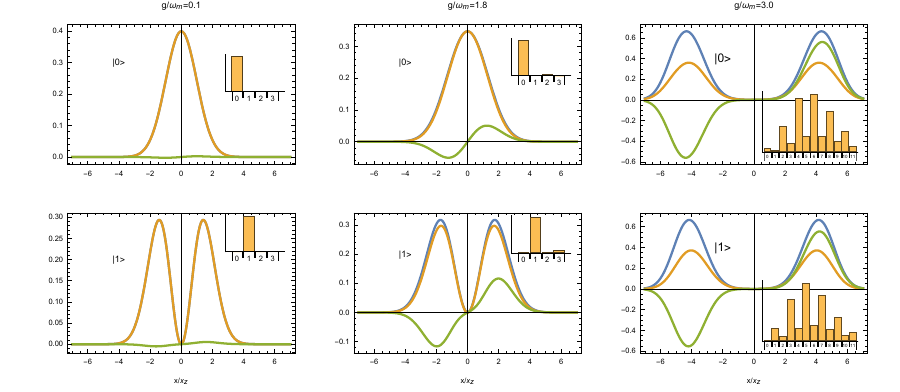}
    \caption{Wavefunctions of the two qubit states
    $|0\rangle$ (upper panels) and $|1\rangle$ (lower panels) 
    for $t/\hbar\omegam=20$, $g/\omegam=0.1$, 1.8, and 3.0.
    We plot -$\langle \sigma_x\rangle(x)$ (yellow), 
    $\langle \sigma_z\rangle(x)$ (green),
    and $\psi_{n+}(x)^2+\psi_{n-}(x)^2$ (blue). 
    Note that for small coupling 
    the yellow and blue lines perfectly overlap.
    The probability of occupation of the first single-harmonic oscillator states are indicated in the insets.
    }
    \label{fig:wave}
\end{figure*}
Resorting again to numerical diagonalization, we present in \refF{fig3} 
a contour plot for the dependence of the anharmonicity on the parameters $t$ and $g$.
The thick contour line for $a=0.05$ defines the 
function $g_{5\%}$, which gives the required coupling 
to obtain a 5\% anharmonicity. 
The region for $t <2 \hbar\omegam$ presents a more complex structure. 
A weaker coupling is required 
to reach the needed anharmonicity.
But in this region the first two levels 
inherit the properties of the double quantum dot to a large extend, so that we will not discuss it further.
Here we explore the mechanical qubit in the parameter range when $t>2\hbar \omegam$, so that the nature of the two lowest energy states of the coupled system remains mechanical. A sizable anharmonicity can only be reached when operating the device near or in the ultra-strong coupling regime, $g>\omegam$, as seen in \refF{fig3}.

\subsection{Eigenstates}%
It is interesting to investigate the nature of the two qubit states 
$|0\rangle$ and $|1\rangle$.
In the position representation, the wavefunction is given by 
$\psi_{n \sigma}(x)=\langle x,\sigma|n\rangle$, 
where 
$|n\rangle$ the Hamiltonian eigenstate and
$|x,\sigma\rangle$ is the 
eigenstate of the displacement $x$
and $\sigma_z$ operators with eigenvalues $x$ and $\sigma$, respectively. 
The wavefunction $\psi_{n\sigma}(x)$ can be chosen to be real-valued.
Instead of looking directly at $\psi_{n\sigma}(x)$,
it is more interesting to consider the averages of the operators $\sigma_i$ 
as a function of 
$x$: 
$\langle \sigma_i \rangle(x)=\sum_{\sigma,\sigma'} \psi_{n\sigma}(x) [\sigma_i]_{\sigma\sigma'} \psi_{n\sigma'}(x)$.
Since by symmetry $\langle \sigma_y \rangle=0$,
only $\langle \sigma_x\rangle=2\psi_{n+}\psi_{n-}$ 
and $\langle \sigma_z\rangle=\psi_{n+}^2-\psi_{n-}^2$ are non-vanishing.

We display in \refF{fig:wave} these two components as well as the total probability for the oscillator displacement $\psi^2=\psi_{n+}^2+\psi_{n-}^2$
(blue curve in \refF{fig:wave}). 
The function $\langle \sigma_z \rangle(x)$ gives the distribution 
of the charge (green curve
in \refF{fig:wave}), while $\langle \sigma_x \rangle(x)$ 
indicates the strength of the coherent superposition of
the two charge-states 
(yellow curve
in \refF{fig:wave}).
These two quantities are in competition. 
From the figure, one sees that for weak coupling 
$\langle \sigma_z \rangle(x) \approx 0$, and the displacement
probability distribution 
coincides with $-\langle \sigma_x \rangle(x)$.
At the value of $g=g_{5\%}$, the distribution of the charge depends on $x$, for both states. 
Finally for the bistable case
with $g/\omegam=3.0$, one reaches the limit where 
$|\langle \sigma_z \rangle(x)|$ is close to the displacement probability, 
indicating a full correlation between the displacement and the charge. 
We also show in the figure the distribution of the harmonic oscillator states. 
One clearly sees that for $g=g_{5\%}$, the two states are still mainly eigenstates of the mechanical oscillator.

\subsection{Mapping in the dispersive regime}

The numerical diagonalization shows that the semiclassical picture 
provides a good qualitative description. 
A natural question is then how far one 
can extend this picture. 
For this reason, we looked for a unitary transformation $U$ that could map the 
Hamiltonian given in \refE{hamiltonian} onto that of a 
simple anharmonic oscillator.  
In the limit of 
$g/|t/\hbar-\omegam| \ll 1 $, known as the dispersive
limit, we find a $U$ such that, at fourth order in $g$, we
can write $H_T=U^\dag H U$ with
\beq
	 H_T= {t\over 2} \sigma_z+{\hbar \omegam \over 4} \left[\alpha_1\hp^2+\alpha_2\hx^2  
	+ \sigma_z (\alpha_3 \hx^2 +\alpha_4 \hx^4) \right]
	\,.
	\label{MappedH}
\eeq
[We discarded the constant $\hbar^3g^2\omegam/(t^2-\hbar^2 \omegam^2)$.]
Here we introduce the quadratures $\hat x=x/\xzp=a^\dag+a$, $\hat p=p/(m \omegam \xzp) = i(a^\dag-a)$, with
$[\hx,\hp]=2i$, 
where $a$ and $a^\dag$ are the creation 
and destruction operators for the harmonic oscillator eigenstates.
The four coefficients read
\beqa
	\alpha_1&=& 1+{128 \hbar^6 g^4 t^2 \omegam^2 \over \Delta^6 \Delta_3^2} ,\qquad
	\alpha_2= 1- {16 \hbar^4 g^4 t^2  \over \Delta^4\Delta_3^2} 
	,
	\label{alpha1}
	\\
	\alpha_3 &=& {4 \hbar t  g^2\over \omegam  \Delta^2} 
	,\qquad 
	\alpha_4 =  
	-{4 \hbar^3t g^4  (3 t^2+\hbar^2\omegam^2)\over 3 \omegam \Delta^6}
	\label{alpha4}
	\,,
\eeqa
where $\Delta^2=t^2-(\hbar\omegam)^2$, 
$\Delta_3^2=t^2-9(\hbar\omegam)^2$.
The derivation and the definition of $U$ are given in Appendix \ref{mapping}. 

Remarkably we find that within this 
approximation, it is possible to map the problem onto a new 
description with two anharmonic oscillators, one for each charge branch. 
The upper branch is unstable if we stop the expansion at $x^4$, since it has a negative quartic term.
This description thus holds for a small but non-zero value of the ratio $\hbar\omegam/t$, giving a more accurate description than the simpler Born-Oppenheimer approach. 
\begin{figure}
    \centering
    \includegraphics[width=8cm]{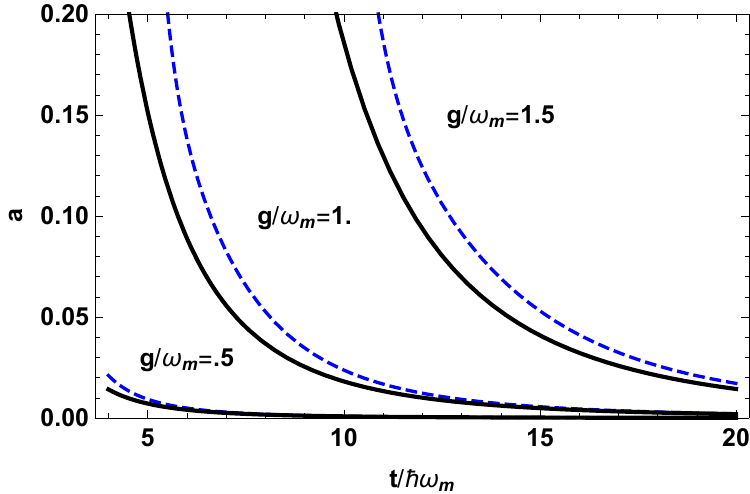}
    \caption{
    Comparison between numerical (full line) and analytical (dashed line) 
    dependence of the anharmonicity parameter $a$ for three values of 
    the coupling $g/\omegam$=0.5, 1, 1.5.
    }
    \label{fig:Anh}
\end{figure}

The anharmonic oscillator is a well-studied problem \cite{hioe_quantum_1978}.
When the quadratic part is positive, 
it is convenient to write the lower branch of \refE{MappedH} in the standard form,
\beq
	H=\hbar \omegam' (\hx^2+\hp^2+\lambda\hx^4 )/4
	\label{stdForm} .
\eeq
This can be done by the scaling $\hx=\xi \hx'$ and $\hp=\hp'/\xi$, so that the commutation relation is preserved $[\hx',\hp']=2i$, 
with
\beq
    \xi=[\alpha_1/(\alpha_2- \alpha_3)]^{1/4} \, .
    \label{xiEq}
\eeq
The renormalized resonant frequency reads
$\omegam'= \omegam [\alpha_1(\alpha_2-\alpha_3)]^{1/2}$
and the quartic coefficient is
\beq
	\lambda = {\alpha_4 \alpha_1^{1/2} \over  (\alpha_2-\alpha_3)^{3/2}}
	\label{OurLambda}
	.
\eeq
Note that we now consider only positive values of $\omegam'$, 
but \refE{MappedH} holds also in the bistable region.
The anharmonicity $a$ defined in \refE{anarmonicity} 
becomes a function of $\lambda$ only.
Using the expression (1.17) 
of Ref.~\onlinecite{hioe_quantum_1978} 
for the eigenvalues in terms of $\lambda$ and \refE{OurLambda}, one can 
obtain an analytical expression for the anharmonicity in terms of the 
parameters $\omegam$, $t$, and $g$ that agrees with the numerics 
with a reasonable accuracy, 
as can be seen in \refF{fig:Anh}.
One finds that the 5\% anharmonicity is achieved for 
$\lambda_c \approx 0.0225$ (the exact numerical result is $\lambda_c=0.0220$). 

\subsection{Operators acting on the qubit}

In order to study the control, readout, and decoherence of the qubit formed by the two states $|0\rangle$ and $|1\rangle$, it is 
necessary to find the projection of the physical operators $\sigma_i$, 
$\hx$, and $\hp$ in the Hilbert space spanned by $\{|0\rangle,|1\rangle\}$. 
In this space, any operator can be written 
as a linear combination of the unit matrix ($\tau_0$) and the three Pauli matrices, that we define here as $\{\tau_x,\tau_y,\tau_z\}$, to distinguish them from the operators $\sigma_i$ acting in the charge space. 
The Hamiltonian of the qubit then simply reads $\hbar \omega_{10} \tau_z/2$.
One can calculate numerically the matrix elements of any operator in the 
qubit sub-space and then obtain its form in
terms of a sum of the four $\tau$-matrices. 
We find for the charge variables [in the representation of \refE{hamiltonian}]
\beq
    \left\{
    \begin{array}{rcl}
    \left.\sigma_x\right|_{\rm qb} &=&\beta_1 \tau_0+\beta_2 \tau_z ,
    \\
    \left.\sigma_y\right|_{\rm qb} &=& \beta_3 \tau_y,
    \\
    \left.\sigma_z\right|_{\rm qb}& =&  \beta_4 \tau_x,
    \end{array}
    \right.
    \label{tauSig}
\eeq
and for the oscillator variables
\beq
    \left\{
    \begin{array}{rcl}
    \left. \hx\right|_{\rm qb} &=& \beta_5 \tau_x \,, 
    \\
    \left. \hp\right|_{\rm qb} &=& \beta_6 \tau_y.
    \end{array}
    \right.
    \label{tauX}
\eeq
The six coefficients can be obtained numerically, 
but it is also interesting to obtain approximate analytical expressions for them. 
This can be achieved using the unitary transformation introduced above
(see Appendix \ref{mapping}):
\beqa
	\beta_1 &=&-1+4 (\hbar g)^2 {(\hbar \omegam)^2-t \hbar\omegam \xi^2+ t^2 \xi^4 
	\over \Delta^4 \xi^2}+g^4\beta_{1,4}\, ,
	\nonumber\\
	&&
	\label{beta1}
	\\
	\beta_2 &=&-2 (\hbar g)^2 {(\hbar\omegam)^2+t^2\xi^4  \over \Delta^4 \xi^2} +g^4\beta_{2,4}\, ,
	\\
	\beta_3 &=&{2 \hbar^2 g\omegam  \over \Delta^2 \xi}
	+g^3\beta_{3,3} \, , 
    \\
	\beta_4 &=& {2 \hbar g  t  \xi   \over \Delta^2}
	+g^3\beta_{4,3} \, ,
	\label{beta4Def}
	\\
	\beta_5 &=&\xi- {2 \hbar^3 g^2  t \omegam  \xi  \over  \Delta^4}
	+g^4\beta_{5,4} \,,
	\\
	\beta_6 &=& 
	{ 1\over \xi}
	-{2\hbar^3 g^2 t \omegam \over \Delta^2 \xi } 
		+g^4\beta_{6,4}
	.\label{beta6}
\eeqa
The coefficients for $g^3$ and $g^4$ are given by 
Eqs.~\refe{beta14}-\refe{beta20} in the Appendix.
\begin{figure}[h]
\begin{center}
\includegraphics[width=8cm]{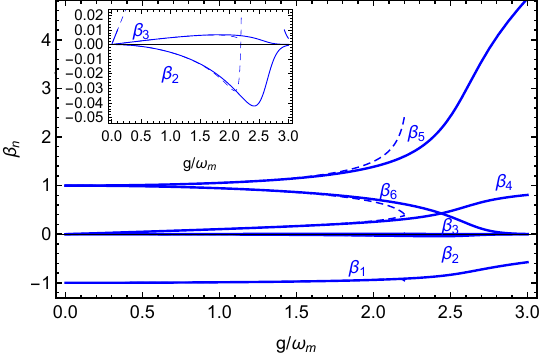}
\end{center}
\caption{
Coefficients $\beta_i$ of the operator projections in
the qubit space, obtained by 
numerical diagonalization (full lines) and from
the analytical approximation to fourth order in $g$ (dashed lines).
The value of $t$ is fixed here to $20 \hbar\omegam$.
}
\label{FigBeta}
\end{figure}

We show in \refF{FigBeta} the behavior of the analytic coefficients as a function 
of $g/\omegam$ for $t/\hbar \omegam=20$, and compare to the exact numerical results.
The analytical expressions again give a good description in the 
interesting range $g<g_{5\%}$.
In particular, these expressions allow us to recognize that 
$\beta_2$ and $\beta_3$ are parametrically small for 
$g\approx \omega_m \ll t/\hbar$. 

Another important result given by the expressions for the $\beta_i$ is the 
charge component of the qubit. 
This can be identified with the value of the $\beta_4$ coefficient,
which gives the projection of the charge operator 
$\sigma_z$ in the qubit space.
This coefficient vanishes linearly in $g$, and it remains small up to
$g\approx \sqrt{\omegam t/\hbar}$ when $t \gg \hbar \omegam$.
In this case, we thus expect that the qubit
has a predominantly mechanical character in its degrees of freedom, 
measured by the $\beta_5$ and $\beta_6$ coefficients,
which remain of the order of unity.

\subsection{Qubit Manipulation}

The values of $\beta$ are also crucial to understanding how to manipulate the qubit. 
This is achieved using a completely classical oscillating 
voltage applied to a nearby wire, 
turned on for some duration with a calibrated amplitude.
The anharmonicity of the system allows this classical signal to 
achieve quantum control.
One can find the effect of an oscillating  voltage on the qubit by considering how this voltage couples to the 
$\sigma_i$ and $\hx$ operators. 
In Appendix \ref{Electrostatics}, we derive these couplings for a potential  $V_{g12}^{\rm AC}$ applied to the two gates controlling
the electrochemical potential of each dot [cf.~\refE{VacCoup}]. 
We find 
that the potential couples to $\sigma_z$ and $\hx$ with the coefficients $\lambda^{\rm ev}$ and $\lambda^{\rm mv} \xzp$, respectively (see Appendix
\ref{Electrostatics} for the explicit expressions).
Since both $\hx$ and $\sigma_z$ 
project on $\tau_x$, we find that the coupling to the 
oscillating field is just 
$\lambda_v \tau_x V_{g12}^{\rm AC} $ with
\beq    
    \lambda_v=\lambda^{\rm mv} \xzp \beta_5+\lambda^{\rm ev}\beta_4 \,.
    \label{lambdav}
\eeq
This indicates that one can use standard methods to manipulate the qubit state,
{\em e.g.} by using nuclear magnetic resonance methods by driving the qubit states at a frequency
$\omega_D$  with pulses that 
induce, in the rotating frame, a term 
$\hbar(\omega_{10}-\omega_D)\tau_z/2+\lambda_v  V_{g12}^{0} \tau_x$ \cite{collin_nmr-like_2004}.
The anharmonicity guarantees that the second excited state will not be populated by these manipulations. 

\subsection{Qubit Readout}
\label{Subsec:readout}

Reading out the state of the qubit can be realised by coupling the system to a microwave superconducting cavity and using a dispersive interaction, analogous to what is done with superconducting qubits \cite{majer_coupling_2007,houck_controlling_2008}. 
The coupling can be obtained from the expression 
of the coupling to an oscillating voltage [cf.~Eqs.~\refe{ElectronVoltage} and \refe{MechanicsVoltage}] with 
the substitution $V^{\rm AC}\rightarrow V_{\rm z} (b+b^\dag)$,
where $b$ is the destruction operator of the photons in the cavity and $V_{\rm z}$ is the zero-point voltage of the cavity.
The coupling Hamiltonian reads 
\beq
    H^{\rm qb-cav} = \hbar g_{v} \tau_x (b+b^\dag)
    \label{Hqb-cav}
\eeq
with $\hbar g_{v}= \lambda_v V_{\rm z}$ [cf.~\refE{lambdav}].
A standard method is then to perform a dispersive
measurement of the superconducting cavity frequency, 
modified by an amount which depends on the qubit state.
By performing a unitary transformation \cite{blais_cavity_2004} 
one can eliminate the term $\tau_x$ from the Hamiltonian  and obtain
for the qubit and cavity Hamiltonian
\beq
    H/\hbar=\omega_{10}\tau_z/2 + 
     (\omega_c +\chi \tau_z)
    b^\dag b 
    \label{DispEq1}
\eeq
where $\omega_c$ is the cavity resonant frequency and 
$\chi= g_{v}^2/(\omega_{10}-\omega_c)$ 
the dispersive frequency shift. 
Since the resonating frequency depends now on the 
qubit state, this allows us to perform an efficient 
quantum non-destructive readout of the qubit state.

This picture remains qualitatively correct, but in 
analogy with what happens in the transmon qubit \cite{koch_charge-insensitive_2007},
when the anharmonicity 
is small, one needs to include the 
other system states to calculate the dispersive coupling correctly.
We present in appendix \ref{Disp} the calculation of 
$\chi$ for the problem at hand by using second order perturbation 
theory in the coupling constant to the cavity.
In this picture the eigenstates can be labeled according to 
the branch ($\sigma=\pm$) 
in Fig.~\ref{figSC} with $|n\sigma\rangle$
and eigenstate energy $E_{n\sigma}$. 
We find that the second excited state, $|2\rangle=|2-\rangle$, 
and two other excited states of the upper branch 
($|0+\rangle$ and $|1+\rangle$)
with an excitation energy of the order of $t$ contribute. 
The parameter $\chi$ entering \refE{DispEq1} reads 
$\chi=\chi_{\rm m}+\chi_{\rm e}$, 
with
\beq
    \chi_{\rm m}(\omega_c) \approx 
    { 
    (g_{\rm ec} \beta_{4,1})^2
    (\omega_{21}-\omega_{10})
    \over
    (\omega_c-\omega_{21})
    (\omega_c-\omega_{10})
    }
    \label{chim2}
\eeq
dominant for $\omega_c \approx \omega_{10}$ and
\beq
    \chi_{\rm e}(\omega_c) \approx 
    {g_{\rm ec}^2
    (\delta_{11}-\delta_{00}) 
    \over  
    2(\omega_c-\delta_{11})
    (
\omega_c-\delta_{00})}
\label{chic2}
,
\eeq
for $\omega_c \approx \delta_{00}$.
Here $\beta_{4,1}=2 \hbar g \xi t/\Delta^2 \ll 1$ is the 
first order contribution to $\beta_4$
[cfr \refE{beta4Def}], 
$\delta_{nm}=(E_{n+}-E_{m-})/\hbar$,
and
$\hbar g_{\rm ec}=\lambda^{\rm ev} V_z$.
One can see that $\chi_{\rm m}$ is proportional to the 
anharmonicity, and thus vanishes in the harmonic case. 
The expression for $\chi_{\rm e}$ 
also vanishes when the coupling constant vanishes, 
 but it does not require an 
anharmonicity:
\beq
\delta_{11}-\delta_{00} \approx {4 g^2 t\over \Delta^2}
.
\label{deltaDif}
\eeq
At lowest order this value is just the difference of the semiclassical 
resonating frequencies of the upper and lower branch.
This dispersive coupling relies on the intrinsic anharmonicity 
of the charge two-level system.

We can further simplify \refE{chim2} by considering 
it for $\omega_c$ close to $\omega_{10}$: 
the small numerator is compensated by a vanishing denominator and one obtains
$\chi_{\rm m} \approx (g_{\rm ec} \beta_{4,1})^2/(\omega_{10}-\omega_c)$,
which remarkably coincides with the standard form of the dispersive coupling.
Even if this looks independent of the anharmonicity, note 
that it is necessary that $|\omega_{21}-\omega_{10}|>g_{\rm ec} \beta_{4,1}$ 
for the calculation to be valid, this condition sets the constraint
on the anharmonicity $a>g_{\rm ec} \beta_{4,1}/\omega_{10}$.
Choosing the detuning to the minimum value allowed by 
second-order perturbation theory: $g_{\rm ec} \beta_{4,1}$,
one obtains $\chi_{\rm m}^{\rm max} \approx g_{\rm ec} \beta_{4,1} < a \omega_{10}$.
Since $\omega_c \approx \omega_{10}$ this shows that 
a quality factor larger than $1/a$ would be largely sufficient 
to detect the qubit state.

Similar arguments can be applied to the expression for $\chi_{\rm e}$,
leading to $\chi_{\rm e}^{\rm max} \approx g_{\rm ec}$.
In this case the limitation is less severe, since 
the condition $|\delta_{11}-\delta_{00}|>g_{\rm ec}$ 
does not involve the anharmonicity.
Using \refE{deltaDif} it gives approximately $g_{\rm ec}<4 g^2/t$.
This result suggests that it may be more convenient to 
tune the cavity to this resonance and exploit 
the $\chi_{\rm e}$ dispersive coupling to readout the qubit state.

\begin{figure}[h]
\begin{center}
\includegraphics[width=4cm]{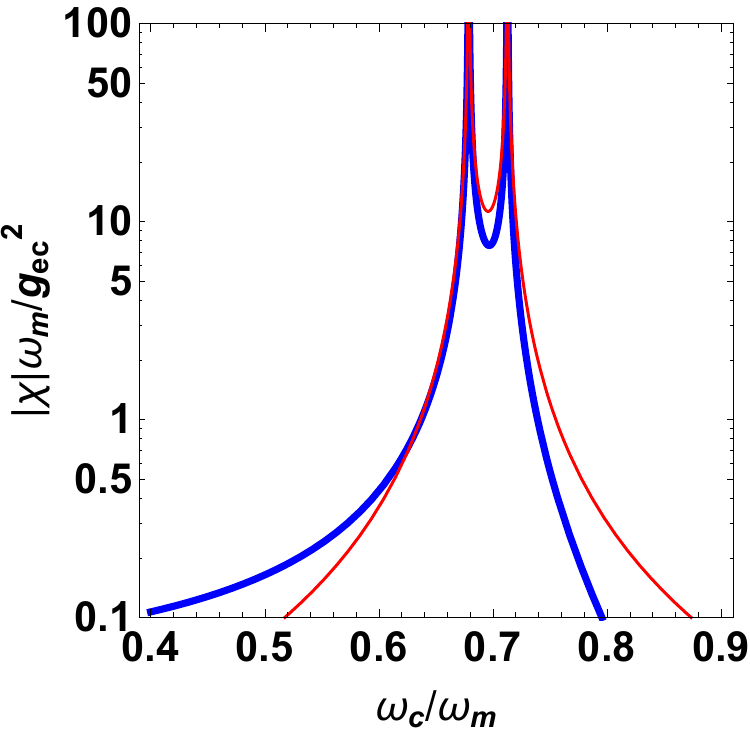}
\includegraphics[width=4cm]{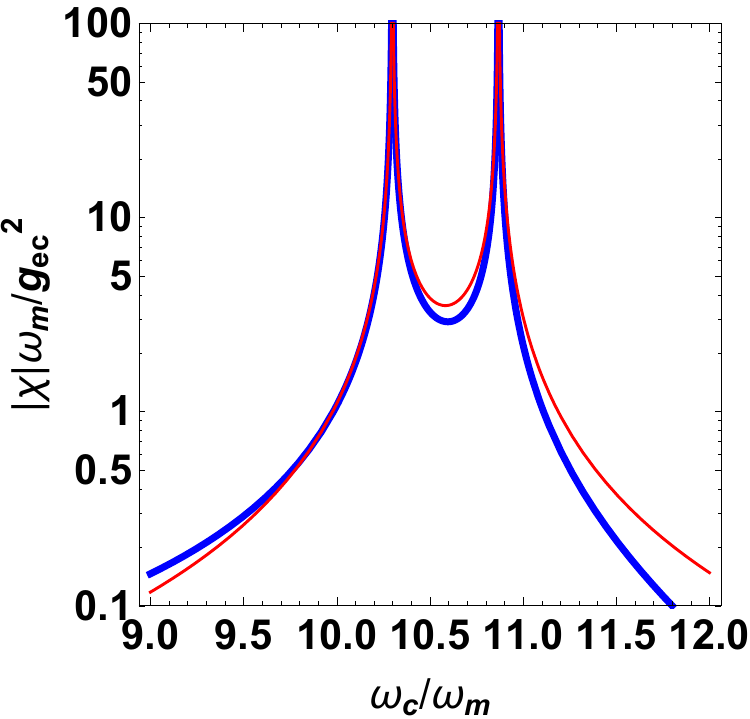}
\end{center}
\caption{
The quantity $|\chi|\omega_m/g_{\rm ec}^2$ for 
$t/\hbar\omega_{\rm m}=10$, $g/\omega_{\rm m}=1.2164$ (for which $a=0.05$) 
as a function of $\omega_c/\omega_{\rm m}$ in two different regions 
of the spectrum: close to $\omega_{10} < \omega_{12}$ 
and  close to $\delta_{00} < \delta_{11}$, left and right panel, respectively.
The thick blue line gives the numerical calculation, the 
thin red line the expressions \refe{chim2} and \refe{chic2},
shown in the left and right panel, respectively. 
}
\label{fig:Chi}
\end{figure}

These analytical expressions are obtained as 
a perturbative expansion in $g/t$, but the expressions 
remain accurate in the range of coupling 
of interest for our purposes as shown as an example 
in Fig.~\ref{fig:Chi}.

\section{Decoherence}
\label{decoherence}

The double quantum dot and the mechanical oscillator are unavoidably coupled to the environment, which induces decoherence and incoherent transitions between energy levels. 
The decoherence rate of the double quantum dot charge qubit is much larger than that of the mechanical resonator, so that it will limit the performances of the mechanical qubit. 
%
%
Best values for the decoherence rate are in the 
MHz range \cite{scarlino_all-microwave_2019}.

In order to study how the nano-mechanical qubit inherits the decoherence of its two sub-system components, we begin by constructing a simple model for the coupling of the sub-systems to the environment. 

We write the coupling Hamiltonian as
\beq
	H_I=  {\hat A}^{\rm c}  \hat E_1+ \hx \hat E_2 \, ,
\eeq
where ${\hat A}^{\rm c}= \sum_{i=x,y,z} v_i \sigma_i = \vec{v} \cdot\vec{\sigma}$ is the most general 
operator in the charge subspace (see for instance \cite{hauss_dissipation_2008}). 
The operators $\hat E_1$ and $\hat E_2$ are given by the sum of operators, themselves involving many degrees of freedom that model
the environment of the charge  and the mechanical oscillator, respectively
(the coupling constant is absorbed in the $\hat E$-operators so that ${\hat A}^{\rm c}$ and $\hx$ are dimensionless).
We assume that we know the correlation functions
$C_i(t)=\langle \hat E_i(t) \hat E_i(0) \rangle $, as well as their Fourier transforms
$ S_i(\omega)=\int dt e^{i\omega t} C_i(t)$,
and that the charge and mechanical environments are independent,
$\langle \hat E_1(t) \hat E_2(0) \rangle=0$. 
If $S_i(\omega)$  is a sufficiently smooth function for $\omega$ close to the qubit resonant frequency, 
the three parameters $v_i$  give a complete 
description of the coupling to the environment of the charge system.
For the mechanical oscillator, we parametrize the coupling to the environment with a single damping rate $\gamma$.

One can then use the standard procedure, integrating out the environmental degrees of 
freedom and finding an equation for the reduced density matrix $\rho$
in the Born-Markov and rotating-wave approximations.
The rate equations have the standard form:
\beqa
	\dot \rho_{nn} &=& -\rho_{nn} \sum_{p\neq n} \Gamma_{n\rightarrow p} 
	+\sum_{p\neq n} \rho_{pp}\Gamma_{p\rightarrow n},
	\\
	\dot \rho_{nm} &=& -\left[	\sum_{p\neq n} \Gamma_{n\rightarrow p}/2 +
	\sum_{p\neq m} \Gamma_{m\rightarrow p}/2+\Gamma^\phi_{nm}\right] \rho_{nm} ,
	\nonumber \\ &&
\eeqa
where $\rho_{nm}=\langle n | \rho | m \rangle$ is the matrix element of $\rho$
in the eigenstate basis $|n\rangle$ of the Hamiltonian \refe{hamiltonian} with eigenvalues $E_n$.
The rates read:
\beqa
	\Gamma_{n\rightarrow m} &=& 2 \pi S_1(\omega_{nm}) |A^{\rm c}_{nm}|^2 +2 \pi S_2(\omega_{nm}) |x_{nm}|^2,
	\nonumber
	\\
	\Gamma^\phi_{nm} &=& 	\pi S_1(0) (A^{\rm c}_{nn}-A^{\rm c}_{mm})^2+\pi S_2(0) (x_{nn}-x_{mm})^2
	\nonumber
\eeqa
where $O_{nm}=\langle n | O| m \rangle$ and $\Gamma^\phi_{nm} $ is the pure dephasing rate.
These equations hold at non-zero temperature $T$, with $S_i(\omega)=S_i(-\omega)e^{\hbar \omega/k_B T}$
and $k_B$ is the Boltzmann constant. 
When only two levels are present one finds 
\beqa
	\dot \rho_{00} &=& -\rho_{00}  \Gamma_{0\rightarrow 1} +\rho_{11}  		\Gamma_{1\rightarrow 0}
	\label{TLS1}
	\\
	\dot \rho_{01}&=&-  \rho_{01} (\Gamma_{0\rightarrow1}+\Gamma_{1\rightarrow 0}+2\Gamma^\phi_{01} )/2 
	\label{TLS2}
	\,.
\eeqa
The last equation defines the coherence time of the qubit $T_2=2/(\Gamma_{0\rightarrow1}+\Gamma_{1\rightarrow 0}+2\Gamma^\phi_{01})$. 
In the following we focus on the two rates $\Gamma_{1\rightarrow 0}$ and
$\Gamma^\phi_{01}$.
(We do not consider the case of equally-spaced levels inducing transfer of coherence
between higher energy states \cite{cohen-tannoudji_atom-photon_1992}.)

\subsection{Non-interacting case}

Let us begin with the non-interacting case ($g=0$) in order to define the rates.
We have two independent systems: the double quantum dot and the mechanical oscillator. 
For the oscillator, one finds $\Gamma^{\rm m}_{1\rightarrow0}=2 \pi S_2(\omegam) = \gamma(1+n_{\rm th})$,
where $n_{\rm th}=1/(e^{\hbar \omegam/k_B T}-1)$ and $\Gamma_{12}^{{\rm m},\phi}=0$.
For the charge system, we begin by diagonalizing the Hamiltonian
$H_0=(\epsilon \sigma_z + t \sigma_x)/2$, 
performing a rotation by an
angle $\theta=\arctan(t/\epsilon)$ around the $y$ axis: $U(\theta)=e^{-i\theta \sigma_y/2}$.
One has
\beqa
	U(\theta)^\dag \sigma_x U(\theta)&=&\cos \theta \sigma_x -\sin \theta \sigma_z ,
	\\	
	U(\theta)^\dag \sigma_z U(\theta)&=&\sin \theta \sigma_x +\cos \theta \sigma_z ,
\eeqa
with $\sigma_y$ invariant.
The charge Hamiltonian coupled to the environment then becomes
\beq
	H'=U^\dag H U =
	 {1\over 2} \sqrt{t^2+\epsilon^2} \sigma_z + \vec{v'} \vec{\sigma}\hat E_1,
\eeq
with 
$	v'_x = \cos \theta v_x +\sin \theta v_z $, 
	$v'_z = -\sin \theta v_x +\cos \theta v_z$,
and $v'_y=v_y$.
This gives the rates
\beqa
	\Gamma^{\rm c}_{1\rightarrow 0}(\theta) &=& 
	2 \pi S_1\left(\sqrt{\epsilon^2+t^2}\right) 
	[(\cos \theta v_x +\sin \theta v_z)^2+{v_y}^2]\,,
	\nonumber\\
	&&\\
	\Gamma^{{\rm c},\phi}_{01}(\theta) &=& 	
	4 \pi S_1(0) (\sin \theta v_x -\cos \theta v_z )^2
	\,.
\eeqa

According to these equations, the pure dephasing and decay rates depend on the value of $\theta$ (i.e. the ratio $\epsilon/t$). 
Since the environmental spectrum depends only on the charge energy splitting, the ratios
\beqa
R^D &\equiv &
{\Gamma^{\rm c} _{0\rightarrow 1}(0) \over 
\Gamma^{\rm c} _{0\rightarrow1} (\pi/2)}
={v_x^2+v_y^2 \over v_z^2+v_x^2} ,
\\
R^\phi
&\equiv &
{\Gamma^{{\rm c},\phi} _{01} (0)\over 
\Gamma^{{\rm c},\phi} _{01}(\pi/2)}
={v_z^2 \over v_x^2} ,
\eeqa
depend only on the values of $v_i$. 
One can then, at least in principle, measure the rates for the same energy splitting $ \sqrt{t^2+\epsilon^2}$
and the two values of $\theta$, 0 and $\pi/2$.
This gives $R^D$ and $R^\phi$ that can be used to express
$v_y$ and $v_z$ in terms of $v_x$:
\beqa
	v_y^2 &=& [R^D(1+R^\phi)-1] v_x^2 
	\,,
	\label{vyEq}
	\\
	v_z^2 &=& R^\phi v_x^2 \,.
	\label{vzEq}
\eeqa
	
\subsection{Interacting case}

We can now consider the interacting case.
We will exploit the fact that the operators $\sigma_i$ and 
$\hx$ in the subspace spanned by $\{|0\rangle,|1\rangle\}$
can be written in terms of the $\tau_i$ operators [\refE{tauSig} and \refE{tauX}].
We neglect the decay rate from and to the third level, which is small as it is only due to oscillator damping and 
vanishes exponentially for $k_B T \ll \hbar \omegam$. 
We obtain then the following results for the decay and decoherence rate of the nanomechanical qubit:
\beqa
	\Gamma_{1\rightarrow0}^{\rm qb}&=&
	2\pi S_1(\omega_{10}) (v_z^2\beta_4^2+v_y^2 \beta_3^2)+
	2\pi S_2(\omega_{10}) \beta_5^2 \,,
	\nonumber
	\\
	\Gamma_{01}^{{\rm qb},\phi}&=&
	4\pi S_1(0)  v_x^2 \beta_2^2
	\,. 
	\nonumber
\eeqa
Using the relations \refe{vyEq}-\refe{vzEq} and assuming that 
$S_i(\omega_{10})\approx S_i(\omegam)$, we find
\beqa
	\Gamma_{1\rightarrow0}^{\rm qb}&=&
	\Gamma_{1\rightarrow0}^{\rm c}(\pi/2) 
	{R_\phi\beta_4^2+[R^D(1+R^\phi)-1] \beta_2^2 \over 1+ R^\phi}
	\nonumber \\
	&&
	+\beta_5^2 \gamma(1+n_{\rm th})
	\label{inverseT2}
	\\
	\Gamma_{01}^{{\rm qb},\phi}&=&
	\beta_2^2 \Gamma_{01}^{{\rm c},\phi}(\pi/2) 
	\,.
\eeqa
In the region of interest, we can use the analytical expressions for $\beta_i$. 
For $\hbar\omegam/t \ll 1$ we can drop the term proportional to $\beta_2^2 \ll \beta_4^2$ and 
obtain
\beqa
	\Gamma_{1\rightarrow0}^{\rm qb}  &\approx&
	{R_\phi \over 1+ R^\phi}
	{4 \hbar ^2 g^2 t^2 \omegam\over\Delta^4 \omegam'}
	\Gamma_{1\rightarrow0}^{\rm c}(\pi/2) 	
	\nonumber \\
	&&
	+{\omegam\over \omegam'}\left[1-{4 \hbar^3 g^2 t\omegam \over \Delta^4}\right]
	 \gamma(1+n_{\rm th}) 	
	 \label{Gamma10Final}
	 \,.
\eeqa
The pure dephasing is controlled by $\beta_2^2 \approx (\hbar g/t)^4  \ll 1$.
The dephasing is thus strongly reduced in the nanomechanical qubit in comparison to the charge system. 

\begin{figure}[h]
\begin{center}
\includegraphics[width=8cm]{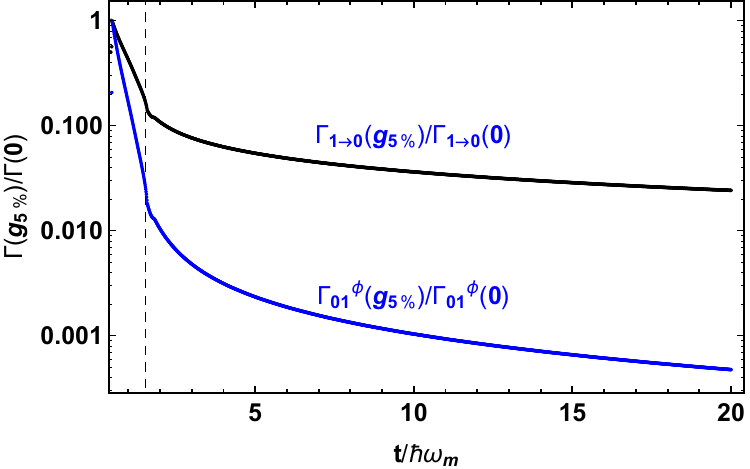}
\end{center}
\caption{
Ratio of the decay rate 
$\Gamma_{1\rightarrow0}^{\rm qb}(g_{5\%})/\Gamma_{1\rightarrow0}^{\rm qb}(g=0)$ 
and pure decoherence rate
$\Gamma_{10}^{\rm qb,\phi}(g_{5\%})/\Gamma_{10}^{\rm qb,\phi}(g=0)$ 
as a function of $t/\hbar \omegam$. 
We assume $R^D=R^\phi=1$ and we neglect oscillator damping ($\gamma=0$).
The vertical dashed line indicates the beginning of the  region where the qubit becomes dominated by the two charge states, i.e. where
$t<1.54\hbar \omegam$ (cf.~also inset of \refF{fig3}). 
}
\label{Fig7}
\end{figure}

We can evaluate numerically the reduction of the decay and 
pure-dephasing rates for the case $R^D=R^\phi=1$.
The result for 
$\Gamma_{1\rightarrow0}^{\rm qb}(g_{5\%})/\Gamma_{1\rightarrow0}^{\rm qb}(g=0)$ 
and 
$\Gamma_{10}^{\rm qb,\phi}(g_{5\%})/\Gamma_{10}^{\rm qb,\phi}(g=0)$ 
is shown  in \refF{Fig7} as a function of $t$ for $\gamma=0$.
As expected from the analytical expressions, 
the larger the value of $t$, the larger the reduction in the 
decoherence. 
This is a natural consequence of the 
mechanical nature of the qubit in this limit.

\section{A two-qubit gate}

We have shown that a carbon-nanotube oscillator can be used as a qubit and how 
manipulation and read out can be performed. 
To use these devices to manipulate quantum information, 
an entangling two-qubit gate is required.
In this section we discuss a possible implementation of the CNOT gate, known to be 
a universal gate.
We follow the idea presented in Ref.~\cite{rigetti_fully_2010} that exploits 
the coupling of two superconducting qubits to the same microwave cavity and that 
has been successfully implemented as reported in Ref.~\cite{chow_simple_2011}.
%

We consider the 
effective coupling generated by a microwave cavity between two nano-mechanical qubits.
In the case of qubits that can be well approximated as two-level systems, 
the coupling to the cavity is of the form of \refE{Hqb-cav}:
$
\hbar g_{v}^{(a)} \tau_x^{(a)} (b+b^\dag)
$, 
where the index $a$ takes the value 1 or 2 to indicate the two qubits.  
One can show that this induces a coupling term in the Hamiltonian 
$J \tau_x^{(1)} \tau_x^{(2)}$.
\rem{
with 
\beq
    J=g_{v}^{(1)}g_{v}^{(2)} \omega_c\left( {1 \over {\omega_{10}^{(1)} }^2--\omega_c^2}+{1 \over {\omega_{10}^{(2)} }^2-\omega_c^2}\right).
\eeq
This coupling is small when the the two qubits frequency splittings are detuned from each other and from the cavity resonance.
}
The driving of the first qubit at the resonant frequency of the second qubit 
can be described by a Hamiltonian term 
$\hbar A \cos(\omega_D t) \tau_x^{(1)} $,
where $A$ is the intensity, and $\omega_D=\omega_{10}^{(2)}$ the driving frequency.
Taking into account the effective coupling induced between the two qubits, this 
translates into the term $ \hbar J_{zx} \tau_z^{(1)} \tau_x^{(2)}$ 
in the rotating frame Hamiltonian with 
\beq
    J_{zx}= 
    { 
    4g_{v}^{(1)} g_{v}^{(2)} A \omega_c\omega_{10}^{(1)} 
    \over 
    ({\omega_{10}^{(1)}}^2-{\omega_{10}^{(2)}}^2) (\omega_c^2-{\omega_{10}^{(2)}}^2)
    }.
    \label{JzxTLS}
\eeq
This is the required gate generating function, $e^{-i t_J J_{zx} \tau_z^{(1)} \tau_x^{(2)}}=
\cos(J_{zx} t_J)-i \sin(J_{zx}t_J)\tau_z^{(1)} \tau_x^{(2)}$,
that allows the CNOT gate to be performed, modulo single-qubit rotations, 
in a time $t_J=\pi/2J_{zx}$.
%

%

\begin{figure}[h]
\begin{center}
\includegraphics[width=8cm]{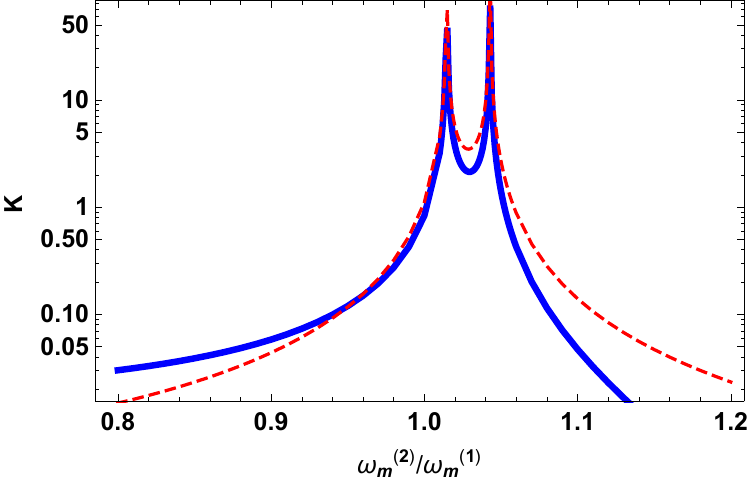}
\end{center}
\caption{
Coefficient $K$ of the contributions to $J_{zx}$ that diverge like $\omega_{m}^{(1)}/(\omega_c-\omega_{10})^{(2)})$ 
[divided by the two coupling constants and the driving intensity 
$g_{\rm ec}^{(1)} g_{\rm ec}^{(2)} A$] as a function of $\omega_{m}^{(2)}$. 
The full line is the numerical result and the dashed line the analytical 
\refE{JzxFull}.
The other parameters are 
$g^{(1)}=g^{(2)}=1.264 \hbar \omega_m^{(1)}$, 
$t^{(1)}=10 \hbar \omega_m^{(1)}$, 
and $t^{(2)}=10.5 \hbar \omega_m^{(1)}$.
}
\label{Fig:Kfact}
\end{figure}

One expects thus that this operation can be applied to the mechanical qubits, 
but since the anharmonicity is not very large, 
we need to investigate the contributions of the higher lying states.
We proceed similarly to what we did for the dispersive coupling in section \ref{Subsec:readout}.
A perturbative calculation is described in Appendix \ref{Disp}. 
It gives 
\beq
    J_{zx} = 
    {
    A g_{\rm ec}^{(1)}  g_{\rm ec}^{(2)} 
    \beta_{1,4}^{(2)} 
    \over 
    {g_{\rm ec}^{(1)} }^2
    (\omega_c-\omega_{10}^{(2)})
    }
    \left[
    \chi_{\rm m}^{(1)}(\omega_{10}^{(2)})
+
    \chi_{\rm e}^{(1)}(\omega_{10}^{(2)})
    \right]
    .
    \label{JzxFull}
\eeq
The expression holds for small $g/t$. 
We note that the coefficient diverges for $\omega_c= \omega_{10}^{(2)}$, while 
in contrast to what is found for the dispersive coupling, 
no divergence is present for $\omega_c$ close to $\delta_{00}$.
We already discussed the functions $\chi_{\rm m}$ and $\chi_{\rm e}$ in \ref{Subsec:readout}, 
we note here that $\chi_{\rm e }^{(1)}$ diverges when its argument equals $\delta_{00}^{(1)}$ or 
$\delta_{11}^{(1)}$. 
Since in general $\omega_{10}^{(2)} \ll \delta_{00}^{(1)} \approx \delta_{11}^{(1)}$, 
the contribution of $\chi_{\rm e}^{(1)}$ is much smaller than that of $\chi_{\rm m}^{(1)}$, 
which diverges when its argument equals $\omega_{10}^{(1)}$ and $\omega_{12}^{(1)}$.
Fig.~\ref{Fig:Kfact} shows the dependence of the factor
$
K=J_{zx} \omega_{\rm m}^{(1)} (\omega_c-\omega_{10}^{(2)})
/
g_{\rm ec}^{(1)} g_{\rm ec}^{(2)} A $
as a function of the ratio $\omega_{\rm m}^{(2)}/\omega_{\rm m}^{(1)}$. 
Both the exact numerical (full line) and analytical expression 
\refE{JzxFull} (dashed line) are shown. 
The double peak corresponds to the values for which 
$\omega_{10}^{(2)}$ equals either $\omega_{10}^{(1)}$
or $\omega_{21}^{(1)}$ [cf.~\refE{chim2}].

This result shows that by driving the qubit 1 it is possible 
to induce a time dependent 
evolution that generates the CNOT gate. 

\rem{
\beq
	J_{zx} = 
	 {8g_{\rm ec}^{(1)} g_{\rm ec}^{(2)} 
			\over 
		\omega_c-\omega_{10}^{(2)}
}
	\left({g_{\rm em}^{(1)} t_1 \xi_-^{(1)} \over \Delta_1^2 } \right)^2
	\left({ g_{\rm em}^{(2)} t_2 \xi_-^{(2)} \over \Delta_2^2 } \right)
	A
	j_{zx}
\eeq

with 
\beq
j_{zx}=
	\left[
	 {\omega_{10}^{(1)} - \omega_{21}^{(1)} \over 
	\left(\omega_{10}^{(2)}-\omega_{21}^{(1)}\right)
	\left(\omega_{10}^{(1)}-\omega_{10}^{(2)}\right)
	}
	+{ \Delta_1^2 \over t_1 {\delta_{00}^{(1)}}^2    {\xi_-^{(1)}}^2 }
	\right]
\eeq
}

\section{Prospect for experimental implementation}

The results found in the two previous sections are very promising for the experimental 
realization of a nanomechanical qubit. 
In this section we discuss possible experimental implementations using currently available technology. 
As discussed in the introduction, the double quantum dot can be realized in a suspendend carbon nanotube and coupled 
to the second mechanical flexural mode of the nanotube.
Such a device has been recently measured at 2~K \cite{khivrich_nanomechanical_2019}, reporting values 
$t/2\pi\hbar=49-96$ GHz with a tunable 
value of $\epsilon$, a second mechanical mode of frequency 
$\omegam/2\pi=327$ MHz with a mechanical quality factor $Q=4\cdot 10^3$
and a coupling constant $g/2\pi=320$ MHz. 
Taking these parameters, we have $t/\hbar \omegam$ up to 150-300,
and $g/\omegam\approx 1$, noting that of course $t$ can be tuned to lower values.
Choosing $t=7\hbar \omegam$, we can operate on the $g_{5\%}$
line (cf.~\refF{fig3}) without changing other parameters. 
At this value of $g$, we already have a sizable reduction of both 
the decoherence and decay rates of the mechanical qubit 
$\Gamma_{1\rightarrow 0}$ and $\Gamma^{\phi}_{10}$ [cf.~\refF{Fig7}] 
compared to that of the charge double quantum dot.
The experiment at 2~K realized with a device fabricated on a Si substrate reports an incoherent tunnelling rate $\Gamma_{1\rightarrow 0}$ 
estimated to  $2\pi\cdot 510$ MHz,
which is clearly too large to use for qubit operations, but 
improvements should be possible, by operating the device at 10 mK to suppress the decoherence induced by low-frequency vibrations (phonon) modes, by producing devices on sapphire substrates that host a minimal number of charge fluctuators, and by current-annealing the nanotube in-situ in the dilution fridge to remove all the contamination adsorbed on the surface of the nanotube \cite{yang_fabry-perot_2020}.
Double-dot structures have been created in non-suspended carbon
nanotubes, and have been coupled to 
superconducting cavities \cite{viennot_coherent_2015}.

One can thus target a mechanical resonator cooled at 10 mK 
with $\omegam/2\pi$ in the range
of 0.6-1 GHz using a nanotube that is shorter and/or is under mechanical tension.
A value of $t/\hbar \omegam=10$ 
will then require a coupling constant 
of the order of $1.1 \hbar\omegam$,
which can be obtained by reducing the nanotube-gate separation and/or increasing the voltage applied on the gate electrode.
With these values the reduction of the 
pure-dephasing decoherence rate of the mechanical qubit 
with respect to that of the double quantum dot 
will be about $10^3$. 
Assuming that the decoherence rate of the order 
of 3 MHz can be obtained, as was achieved in GaAs coupled double quantum dots \cite{scarlino_all-microwave_2019} and that it is mainly limited by pure dephasing, it should be possible to implement most of the 
standard protocols for quantum computation
using  a mechanical qubit with a 3 kHz decoherence rate. 
Note that we did not consider the decoherence induced by the mechanical damping.
Assuming a $Q$ of $10^6$, that has been  experimentally observed in suspended carbon nanotubes \cite{urgell_cooling_2020}, this would give a decoherence rate of only 500 Hz.
Another possible implementation consists in using a nonsuspended GaAs double quantum dot with 3 MHz charge decoherence rate coupled to a suspended metal beam, such as a carbon nanotube. 

With these parameters, one could implement a CNOT gate by choosing
$\omega_{10}^{(1)}/2\pi=500$ MHz, 
$\omega_{10}^{(2)}/2\pi=550$ MHz 
(these values are reduced with respect to the oscillator mechanical frequencies), 
and tune the cavity to $\omega_c/2\pi=475$ MHz.
For $t^{(a)}/\hbar \omega_m^{(a)}\approx 10.$,
one obtains $K$ of the order of 1. 
We assume a coupling constant $g_{\rm ec}/2\pi=50$ MHz of the order of what reported 
in Ref.~\cite{cubaynes_highly_2019} for carbon nanotubes coupled to superconducting 
cavities. 
With these values and a drive $A/2\pi$ also of the order of 50 MHz, which is 
the detuning between the two qubits frequencies, one finds that 
$J_{zx}/2 \pi \approx 2.5$ MHz, that is of the same order of what used in 
Ref.~\cite{chow_simple_2011} to implement the CNOT gate in superconducting qubits.

With the chosen value of $\omega_m$ the typical range for $\omega_{01}/2\pi$ does not 
exceed 500 MHz.
This is sufficient to perform single and two qubit operations, but error correction could be difficult since very low level of thermal occupation 
is required. 
In the long term, 
it seems feasible to increase the mechanical frequency to higher values, a qubit splitting of 1GHz is the target for implementing error correction.

\section{Quantum sensing of a static force with the nanomechanical qubit}

As an important application, we discuss here the possibility of using the nanomechanical qubit for quantum sensing. 
A mechanical oscillator can couple to a variety of forces;
independently of the nature of the force, 
the additional term in the Hamiltonian describing this 
coupling can be written as 
$ H_F= F x $, with $F$ the external force. 
In terms of the nanomechanical qubit operators this gives 
$H_F= F \gamma_F \tau_x/2$, with $\gamma_F= 2 \xzp \beta_5$ 
[cf.~\refE{tauX} and we introduced a factor of two for convenience in the notation]. 
One can then use the protocols for qubit preparation and read-out in order to measure $F$ with great sensitivity. 

As a relevant example we consider here the Rabi measurement protocol, as described in Ref.~\cite{degen_quantum_2017} 
sec. IV.D.  
In a nutshell it consists in preparing the qubit in the ground state, and then let it evolve in the presence of the static force $F$ according to the Hamiltonian 
\beq
    H=\hbar (\omega_{10} \tau_z+\omega_F \tau_x)/2,
\eeq
with $\omega_F=F \gamma_F/\hbar$. 
This induces a Larmor-like precession with a Rabi
frequency $\omega_{\rm R}=\sqrt{\omega_{10}^2+\omega_F^2}$ 
of the pseudo spin representing
the qubit state in the Bloch sphere around the 
direction of the effective magnetic field 
vector $(\omega_F, 0, \omega_{10})$. 
The probability $P_1$ of measuring the qubit in the 
excited state oscillates as
\beq
    P_1 = {\omega_F^2\over \omega_{\rm R}^2}\sin^2(\omega_{\rm R} t/2)
    \,.
\eeq
For large $t$ the sine part of the expression 
is very sensitive to a small variation of $\omega_{\rm R}$
and thus of the force.
For a detection time $t_d$ such that
$\omega_{\rm R} t_d=\pi/2+k\pi$, with $k$ a large integer
one finds
\beq
    \delta P_1 \approx \left(\omega_F\over \omega_{\rm R}\right)^3 {\gamma_F t_d\over 2\hbar } F
    \,.
\eeq
The sensitivity thus increases with the oscillation time
$t_d$. This is mainly limited by the coherence time of the qubit.
One also sees that in order to have a large signal it is better to have 
$\omega_F$ of the same order or larger than $\omega_{10}$.
In our case this could be achieved using the 
gate voltage that 
generates an additional controllable static force to the oscillator.
The most fundamental source of uncertainty in quantum sensing is the  binomial fluctuation of the qubit readout outcome.
Following Ref.~\cite{degen_quantum_2017} a rough estimate of the signal to noise that can be achieved with this method gives that the minimum detectable static force per 
unit bandwidth is
\beq
	\delta F_{\rm min}  \approx {\hbar \over \gamma_F \sqrt{T_2} } , 
	\label{QSlimit}
\eeq
where $T_2$ is the coherence time. 
Using typical values for carbon nanotube resonators 
$\omegam=2\pi \cdot 600$ MHz, $m=10^{-21}$ Kg, one has 
$\xzp \approx 4 \cdot 10^{-12}$ m.
Using $T_2\sim 50 \mu$s from the 3~kHz decoherence rate for the nanotube mechanical qubit estimated in the last section, the static force sensitivity is $\sim 10^{-21}$ N/Hz$^{1/2}$.
For comparison, the resolution in static force measurements is $10^{-17}$~N using optically levitated particles \cite{hebestreit_sensing_2018} and $10^{-12}$~N with atomic force cantilevers in high vacuum and at low temperatures \cite{hug_low_1999}, while a sensitivity of $10^{-15}$~N/Hz$^{1/2}$ can be achieved using optical tweezers in liquids  \cite{ribezzi-crivellari_counter-propagating_2013}.
One finds that when the electronic contribution to the 
decoherence can be neglected with respect to the 
mechanical part, then quantum sensing can reach sensitivities of the order of the 
standard quantum limit \cite{clerk_introduction_2010}.

\section{Conclusions}

In conclusion we have shown that coupling a double quantum 
dot capacitively to the second flexural mode of a suspended carbon nanotube, 
and appropriately tuning the hopping amplitude between the two charge states of the quantum dot, one can introduce a strong anharmonicity in the spectrum of the mechanical mode.
This enables one to address directly the first two energy quasi-mechanical eigenstates without populating the third state (cf.~\refF{fig3}).
These two states form a qubit with mainly a mechanical character.
Manipulation and read-out is then possible with standard techniques, but at the same time, we found that the coupling to the environment is strongly 
reduced. The main benefit is the reduction by up to 3-4 orders of magnitude of the 
pure-dephasing rate, with respect to 
the double quantum dot.
Combined with the expectation of improved dephasing times, this suggests the potential for nanomechanical qubits with very long coherence times.
Furthermore,  the production of mechanical devices using conventional microfabrication techniques is promising for scalability.

The mechanical qubit can be used to couple to a wide number of modalities for external fields,
including acceleration, magnetic forces or other forces.
We have shown that any fields that induce forces on the mechanical oscillator can be detected with unprecedented sensitivity, using quantum preparation and detection protocols. 

We have shown that the nanomechanical qubits can be coupled to each other by microwave cavities, allowing the implementation of a CNOT gate with purely microwave control. 
In principle all other operations involving multiple qubits can be obtained by applying the CNOT gate and single qubit operations.

On the more technical side, we also found a unitary transformation, valid in the 
dispersive limit of $g/|t/\hbar-\omegam| \ll 1$, that maps the problem to the anharmonic oscillator, giving the explicit expressions of the 
main physical operators in the qubit subspace.

\section*{Acknowledgements}

F.P. acknowledges support from the French {\em Agence Nationale de la Recherche} (grant SINPHOCOM ANR-19-CE47-0012) and {\em Idex Bordeaux} 
(grant Maesim Risky project 2019 of the LAPHIA Program).
A.N.C. acknowledges support from the Army Research Laboratory, the DOE, Office of Basic Energy Sciences, and from the UChicago MRSEC (NSF DMR-1420709).
A.B. acknowledges ERC advanced grant number 692876, AGAUR (grant number 2017SGR1664), MICINN grant number RTI2018-097953-B-I00, the Fondo Europeo de Desarrollo, the Spanish Ministry of Economy and Competitiveness through the “Severo Ochoa” program for Centres of Excellence in R\&D (CEX2019-000910-S), Fundacio Privada Cellex, Fundacio Mir-Puig, and Generalitat de Catalunya through the CERCA program.

\appendix

%
%
%
%
%

\section{Electrostatics and derivation of the coupling constants}
\label{Electrostatics}

We give here a derivation of the Hamiltonian. 
For this  we need to calculate the electrostatic energy of the system. 
The only subtle point is the contribution of the voltage sources, as it is 
well known for the Coulomb blockade problem \cite{grabert_single_2013}. 
One needs the electrostatic energy as a function of the charges in the system, and not of the voltages; this is particularly 
important for the expression of the mechanical force.
Following Ref. \onlinecite{van_der_wiel_electron_2002} (appendix A) the 
electrostatic problem of $N$ conductors plus a ground conductor can be treated by introducing a capacitance matrix $C^{(0)}_{ij}$ for which
the charges on the conductor $i$ can be related to the potentials of the other conductors:
\beq
	Q_i=\sum_{j=0}^N C^{(0)}_{ij} V_j \,.
	\label{QeCV}
\eeq
Here $C^{(0)}_{ii}=\sum_{i\neq j} c_{ij}$ and $C^{(0)}_{ij}=-c_{ij}$, where $c_{ij}$ is the capacitance between conductor $i$ and $j$
and clearly $^tC=C$. 
We include in the list of conductors the ground with the index 0.
The relation given by \refE{QeCV} cannot be inverted, since the capacitance matrix has vanishing determinant. 
This just indicates that one can shift all the potential by a constant. 
One can then set one of the potential to 0, say the ground,  and eliminate one line of the matrix, which we choose to be that related to the 
charge on the ground. 
\newcommand{\veca}[1]{{#1}}
The $N\times N$ capacitance  matrix obtained in this way, $C_{ij}$, is then invertible and one can write 
\beq
	V_i=\sum_{j=1}^N \left(C^{-1}\right)_{ij} Q_j \,.
\eeq
The total energy of the system is $U=\sum_{i=0}^N V_i Q_i/2$. With our choice of $V_0=0$, it reduces to
$U=\sum_{i=1}^N V_i Q_i/2=    { ^t\veca{V}}  \veca{Q}/2$, where we introduced the vector notation for the charge and the potentials. 
Using the capacitance matrix we have 
\beq
	U= {1\over 2} {^t \veca{V}} C \veca{V} = {1\over 2} {^t \veca{Q}} C^{-1} \veca{Q}
	.
\eeq
In typical problems one needs to include potential sources. 
These can be modeled with metallic leads with a macroscopic capacitance 
to the ground $C_B\rightarrow \infty$, and the charge on this island $Q_B\rightarrow \infty$ with 
$Q_B/C_B=V_B$ constant. 
In the following, without loss of generality, we will assume that the capacitances of all sources have the same value $C_B$. 

The relevant energy for the problem at hand is the energy expressed as a function of the charges in the metallic islands and leads. 
The mechanical displacement $x$ of any mechanical element of the circuit induces a change in the capacitance matrix, which acquires a dependence on the displacement $C(x)$. 
(For simplicity we consider a single mechanical mode whose displacement is parametrized by the variable $x$; generalization to several modes 
is straightforward.)

The expression for the potential energy is thus 
\beq
	U(Q,x)= {1\over 2} {^t \veca{Q}} {C(x)}^{-1} \veca{Q}  \,.
	\label{UQ}
\eeq
From this expression we can find the expression of the potential energy as a function of the charges in the dots and $x$. 
We can then eliminate the charges  in the leads by using their potentials. 
For this we need to invert the matrix $C$ exploiting the large $C_B$ limit. 
Following Ref.~\onlinecite{van_der_wiel_electron_2002} we first divide the indices in $c$ and $v$, for charge nodes and voltage sources, respectively. We can write
\beq
	C= 
	\left( 
	\begin{array}{cc} 
	C_{cc} & C_{cv} \\ C_{vc} & C_{vv} 
	\end{array}
	\right) \, . 
\eeq
The inverse of this matrix can be written  as follows
\beqa
	\left(C^{-1} \right)_{cc}&=&C_{cc}^{-1} + C_{cc}^{-1} C_{cv} D C_{vc} C_{cc}^{-1}  \\
	\left(C^{-1} \right)_{vc}&=& -D C_{vc} C^{-1}_{cc} \\
	\left(C^{-1} \right)_{vv}&=& C_{vv}^{-1}(1-C_{vc}C_{cc}^{-1} C_{cv}D)
\eeqa
where $D=(C_{vv}-C_{vc}C_{cc}^{-1}C_{cv})^{-1}$.		
Since we eliminated the ground metal island, the only macroscopic matrix elements left are in the diagonal part 
of $C_{vv}\sim C_B$ (cf.~\refE{Csingle} in the following). We can then simplify greatly the inverse since to leading order in $C_B$ one has $D=1/C_B$, 
\beqa
	\left(C^{-1} \right)_{cc}&=&C_{cc}^{-1} , \\
	\left(C^{-1} \right)_{vc}&=& - C_{vc} C^{-1}_{cc}/C_B, \\
	\left(C^{-1} \right)_{vv}&=& 1/C_B .
\eeqa
This allows to express the energy as follows:
\beq
	U={1\over 2} {^t Q_c} 	C_{cc}^{-1} Q_c
		-{^t  Q_c} C_{cc}^{-1} C_{cv} Q_v/C_B
		+ {1\over 2} {^t  Q_v} Q_v/C_B ,
\eeq
but $Q_v/C_B=V_v$ are the sources voltages and the last term is independent of $Q_c$. We thus have
\beq
	U={1\over 2} {^t Q_c} C_{cc}^{-1} Q_c
		-{^t  Q_c} C_{cc}^{-1} C_{cv} V_v
		\,.
		\label{finalU}
\eeq

\subsection{Couplings}
From this expression we can derive the coupling to the mechanical displacement and to 
the voltage applied to a nearby gate electrode. For this, we include the $x$ dependence of the capacitances and the substitution 
$V_v=V_v^{\rm DC}+V_v^{\rm AC}$, where $V_v^{\rm DC}$ is the static part 
and $V_v^{\rm AC}$ the oscillating part of the voltage.
If a gate electrode is part of an electromagnetic cavity, one can obtain the 
coupling to the photon creation and destruction operators via the substitution
$V_v^{\rm AC}=V_v^{\rm z}(b_v+b_v^\dag)$, where
$V_v^{\rm z}$ is the zero-point voltage of the cavity and $b_v$ the destruction 
operator for the photons.

We now need a description in terms of the charge fields. 
Let us associate to each charge variation $\delta q^i_c$ 
the occupation operator $n_i$ with eigenvalues 0 or 1 so 
that the operator for the total number of charges 
can be written as $Q_c = Q_c^0+\sum_i n_i \delta q^i_c$.
The index $i$ can take into account spin or other degrees of freedom and we included a back-ground frozen charge $Q_c^0$.
By including this expression into \refe{finalU}, at lowest order in $x$ we obtain
\beqa
	U&=&U_C+x \sum_i n_i \left(\lambda^{\rm em}_{i}  + \sum_{j \neq i}  n_j \lambda^{\rm em}_{ij}\right)
	\nonumber \\
	&& +
	\sum_i n_i \lambda^{\rm ev}_{iv} V^{AC}_v
	+x  \lambda^{\rm mv}_v V^{AC}_v 
\eeqa
where
\beq
	U_C=\sum_i n_i {^t \delta q^i_c} C_{cc}^{-1}\left (Q_c^0-C_{cv} V_v^{DC}
	+\sum_{j} {1\over 2}  \delta q_c^j  n_j \right)
\eeq
is the pure Coulomb part and the other three terms describe the interaction between the three degrees of freedom 
$x$, $V^{\rm AC}$, and $n_i$, which are associated with the indices m, v, and e, respectively.
(We discarded the constant $U_0= {^t Q^0_c} C_{cc}^{-1} Q^0_c/2-{^t  Q^0_c} C_{cc}^{-1} C_{cv} V_v^{DC}$.)
Here 
\beq
	\lambda^{\rm em}_{i} = 
	\partial_x     
	\left[\left({^t Q^0_c}+{^t \delta q_c^i \over 2} \right) C_{cc}^{-1}- V^{DC}_v C_{vc} C_{cc}^{-1}\right]{\delta q_c^i}
\eeq	
and 
$\lambda^{\rm em}_{ij} =  {^t \delta q_c^i} \partial_x C_{cc}^{-1} {\delta q_c^j}/2$
are the electromechanical couplings,
$\lambda^{\rm ev}_{iv}=- {^t \delta q_c^i} C_{cc}^{-1} C_{cv} $ 
the voltage-electron coupling,  and 
$\lambda^{\rm mv}_v=-{^t Q^0_c} \partial_x( C_{cc}^{-1} C_{cv})$ the mechanical oscillator-voltage coupling. 
\begin{figure}[h]
\begin{center}
\includegraphics[width=8cm]{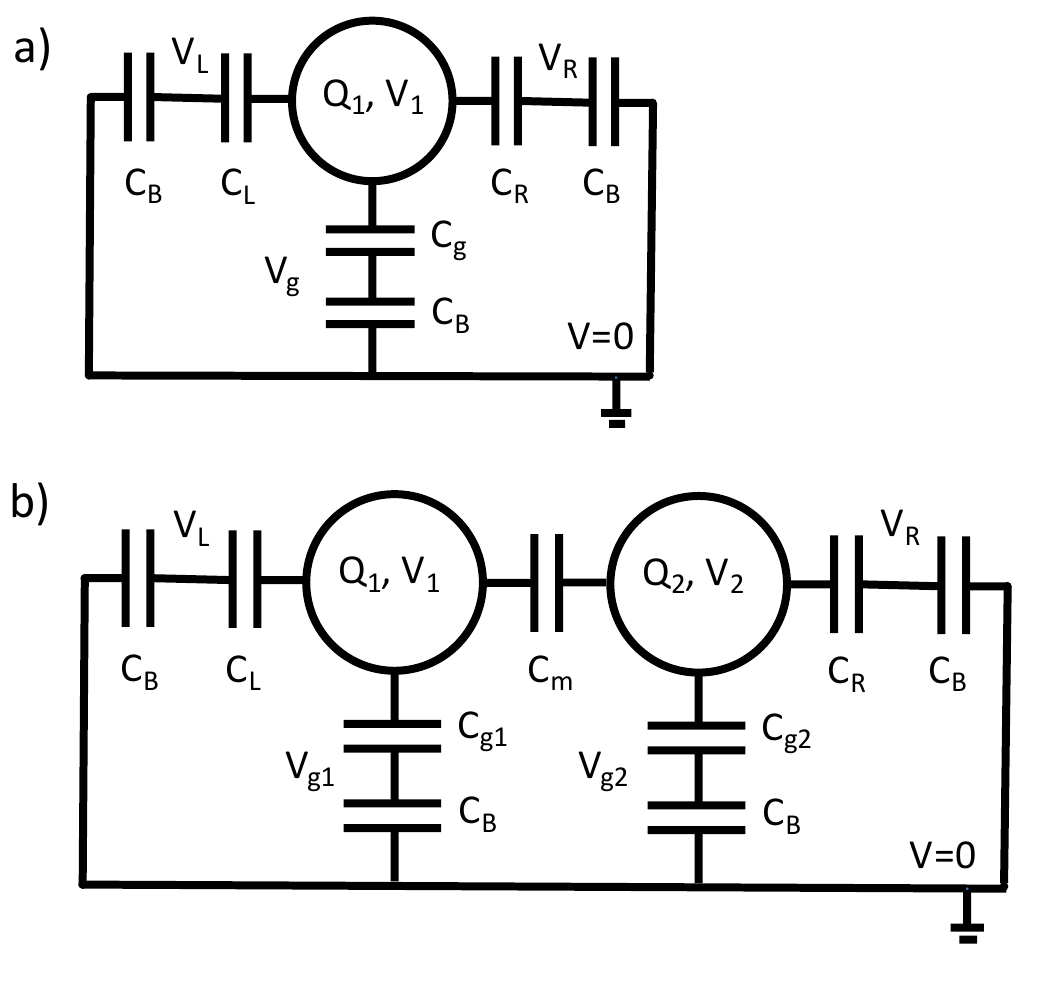}
\end{center}
\caption{
Network of capacitances representing the (a) single- and (b) double-dot circuit.
The capacitances $C_B$ are used to model the voltage sources. 
}
\label{Network}
\end{figure}

\subsection{Single- and double-dot cases}
We now consider two examples. 

({\em i}) The single dot. 
In this case we have 4 metallic entities, one for the dot, 3 for the left, right and gate leads
[cf.~\refF{Network} (a)].
The matrix $C$ reads:
\beq
	C=\left( 
	\begin{array}{cccc}
	C_1 & -C_R & -C_L & -C_g  \\
	-C_R & C_B + C_R& 0& 0 \\
	-C_L & 0 & C_B+C_L & 0 \\
	-C_g & 0 & 0 & C_B+ C_g	
	\end{array}	
	\right)
	\label{Csingle}
\eeq
with obvious notation for the capacitances and with $C_1=C_L+C_R+C_g$. 
This gives $C_{cc}=C_1$, $C_{cv}=-(C_R , C_L , C_g)$, and $C_{vv}= C_B+{\rm diag}(C_R,C_L,C_g)$. 
We assume that only $C_g$ depends on $x$, this gives $\partial_x C_{cc}=\partial_x C_g = C_g'$ 
and $\partial_x C_{cv}=-C_g' (0 ,0 , 1)$. 
We also have $\delta q_c^i=-e$ (with 
$e$ the electron charge) and for simplicity we report the expressions for 
$V_L=V_R=0$.
We then have for the couplings 
\beq
	\lambda^{\rm em}_i=e C_g' [Q_0-(C_1-C_g)V_g -e/2]/C_1^2 ,
\eeq
$\lambda^{\rm em}_{ij}=-e^2C_g'/(2C_1^2)$, 
$\lambda^{\rm mv}_{v}=-C_g'  Q_0(C_R,C_L,C_g-C_1)/C_1^2$.
The last coupling constant is related to $\lambda^{\rm em}$.
Using the value of $Q_0$ that minimizes the electrostatic energy: 
$Q_0=-C_g V_g$ and assuming $|Q_0|\gg e$ one obtains for the 
single-dot coupling constant
$\lambda^{\rm em}_i=-e C_g'V_g/C_1$. 
Note also that in this limit
$\lambda^{\rm em}_{ij}/\lambda^{\rm em}_{i} = e/2 C_1 V_g \ll1$.

({\em ii}) Double dot. 
Let us consider a double dot, with each dot coupled to a gate voltage
[cf.~\refF{Network} (b)].
The capacitance matrix is: 
\beqa
	C_{cc} &=&
	\left( 
	\begin{array}{cc}
	C_1 & -C_m  \\
	-C_m & C_2 
	\end{array}	
	\right) \, , 
	\\
	C_{cv}&=&-\left( 
	\begin{array}{cccc}
	C_L & C_{g1} & 0 & 0  \\
	0 & 0 & C_{g2} & C_R 
	\end{array}	
	\right)	\, , 
\eeqa
and $C_{vv}=C_B+{\rm diag}(C_L,C_{g1},C_{g2},C_R)$. 
Here $C_1=C_L+C_m+C_{g1}$ and $C_2=C_R+C_m+C_{g2}$
We can distinguish two types of $n$ operators, one for the dot 1 ($n_1$) and the other for the dot 2 ($n_2$).
We have 
$\delta q_c^{1} =(-e,0)$ and $\delta q_c^{2} =(0,-e)$. 
For simplicity in the following we assume a symmetric situation $C_L=C_R=C$, $V_L=V_R=0$, $C_1=C_2=C_S$
$Q_c^0=(Q_0,Q_0)$, and $V_v=(0,V_{g1},V_{g2},0)$. 
For our specific problem, for which the interesting mechanical mode is the second one, we assume 
that $C_{g1}(x)=C_{g2}(-x)$ by symmetry, so that $C_{g1}'    =-C_{g2}'$. 
With this hypothesis we find for the coupling constants:
\beqa
	\lambda^{\rm em}_1 +\lambda^{\rm em}_2 &=& 
	-{e C_g'(C+2C_m) (V_{g1}-V_{g2})\over C_S^2-C_m^2}
	\nonumber
\\
	\lambda^{\rm em}_1 -\lambda^{\rm em}_2 &=& 
	{ e C_g'   [2Q_0-e-C( V_{g1}+V_{g2})]\over C_S^2-C_m^2}
	\nonumber
	\\
	\lambda^{\rm ev}_{1v}  &=&- e{(C C_S, C_g C_S, C_g C_m, C C_m )\over C_S^2-C_m^2}
	\nonumber
	\\
	\lambda^{\rm ev}_{2v}  &=&- e{(C C_m, C_g C_m, C_g C_S, C C_S ) \over C_S^2-C_m^2}
	\nonumber
	\\
	\lambda^{\rm mv}_{v}  &=&Q_0 C_g'  {(-C , 2C_m+C, -2C_m-C, C ) \over C_S^2-C_m^2}
	\nonumber
\eeqa
and $\lambda^{\rm em}_{12}=0$. 
For $V_{g1}=V_{g2}=V_g$ $\lambda^{\rm em}_1 =-\lambda^{\rm em}_2=\lambda^{\rm em}$ leading to the Hamiltonian term 
that we used in the main text: $\lambda^{\rm em} x (n_1-n_2)$.
When we reduce the Hilbert space to the two charge states 
$(1,0)$ and $(0,1)$, this Hamiltonian term can be written as $\lambda^{\rm em} x \sigma_z \equiv - \hbar g (x/\xzp) \sigma_z$. 
In this basis $n_1=(\sigma_z+\mathds{1})/2$ and $n_2=(\mathds{1}-\sigma_z)/2$.
This gives
\beq
    g={e C'_g[2C V_{g}+e-2Q_0] x_z \over 2 \hbar (C_S^2-C_m^2)}
    \,. 
\eeq
For the case $Q_0=-C_g V_g$, $|Q_0|\gg e$, and 
$C_m\ll C_S$, we obtain
$g=e C_g'V_g x_z/\hbar C_S$, that coincides with the 
single dot coupling constant.
We also have the coupling of the charge of the dots to the voltages of the gate electrodes:
\beq
	H^{\rm ev}={eC_g \over 2 }
	\left[\mathds{1} { V_{g1}^{\rm AC}+V_{g2}^{ \rm AC} \over C_S-C_m}  
	+
	\sigma_z {V_{g1}^{\rm AC}-V_{g2}^{\rm AC} \over C_S+C_m}\right]
	\label{ElectronVoltage}
	\,.
 \eeq
Finally the direct coupling between the mechanical oscillator and the voltages of the gate electrodes is:
 \beq
 	H^{\rm mv}=
 	Q_0C_g'  {C+2 C_m \over C_S^2 -C_m^2} x(V_{g1}^{\rm AC}-V_{g2}^{\rm AC})
 	\label{MechanicsVoltage}
 	\,.
 \eeq
 In order to compare the last two coupling constants we can write this part of 
 the Hamiltonian as follows 
 \beq
 	H = \left[\lambda^{\rm ev} \sigma_z +  \lambda^{\rm mv} \xzp(a+a^\dag)\right] V_{g12}^{AC}
 	\label{VacCoup}
\eeq
 with $\lambda^{\rm ev}=eC_g/[2(C_S+C_m)]$, 
 $\lambda^{\rm mv }=Q_0 C_g'  (C+2C_m)/(C_S^2-C_m^2)$, 
 $V_{g12}^{AC}=V_{g1}^{\rm AC}-V_{g2}^{\rm AC}$, and we used $x= \xzp(a+a^\dag)$.
 The ratio of the two coupling constant is then of the order of 
 \beq
 { \lambda^{\rm mv} \xzp \over \lambda^{\rm ev} }
 =
 {Q_0\over e} {C_g'\over 2C_g} \xzp {C+2C_m \over C_S-C_m}
 \,.
\eeq
In general this ratio is small $\sim (Q_0/e)(\xzp/L)$ where $L=C_g/C_g' $ is typically  of the order of the distance of the nanotube from the gate.
Thus the oscillating voltage field couples mainly to the charge
degree of freedom.

%
%
%
%
%

\section{Mapping of the Hamiltonian on the anharmonic oscillator in the dispersive regime}
\label{mapping}

In this Appendix 
we show that the Hamiltonian for the system we are considering given by 
\refE{hamiltonian} can be mapped in the 
dispersive regime on the Hamiltonian of an anharmonic oscillator.
We begin by considering $H$ for $\epsilon=0$. 
It reduces to $H=t\sigma_x/2+\hbar \omegam a^\dag a -\hbar  g (a+a^\dag) \sigma_z $.
Performing a rotation of $\pi/2$ around the $y$-axis in the charge space with the operator 
$U_{r}=e^{-i\pi\sigma_y/4}=(1-i\sigma_y)/\sqrt{2}$, 
one has that $U^\dag_r\sigma_xU_r=\sigma_z$ and $U^\dag_r\sigma_zU_r=-\sigma_x$, with 
$\sigma_y$ left unchanged. 
The Hamiltonian is then in the standard form for the Rabi model:
\beq
	H_1= U_r^\dag H U_r=  {t\over 2} \sigma_z+\hbar \omegam a^\dag a + \hbar g (a+a^\dag) \sigma_x .
\eeq 
This model has a long history describing the coupling of electromagnetic radiation to a two-level system, 
but only very recently it has been diagonalized analytically \cite{braak_integrability_2011}.
In practice it is difficult to make use of this solution, but for the case considered in the present paper,
an approximate solution, which holds in the so called dispersive limit of  $|t-\hbar \omegam| \ll g$, could be sufficient 
to obtain an accurate description of the system.
As described in Ref.~\onlinecite{zueco_qubit-oscillator_2009}, it exists a unitary transformation $D_1$ such that
\beq
	H_2
	=D_1^\dag H_1 D_1
	=t {\sigma_z \over 2}+
	{\hbar \omegam\over 4} (\hp^2+\hx^2)   + \sigma_z \hx^2  { t \hbar g^2 \over \Delta^2}  + \dots
	\label{H2Def}
\eeq
where we recall $\Delta^2=t^2-(\hbar \omegam)^2$, $\hx=a^\dag+a$, and $\hp=i(a^\dag-a)$, with $[\hx,\hp]=2i$. 
The Hamiltonian is quadratic in $\hx$ and $\hp$ and commutes with $\sigma_z$.
It can thus be diagonalized 
\beq
	H_2=t\sigma_z/2 + \sum_{\sigma=\pm} [\hbar \omega_\sigma   \pi_\sigma (1/2+a_\sigma^\dag a_\sigma)],
\eeq
where 
\beq
    \hx= 
    \sum_\sigma \xi_\sigma 
      ( a_\sigma^\dag+a_\sigma) \pi_\sigma, 
    \quad
    \hp =  
    \sum_\sigma\xi_\sigma^{-1} 
    i(a_\sigma^\dag-a_\sigma)\pi_\sigma, 
\eeq
with
\beq
    \omega_\sigma=\omegam 
    [1+4\sigma  t \hbar g^2/\omegam \Delta^2]^{1/2}
\eeq
the mechanical frequency of each branch,
$\pi_\sigma=(1+\sigma \sigma_z)/2$ 
the projector on the $\sigma$ branch, 
$\xi_\sigma=(\omegam/\omega_\sigma)^{1/2}$, 
and $[a_\sigma,a^\dag_{\sigma'}]=\delta_{\sigma,\sigma'}$.
Note that this result reduces to the Born-Oppenheimer picture for 
$\hbar \omegam /t  \rightarrow 0$. 
It describes two harmonic oscillators, with different resonating frequencies, 
the lower branch being softened and the upper being hardened by the interaction.

The transformation found in 
Ref.~\cite{zueco_qubit-oscillator_2009} allows to simplify the Hamiltonian 
only at order 2 in  $\hbar g/|t-\hbar \omegam|$. 
For our purposes we need a transformation allowing to obtain the form of the Hamiltonian up 
to the quartic terms in $\hx$. 
For this reason we look for an higher-order unitary transformation $D$ that 
allows to map $H_1$ to $H_T=D^\dag H_1 D$
(the full unitary transformation acting 
on $H$ includes the rotation $U=U_r D$) with $H_T$ given by \refE{MappedH} of the main
text valid at order four in 
$g/(t/\hbar-\omegam)$.

In general one can express any unitary transformation as $D=e^A$, where $A=-A^\dag$. 
We begin by expressing the transformation of Ref.~\cite{zueco_qubit-oscillator_2009} 
in terms of the operators $\hx$ and $\hp$:
\beq
	A_1={i\hbar g \over \Delta^2} (t \sigma_y x +\hbar \omegam \sigma_x p)
	\,.
\eeq
The transformed operators can be found using the standard relation:
\beq
	e^{A} O e^{-A}= \sum_n {1\over n!} C^{O}_{n} \, , 
\eeq
with $C^{O}_{n} = [A,C^{O}_{n-1}]$, and $C^{O}_{0}=O$.
Performing the expansion at order 2 for $O=H_1$ and $A=A_1$ one obtains the expression for $H_2$. 
Performing the expansion at order 4 generates the sought terms $x^4$, but also other terms proportional
 to $x^3 \sigma_x$, $xpx \sigma_y$ and $x^2p^2 \sigma_z$. 
In order to eliminate these terms we add two terms to the $A_1$ operator so that $A=A_1+g^3 A_3+g^4 A_4$. 
By inspection of the terms generated one can realize that $A_3$ should involve only cubic terms in $\hx$ and $\hp$, while $A_4$ only quartic terms.
These terms are multiplied by any of the three Pauli matrices and the unit matrix. 
This leaves 12 free parameters for $A_3$ and 15 free parameters for $A_4$. 
By imposing that the cubic and quartic terms (apart from $ x^4$) vanish, 
we find an explicit expression for $A_3$ and $A_4$
\beqa
	A_3 &=& {4 it \hbar^3 \over 3\Delta_3^2\Delta^6}   \left[
	4\sigma_x t\hbar \omegam[ \hx\hp\hx (3\hbar^2\omegam^2-t^2)+2 \hbar^2\omegam^2 \hp^3]
	\right.
	\nonumber
	\\
	&&
	\left.
	+\sigma_y [8 t^2 \hbar^2 \omegam^2 \hp\hx\hp+\hx^3(-t^4+6t^2 \hbar^2\omegam^2+3\hbar^4\omegam^4)]
	\right]  
	\nonumber \\
	&&
	\\
	A_4 &=& {i\sigma_z(\hx^3\hp+\hp \hx^3) t \hbar^5 \omegam (11t^2-3 \hbar^2\omegam^2) 	
	\over 
	6 \Delta_3^2\Delta^6} 
	\,.
\eeqa
This leads to the Hamiltonian \refe{MappedH} with the coefficients
given by Eqs.~\refe{alpha1}-\refe{alpha4}.
Note that the coefficients $\alpha_1$ and $\alpha_2$ 
%
are very close to one in the limit 
$\hbar\omegam/t \ll 1$ since the correction 
scale like $(\hbar g/t)^4$ and $\hbar^4g^4 \omegam^2/t^6$. 

We thus have shown that 
the Born-Oppenheimer picture gives a qualitatively 
correct description of the problem, even deep in the quantum 
regime when $\hbar \omegam$ is not negligible in front of $t$.
This implies a non-trivial unitary transformation that, 
in contrast with the Born-Oppenheimer picture, mixes 
the mechanical and charge degrees of freedom.
The second important difference is that the coefficients for the quadratic and quartic terms differs from the ones of the semiclassical case. 
These are of course important if a quantitative description of the anharmonicity is needed.

\subsection{Form of the operators in the qubit Hilbert space}
\label{TransformOperators}

In order to study the decoherence and the way in which the mechanical qubit can be manipulated it is important to obtain the projection of the main operators on the Hilbert subspace formed by the lowest two Hamiltonian eigenstates. 
This of course can be done numerically in a straightforward way, but it is also useful to have simple, though approximate, expressions for the form of the operators. 
For this purpose one can apply the unitary transformations $U=U_r D$, introduced above,
to find
the expression of the relevant operators in the base for which the Hamiltonian reduces to the form 
\refe{MappedH} at order $g^4$.
We are interested by the Pauli matrices for the 
charge sector and the $\hx$ and $\hp$ operators, for the oscillator sector. 
Let's define $O^T = U^\dag O U $.
\begin{widetext}
We obtain:
\beqa
\sigma_x^T & =&
	\sigma_z+2\hbar g 
	{\hp \sigma_y \hbar\omegam-
	\sigma_x \hx t \over \Delta^2}
	-2\hbar^2 g^2 {\sigma_z \hx^2 t^2+2 t \hbar \omegam + \hp^2 \sigma_z (\hbar\omegam)^2 \over \Delta^4} + o(g^3) ,
	\\
\sigma_y^T & =&
	\sigma_y-2\hbar^2 g\omegam  {\hp \sigma_x  \over \Delta^2}
	+\hbar^3 g^2\omegam {\sigma_x t (\hx\hp+\hp\hx)-2\hp^2 \sigma_y \hbar \omegam
	\over \Delta^4} + o(g^3) ,
	\\
\sigma_z^T &=& 
	-\sigma_x-2\hbar g t {\hx \sigma_z  \over \Delta^2}
	+\hbar^2 g^2 t  {2\sigma_z \hx^2 t-\sigma_y \hbar \omegam(\hx\hp+\hp\hx) \over \Delta^4} + o(g^3) ,
	\label{sigmaZT}
	\\
\hx^T & =&
	\hx 
	+2\hbar^2 g\omegam  {\sigma_x  \over \Delta^2}
	+2\hbar^3 g^2 {\sigma_z \hx t  \omegam \over \Delta^4} + o(g^3)  \,,
	\\
\hp^T  &=&
	\hp 
	-2 \hbar g t  {\sigma_y  \over \Delta^2}
	+2 \hbar^3 g^2 {\sigma_z \hp t  \omegam \over \Delta^4} + o(g^3)  \,.
\eeqa
\end{widetext}
The projection in the subspace of the first two-excited states can be readily calculated
by neglecting the quartic term of the Hamiltonian given by Eq.~\refe{MappedH}. 
This implies a scaling of the $\hx$ and $\hp$ operators by the factor $\xi=\xi_-$ defined by \refE{xiEq}: $\hx\rightarrow \xi \hx $ and $\hp \rightarrow \hp/\xi$.
The result at order 4 in $g$ gives that only 6 components are non-vanishing, out of the possible 16. 
These are given by \refE{tauSig} and \refE{tauX} in the main text. 
The expression for the $\beta$ coefficients is given in the 
main text \refe{beta1}-\refe{beta6} to order $g^2$.
\rem{
The explicit expressions for the $\beta$ coefficients to order 4 read 
\beqa
	\beta_1 &=&-1+4 g^2 {\omegam^2-t \omegam \xi^2+ t^2 \xi^4 \over \Delta^4 \xi^2}+g^4 \beta_{14}
 \label{beta1}
	\\
	\beta_2 &=&-2 g^2 {\omegam^2+t^2\xi^4  \over \Delta^2 \xi^2} +g^4 \beta_{24},
	\\
	\beta_3 &=&{2 g \omegam  \over \Delta^2 \xi}+g^3 \beta_{33}, 
	\\
	\beta_4 &=&{2 g  t  \xi   \over \Delta^2}+g^3 \beta_{43},
	\\
	\beta_5 &=&\xi- {2 g^2  t \omegam  \xi  \over  \Delta^4 }
	+g^4 \beta_{54}
	\\
	\beta_6 &=&{1\over \xi}-{2 g^2  t \omegam    \over \xi \Delta^2 }
	+g^4 \beta_{64}
\label{beta5}
\eeqa
}
From these expression one can see how the different degrees of freedom are 
mixed by the interaction. 
For instance, the displacement acquires a $\sigma_x$ component, which in this basis is 
the charge operator. On the other side the charge operator $\sigma_z$ 
acquires a component of the displacement operator. 
We give here the $g^3$ and $g^4$ terms (we use $\hbar=1$ in these expressions):
\begin{widetext}
\beqa
    \beta_{14} 
    &=&\frac{16 t \left(6 \omegam ^2 t^3 \left(9 \xi ^4+2\right) \xi ^4-4 \omegam ^3 t^2 \left(15
   \xi ^4+16\right) \xi ^2+9 \omegam ^4 t \left(3 \xi ^8+4 \xi ^4+8\right)-18 \omegam ^5
   \xi ^6+14 \omegam  t^4 \xi ^6-9 t^5 \xi ^8\right)}
   {3 \Delta^8 \Delta_3^2 \xi^4} 
   \label{beta14}
   \\
    \beta_{24} &=& \frac{16 t \left(-4 \omegam ^2 t^3 \left(9 \xi ^4+2\right) \xi ^4+2 \omegam ^3 t^2 \left(15
   \xi ^4+16\right) \xi ^2-6 \omegam ^4 t \left(3 \xi ^8+4 \xi ^4+8\right)+9 \omegam^5 \xi
   ^6-7 \omegam  t^4 \xi^6+6 t^5 \xi ^8\right)}{3 \Delta^8 \Delta_3^2 \xi^4}
   \\
   \beta_{33} &=&
   \frac{96 \omegam ^3 t^2 \left(\xi ^4+2\right)-32 \omegam  t^4 \xi ^4}{3 \Delta^6
   \Delta_3^2 \xi^3}
   \\
    \beta_{43} &=&
    \frac{8 \left(2 \omegam ^2 t^3 \left(9 \xi ^4+4\right)+9 \omegam ^4 t \xi ^4-3 t^5 \xi
   ^4\right)}{3 \Delta^6 \Delta_3^2 \xi }
   \\
      \beta_{54} &=&
  \frac{2 \omegam  t \left(3 \xi ^4 \left(-58 \omegam ^2 t^2-15 \omegam ^4+9 t^4\right)-64
   \omegam ^2 t^2-96 \omegam  t \xi ^2 (t-\omegam ) (\omegam +t)\right)}{3 \Delta^8
   \Delta_3^2 \xi }
   \\
     \beta_{64} &=&
     \frac{2 \omegam  t \left(\xi ^4 \left(-66 \omegam ^2 t^2-27 \omegam ^4+29 t^4\right)-192
   \omegam ^2 t^2+96 \omegam  t \xi ^2 (t-\omegam ) (\omegam +t)\right)}{3 \Delta^8
   \Delta^2 \xi ^3}
      \label{beta20}
   .
\eeqa
\end{widetext}

\section{Microwave cavity coupled to one and two qubits}
\label{Disp}

Let us consider a generic system coupled linearly through the operator $S$ to 
a microwave cavity. 
The Hamiltonian can be written as:
\beq
    H/\hbar=H_S/\hbar+\omega_c b^\dag b + S (b^\dag +b)
    ,
\eeq
where $b$ are the photon destruction operators, $\omega_c$ the cavity 
resonating angular velocity, $H_S$ the unspecified system Hamiltonian. 
We assume that $S$ acts only in the system Hilbert space. 
Let us also define the energy eigenvalues of $H_S$: 
$\hbar \epsilon_i$ with eigenstates 
$|i\rangle$ such that 
$H_S|i\rangle = \hbar \epsilon_i |i\rangle$.

Assuming that $S$ is small we 
find the modification of the eigenvalues and eigenvectors of the full system
by standard second-order perturbation theory. 
The unperturbed eigenvectors of the system plus cavity 
are $|im\rangle$ with eigenvalue $\varepsilon_{im}^{(0)}=\epsilon_i+m \omega_c$.
The first order correction vanishes. 
The second order reads:
\beq
    \varepsilon^{(2)}_{im}
    =
    \sum_j |S_{ij}|^2 
    \left[
    {m\over \epsilon_{ij}+\omega_c}
    +
    {m+1 \over \epsilon_{ij}-\omega_c}
    \right],
\eeq
 with $\epsilon_{ij}=\epsilon_i-\epsilon_j$.
The linear part in $m$ of this expression 
gives the renormalization of the resonator 
frequency. 
It normally depends on the system state $i$:
\beq
    \Delta \omega_i 
    = 
    \sum_j |S_{ij}|^2 {2 \epsilon_{ij} \over (\epsilon_{ij}^2-\omega_c^2)}
    .
    \label{EqXX}
\eeq
Thus the dispersive coupling $\chi$ [cf.~\refE{DispEq1}] 
defined as half the variation of the resonating frequency for 
a transition from the ground to the first excited state 
of the system is:
\beq
    \chi=(\Delta \omega_1-\Delta \omega_0)/2
    \label{ChiDef}
    .
\eeq

\subsection{Dispersive coupling for a single qubit}

As a simple example one can consider the case 
$H_S/\hbar = \epsilon_{10} \tau_z/2$ and 
$S = g_{\rm v} \tau_x$.
One finds 
$\Delta \omega_1=-\Delta \omega_0 =  
2 g_{\rm v}^2 \epsilon_{10}/(\epsilon_{10}^2-\omega_c^2)$.
For $\omega_c$ close to $\epsilon_{10}$ one then 
recovers the value of 
$\chi=g_{\rm v}^2/(\epsilon_{10}-\omega_c^2)$
entering \refE{DispEq1}.

Using \refE{EqXX} we can now find the dispersive 
coupling for the nanomechanical qubit.
We perform the unitary transformation given by $D_1 U_r$ and we use for $H_S$ the quadratic Hamiltonian $H_2$ given in \refE{H2Def}.
In this case the eigenvectors are $|n\sigma\rangle$ with eigenvalues $E_{n\sigma}=\hbar n \omega_\sigma+t_R\sigma/2$
 [here $t_R=t+\hbar (\omega_+-\omega_-)$ is the 
hopping amplitude renormalized by the zero point energies]. 
The system couples to the cavity through the charge and the displacement operators, but since the latter coupling is much smaller than the former, we consider in the following 
only the charge operator $\sigma_z$. 
We write the coupling operator in the new basis:
$ S = g_{\rm ec} D_1^\dag U_r^\dag  \sigma_z U_r D_1$.
At lowest order it reads (cf.~\refE{sigmaZT}):
\beq
    {S\over g_{\rm ec}}  
    =
    \sigma_x
    + {2 \hbar g t \over \Delta^2} 
    \left(\begin{array}{cc} (a_+ +a_+) \xi_+ & 0 \\ 0 & (a_- +a_-) \xi_- \end{array}\right)
    +\dots
    .
\eeq
Substituting $S$ into \refE{EqXX} and \refE{ChiDef} 
with the two lowest lying states $|0-\rangle$ and $|1-\rangle$, 
we find $\chi=\chi_{\rm m}+\chi_{\rm e}$ with
\beq
    \chi_{\rm m}= 2 g_{\rm ec}^2 \beta_{4,1}^2 
    {
    (\omega_{21}-\omega_{10})(\omega_c^2+\omega_{10}\omega_{21})
    \over
    (\omega_c^2-\omega_{21}^2)(\omega_c^2-\omega_{10}^2)
    }
    \label{chim}
\eeq
and 
\beq
    \chi_{\rm e} = g_{ec}^2 
    {(\delta_{11}-\delta_{00})
    (\omega_c^2+\delta_{11} \delta_{00}) 
    \over
    (\omega_c^2 -\delta_{11}^2)
    (\omega_c^2-\delta_{00}^2)
    },
    \label{chic}    
\eeq
where we recall that $\beta_{4,1}=2 \hbar gt \xi_-/\Delta^2$ and 
$\delta_{nm}=(E_{n+}-E_{n-})/\hbar$.
Note that the expression in \refE{chim} vanishes if
the lowest order approximation for the energy eigenvalues is used.
A non-linearity is needed in order to have a finite dispersive 
coupling. 
For this reason we do not specify the values 
of $\omega_{nm}$ and $\delta_{nm}$ for the moment.
Both expressions have a divergent behaviour:
$\chi_{\rm m}$ for $\omega_c$ close to either 
$\omega_{01}$ or $\omega_{21}$,
$\chi_{\rm e}$ for $\omega_c$ 
close to either $\delta_{00}$ or 
$\delta_{11}$.
This allows us to write the approximate 
Eqs.~\refe{chim2} and \refe{chic2} in the main text.

\subsection{Coupling two-qubits via the cavity}

We apply now this approach to study two nanomechanical qubits 
coupled to the same microwave cavity. 
Our main goal is to find the expression of a system operator $F$, acting only in 
the system Hilbert space, on the eigenvectors basis of the coupled system 
of the two qubits plus the microwave cavity. 
We are looking at the $m$-independent part, that gives the change of the operator in the 
system subspace.
Applying second order perturbation theory with the same notation of before we obtain
\begin{eqnarray}
    && 
    \left\langle i'm|F|  im \right\rangle
	=
	F_{i'i}
	+\sum_{k,l\neq i'} {S_{i'k} S_{kl}  F_{li} \over (\epsilon_{i'k}-\omega_c) \epsilon_{i'l} }
	\nonumber \\
	&&
	+\sum_{k,l\neq i} {F_{i'l } S_{lk} S_{ki} \over (\epsilon_{ik}-\omega_c)\epsilon_{il}}
	+\sum_{kk'} {S_{i'k} F_{kk'} S_{k'i} \over (\epsilon_{i'k}-\omega_c)(\epsilon_{ik'}-\omega_c)}
	\nonumber \\
	&&	
	-{F_{i'i}\over 2} \left[\sum_{j\neq i} {|S_{ij}|^2 \over (\epsilon_{ij}-\omega_c)^2}
	+ \sum_{j\neq i'} {|S_{i'j}|^2 \over (\epsilon_{i'j}-\omega_c)^2}\right] .
	\label{AnaJzx}
\end{eqnarray}
As a simple application we can consider a system composed of two pure 
two-level systems qubits: 
$H_S=\sum_{a=1,2}\hbar \epsilon_{10}^{(a)} \tau_z^{(a)}/2$,
with
$S=\sum_{a=1,2}  g_{\rm v}^{(a)} \tau_x^{(a)}$.
When a drive is applied to qubit 1 this can be modeled by 
a term in the Hamiltonian $ \hbar A \cos(\omega_D t) \tau_x^{(1)}$.
We thus look how $F=\tau_x$ reads in the Hamiltonian eigenvector basis.
Using \refE{AnaJzx} we find that 
\beq
    F=F_{x0} \tau_x^{(1)}+F_{0x} \tau_x^{(2)}+F_{zx} \tau_x^{(1)}\tau_x^{(2)} ,
    \label{Fform}
\eeq
with $F_{zx}$ given by the expression \refe{JzxTLS} for $J_{zx}$ 
with $\omega_{10}\rightarrow \epsilon_{10}$ and $A \rightarrow 1$. 

We consider now the case of a nanomechanical qubit. 
To evaluate \refE{AnaJzx} we use the same method applied for the single qubit. 
The coupling operator is now $S=\sum_{a=1,2} g_{\rm ec}^{(a)} \sigma_z^{(a)}$.
The eigenstates of the composite system can be labeled with the four indices
$\{n_1,\sigma_1;n_2 \sigma_2\}$ with eigenvalues 
$E_{n_1,\sigma_1}+E_{n_2,\sigma_2}$. 
As before we assume we have the exact expressions for the eigenvalues and we 
use the matrix elements given by the quadratic Hamiltonian. 
We look for the contributions leading to 
the operator $\tau_z^{(1)} \tau_x^{(2)}$.
We find that also in this case  
$F$ has the form of \refE{Fform}.

At lowest order in the electromechanical coupling constants these terms 
are generated by selecting the contribution of two $\sigma_x^{(1)}$ and one ${\hat x}^{(2)}$ 
operators entering the matrix elements of $F$ and $S$. 
They have dominant divergent terms in $1/(\omega_c-\omega_{10}^{(2)})$.
Collecting them one obtains:
\beq
    F_{zx}^{\rm e} 
    =  {g_{\rm ec}^{(1)} g_{\rm ec}^{(2)}  \beta_{4,1}^{(2)} \over \omega_{10}^{(2)}-\omega_c}
	\left[
	{ \delta_{11}^{(1)}
	\over
	{\delta_{11}^{(1)}}^2-{\omega_{10}^{(2)}}^2
	}
	-
	{
	\delta_{00}^{(1)}
	\over
	{\delta_{11}^{(1)}}^2-{\omega_{10}^{(2)}}^2
	}
	\right]
\eeq
that close to the resonance can be written as 
\beq
	F_{zx}^{\rm e} \approx 
	 -{g_{\rm ec}^{(1)}g_{\rm ec}^{(2)} \beta_{1,4}^{(2)} \over \omega_{10}^{(2)}-\omega_c}
	 {( \delta_{11}^{(1)}-\delta_{00}^{(1)}) 
	 ({\delta_{00}^{(1)}}^2 +{\omega_{10}^{(2)}}^2)
	 \over  
	({\delta_{00}^{(1)}}^2-{\omega_{10}^{(2)}}^2)^2
	}	
	\label{Fzxc}
	.
\eeq 
Even if this term appears to be a first order contribution in $g^{(2)}$, 
we know that the numerator is of order  ${g^{(1)}}^2$ 
[cf. \refE{deltaDif}].
We thus need to evaluate also the next order contributions 
in \refE{AnaJzx} that imply for the operators $F$ and $S$ 
two ${\hat x}^{(1)}$ and one ${\hat x}^{(2)}$ operators. 
These terms are of order $g^{(2)} { g^{(1)}}^2$.
Collecting the divergent contribution as before and 
evaluating it close to the divergence we have:
\beq
	F_{zx}^{\rm m} =  {8 g_{\rm ec}^{(1)} g_{\rm ec}^{(2)} 
			\over 
		\omega_c- \omega_{10}^{(2)}
}
{\beta_{1,4}^{(1)}}^2 {\beta_{1,4}^{(2)}}
	 {\omega_{10}^{(1)} - \omega_{21}^{(1)} \over 
	\left(\omega_{10}^{(2)}-\omega_{21}^{(1)}\right)
	\left(\omega_{10}^{(1)}-\omega_{10}^{(2)}\right)
	}.
\eeq
The two terms $F_{zx}^{\rm m}$ and $F_{zx}^{\rm c}$ can be combined in the form given 
by \refE{JzxFull} in the main text and written using the results obtained for 
the dispersive shifts $\chi_{\rm e}$ and $\chi_{\rm e}$ as defined in Eqs.~\refe{chim} and \refe{chic}.

\bibliography{FullBIB}

\begin{thebibliography}{60}%
\makeatletter
\providecommand \@ifxundefined [1]{%
 \@ifx{#1\undefined}
}%
\providecommand \@ifnum [1]{%
 \ifnum #1\expandafter \@firstoftwo
 \else \expandafter \@secondoftwo
 \fi
}%
\providecommand \@ifx [1]{%
 \ifx #1\expandafter \@firstoftwo
 \else \expandafter \@secondoftwo
 \fi
}%
\providecommand \natexlab [1]{#1}%
\providecommand \enquote  [1]{``#1''}%
\providecommand \bibnamefont  [1]{#1}%
\providecommand \bibfnamefont [1]{#1}%
\providecommand \citenamefont [1]{#1}%
\providecommand \href@noop [0]{\@secondoftwo}%
\providecommand \href [0]{\begingroup \@sanitize@url \@href}%
\providecommand \@href[1]{\@@startlink{#1}\@@href}%
\providecommand \@@href[1]{\endgroup#1\@@endlink}%
\providecommand \@sanitize@url [0]{\catcode `\\12\catcode `\$12\catcode
  `\&12\catcode `\#12\catcode `\^12\catcode `\_12\catcode `\%12\relax}%
\providecommand \@@startlink[1]{}%
\providecommand \@@endlink[0]{}%
\providecommand \url  [0]{\begingroup\@sanitize@url \@url }%
\providecommand \@url [1]{\endgroup\@href {#1}{\urlprefix }}%
\providecommand \urlprefix  [0]{URL }%
\providecommand \Eprint [0]{\href }%
\providecommand \doibase [0]{https://doi.org/}%
\providecommand \selectlanguage [0]{\@gobble}%
\providecommand \bibinfo  [0]{\@secondoftwo}%
\providecommand \bibfield  [0]{\@secondoftwo}%
\providecommand \translation [1]{[#1]}%
\providecommand \BibitemOpen [0]{}%
\providecommand \bibitemStop [0]{}%
\providecommand \bibitemNoStop [0]{.\EOS\space}%
\providecommand \EOS [0]{\spacefactor3000\relax}%
\providecommand \BibitemShut  [1]{\csname bibitem#1\endcsname}%
\let\auto@bib@innerbib\@empty
\bibitem [{\citenamefont {Barzanjeh}\ \emph {et~al.}(2011)\citenamefont
  {Barzanjeh}, \citenamefont {Vitali}, \citenamefont {Tombesi},\ and\
  \citenamefont {Milburn}}]{barzanjeh_entangling_2011}%
  \BibitemOpen
  \bibfield  {author} {\bibinfo {author} {\bibfnamefont {S.}~\bibnamefont
  {Barzanjeh}}, \bibinfo {author} {\bibfnamefont {D.}~\bibnamefont {Vitali}},
  \bibinfo {author} {\bibfnamefont {P.}~\bibnamefont {Tombesi}},\ and\ \bibinfo
  {author} {\bibfnamefont {G.~J.}\ \bibnamefont {Milburn}},\ }\bibfield
  {title} {\bibinfo {title} {Entangling optical and microwave cavity modes by
  means of a nanomechanical resonator},\ }\href
  {https://doi.org/10.1103/PhysRevA.84.042342} {\bibfield  {journal} {\bibinfo
  {journal} {Phys. Rev. A}\ }\textbf {\bibinfo {volume} {84}},\ \bibinfo
  {pages} {042342} (\bibinfo {year} {2011})}\BibitemShut {NoStop}%
\bibitem [{\citenamefont {Palomaki}\ \emph {et~al.}(2013)\citenamefont
  {Palomaki}, \citenamefont {Harlow}, \citenamefont {Teufel}, \citenamefont
  {Simmonds},\ and\ \citenamefont {Lehnert}}]{palomaki_coherent_2013}%
  \BibitemOpen
  \bibfield  {author} {\bibinfo {author} {\bibfnamefont {T.~A.}\ \bibnamefont
  {Palomaki}}, \bibinfo {author} {\bibfnamefont {J.~W.}\ \bibnamefont
  {Harlow}}, \bibinfo {author} {\bibfnamefont {J.~D.}\ \bibnamefont {Teufel}},
  \bibinfo {author} {\bibfnamefont {R.~W.}\ \bibnamefont {Simmonds}},\ and\
  \bibinfo {author} {\bibfnamefont {K.~W.}\ \bibnamefont {Lehnert}},\
  }\bibfield  {title} {\bibinfo {title} {Coherent state transfer between
  itinerant microwave fields and a mechanical oscillator},\ }\href
  {https://doi.org/10.1038/nature11915} {\bibfield  {journal} {\bibinfo
  {journal} {Nature}\ }\textbf {\bibinfo {volume} {495}},\ \bibinfo {pages}
  {210} (\bibinfo {year} {2013})}\BibitemShut {NoStop}%
\bibitem [{\citenamefont {Andrews}\ \emph {et~al.}(2014)\citenamefont
  {Andrews}, \citenamefont {Peterson}, \citenamefont {Purdy}, \citenamefont
  {Cicak}, \citenamefont {Simmonds}, \citenamefont {Regal},\ and\ \citenamefont
  {Lehnert}}]{andrews_bidirectional_2014}%
  \BibitemOpen
  \bibfield  {author} {\bibinfo {author} {\bibfnamefont {R.~W.}\ \bibnamefont
  {Andrews}}, \bibinfo {author} {\bibfnamefont {R.~W.}\ \bibnamefont
  {Peterson}}, \bibinfo {author} {\bibfnamefont {T.~P.}\ \bibnamefont {Purdy}},
  \bibinfo {author} {\bibfnamefont {K.}~\bibnamefont {Cicak}}, \bibinfo
  {author} {\bibfnamefont {R.~W.}\ \bibnamefont {Simmonds}}, \bibinfo {author}
  {\bibfnamefont {C.~A.}\ \bibnamefont {Regal}},\ and\ \bibinfo {author}
  {\bibfnamefont {K.~W.}\ \bibnamefont {Lehnert}},\ }\bibfield  {title}
  {\bibinfo {title} {Bidirectional and efficient conversion between microwave
  and optical light},\ }\href {https://doi.org/10.1038/nphys2911} {\bibfield
  {journal} {\bibinfo  {journal} {Nat. Phys.}\ }\textbf {\bibinfo {volume}
  {10}},\ \bibinfo {pages} {321} (\bibinfo {year} {2014})}\BibitemShut
  {NoStop}%
\bibitem [{\citenamefont {Lecocq}\ \emph {et~al.}(2016)\citenamefont {Lecocq},
  \citenamefont {Clark}, \citenamefont {Simmonds}, \citenamefont {Aumentado},\
  and\ \citenamefont {Teufel}}]{lecocq_mechanically_2016}%
  \BibitemOpen
  \bibfield  {author} {\bibinfo {author} {\bibfnamefont {F.}~\bibnamefont
  {Lecocq}}, \bibinfo {author} {\bibfnamefont {J.~B.}\ \bibnamefont {Clark}},
  \bibinfo {author} {\bibfnamefont {R.~W.}\ \bibnamefont {Simmonds}}, \bibinfo
  {author} {\bibfnamefont {J.}~\bibnamefont {Aumentado}},\ and\ \bibinfo
  {author} {\bibfnamefont {J.~D.}\ \bibnamefont {Teufel}},\ }\bibfield  {title}
  {\bibinfo {title} {Mechanically {{Mediated Microwave Frequency Conversion}}
  in the {{Quantum Regime}}},\ }\href
  {https://doi.org/10.1103/PhysRevLett.116.043601} {\bibfield  {journal}
  {\bibinfo  {journal} {Phys. Rev. Lett.}\ }\textbf {\bibinfo {volume} {116}},\
  \bibinfo {pages} {043601} (\bibinfo {year} {2016})}\BibitemShut {NoStop}%
\bibitem [{\citenamefont {Vainsencher}\ \emph {et~al.}(2016)\citenamefont
  {Vainsencher}, \citenamefont {Satzinger}, \citenamefont {Peairs},\ and\
  \citenamefont {Cleland}}]{vainsencher_bi-directional_2016}%
  \BibitemOpen
  \bibfield  {author} {\bibinfo {author} {\bibfnamefont {A.}~\bibnamefont
  {Vainsencher}}, \bibinfo {author} {\bibfnamefont {K.~J.}\ \bibnamefont
  {Satzinger}}, \bibinfo {author} {\bibfnamefont {G.~A.}\ \bibnamefont
  {Peairs}},\ and\ \bibinfo {author} {\bibfnamefont {A.~N.}\ \bibnamefont
  {Cleland}},\ }\bibfield  {title} {\bibinfo {title} {Bi-directional conversion
  between microwave and optical frequencies in a piezoelectric optomechanical
  device},\ }\href {https://doi.org/10.1063/1.4955408} {\bibfield  {journal}
  {\bibinfo  {journal} {Appl. Phys. Lett.}\ }\textbf {\bibinfo {volume}
  {109}},\ \bibinfo {pages} {033107} (\bibinfo {year} {2016})}\BibitemShut
  {NoStop}%
\bibitem [{\citenamefont {Bochmann}\ \emph {et~al.}(2013)\citenamefont
  {Bochmann}, \citenamefont {Vainsencher}, \citenamefont {Awschalom},\ and\
  \citenamefont {Cleland}}]{bochmann_nanomechanical_2013}%
  \BibitemOpen
  \bibfield  {author} {\bibinfo {author} {\bibfnamefont {J.}~\bibnamefont
  {Bochmann}}, \bibinfo {author} {\bibfnamefont {A.}~\bibnamefont
  {Vainsencher}}, \bibinfo {author} {\bibfnamefont {D.~D.}\ \bibnamefont
  {Awschalom}},\ and\ \bibinfo {author} {\bibfnamefont {A.~N.}\ \bibnamefont
  {Cleland}},\ }\bibfield  {title} {\bibinfo {title} {Nanomechanical coupling
  between microwave and optical photons},\ }\href
  {https://doi.org/10.1038/nphys2748} {\bibfield  {journal} {\bibinfo
  {journal} {Nat. Phys.}\ }\textbf {\bibinfo {volume} {9}},\ \bibinfo {pages}
  {712} (\bibinfo {year} {2013})}\BibitemShut {NoStop}%
\bibitem [{\citenamefont {{Ockeloen-Korppi}}\ \emph {et~al.}(2016)\citenamefont
  {{Ockeloen-Korppi}}, \citenamefont {Damsk{\"a}gg}, \citenamefont
  {Pirkkalainen}, \citenamefont {Clerk}, \citenamefont {Woolley},\ and\
  \citenamefont {Sillanp{\"a}{\"a}}}]{ockeloen-korppi_quantum_2016}%
  \BibitemOpen
  \bibfield  {author} {\bibinfo {author} {\bibfnamefont {C.~F.}\ \bibnamefont
  {{Ockeloen-Korppi}}}, \bibinfo {author} {\bibfnamefont {E.}~\bibnamefont
  {Damsk{\"a}gg}}, \bibinfo {author} {\bibfnamefont {J.-M.}\ \bibnamefont
  {Pirkkalainen}}, \bibinfo {author} {\bibfnamefont {A.~A.}\ \bibnamefont
  {Clerk}}, \bibinfo {author} {\bibfnamefont {M.~J.}\ \bibnamefont {Woolley}},\
  and\ \bibinfo {author} {\bibfnamefont {M.~A.}\ \bibnamefont
  {Sillanp{\"a}{\"a}}},\ }\bibfield  {title} {\bibinfo {title} {Quantum
  {{Backaction Evading Measurement}} of {{Collective Mechanical Modes}}},\
  }\href {https://doi.org/10.1103/PhysRevLett.117.140401} {\bibfield  {journal}
  {\bibinfo  {journal} {Phys. Rev. Lett.}\ }\textbf {\bibinfo {volume} {117}},\
  \bibinfo {pages} {140401} (\bibinfo {year} {2016})}\BibitemShut {NoStop}%
\bibitem [{\citenamefont {Rabl}\ \emph {et~al.}(2010)\citenamefont {Rabl},
  \citenamefont {Kolkowitz}, \citenamefont {Koppens}, \citenamefont {Harris},
  \citenamefont {Zoller},\ and\ \citenamefont {Lukin}}]{rabl_quantum_2010}%
  \BibitemOpen
  \bibfield  {author} {\bibinfo {author} {\bibfnamefont {P.}~\bibnamefont
  {Rabl}}, \bibinfo {author} {\bibfnamefont {S.~J.}\ \bibnamefont {Kolkowitz}},
  \bibinfo {author} {\bibfnamefont {F.~H.~L.}\ \bibnamefont {Koppens}},
  \bibinfo {author} {\bibfnamefont {J.~G.~E.}\ \bibnamefont {Harris}}, \bibinfo
  {author} {\bibfnamefont {P.}~\bibnamefont {Zoller}},\ and\ \bibinfo {author}
  {\bibfnamefont {M.~D.}\ \bibnamefont {Lukin}},\ }\bibfield  {title} {\bibinfo
  {title} {A quantum spin transducer based on nanoelectromechanical resonator
  arrays},\ }\href {https://doi.org/10.1038/nphys1679} {\bibfield  {journal}
  {\bibinfo  {journal} {Nat. Phys.}\ }\textbf {\bibinfo {volume} {6}},\
  \bibinfo {pages} {602} (\bibinfo {year} {2010})}\BibitemShut {NoStop}%
\bibitem [{\citenamefont {Stannigel}\ \emph {et~al.}(2010)\citenamefont
  {Stannigel}, \citenamefont {Rabl}, \citenamefont {S{\o}rensen}, \citenamefont
  {Zoller},\ and\ \citenamefont {Lukin}}]{stannigel_optomechanical_2010}%
  \BibitemOpen
  \bibfield  {author} {\bibinfo {author} {\bibfnamefont {K.}~\bibnamefont
  {Stannigel}}, \bibinfo {author} {\bibfnamefont {P.}~\bibnamefont {Rabl}},
  \bibinfo {author} {\bibfnamefont {A.~S.}\ \bibnamefont {S{\o}rensen}},
  \bibinfo {author} {\bibfnamefont {P.}~\bibnamefont {Zoller}},\ and\ \bibinfo
  {author} {\bibfnamefont {M.~D.}\ \bibnamefont {Lukin}},\ }\bibfield  {title}
  {\bibinfo {title} {Optomechanical {{Transducers}} for {{Long}}-{{Distance
  Quantum Communication}}},\ }\href
  {https://doi.org/10.1103/PhysRevLett.105.220501} {\bibfield  {journal}
  {\bibinfo  {journal} {Phys. Rev. Lett.}\ }\textbf {\bibinfo {volume} {105}},\
  \bibinfo {pages} {220501} (\bibinfo {year} {2010})}\BibitemShut {NoStop}%
\bibitem [{\citenamefont {Satzinger}\ \emph {et~al.}(2018)\citenamefont
  {Satzinger}, \citenamefont {Zhong}, \citenamefont {Chang}, \citenamefont
  {Peairs}, \citenamefont {Bienfait}, \citenamefont {Chou}, \citenamefont
  {Cleland}, \citenamefont {Conner}, \citenamefont {Dumur}, \citenamefont
  {Grebel}, \citenamefont {Gutierrez}, \citenamefont {November}, \citenamefont
  {Povey}, \citenamefont {Whiteley}, \citenamefont {Awschalom}, \citenamefont
  {Schuster},\ and\ \citenamefont {Cleland}}]{satzinger_quantum_2018}%
  \BibitemOpen
  \bibfield  {author} {\bibinfo {author} {\bibfnamefont {K.~J.}\ \bibnamefont
  {Satzinger}}, \bibinfo {author} {\bibfnamefont {Y.~P.}\ \bibnamefont
  {Zhong}}, \bibinfo {author} {\bibfnamefont {H.-S.}\ \bibnamefont {Chang}},
  \bibinfo {author} {\bibfnamefont {G.~A.}\ \bibnamefont {Peairs}}, \bibinfo
  {author} {\bibfnamefont {A.}~\bibnamefont {Bienfait}}, \bibinfo {author}
  {\bibfnamefont {M.-H.}\ \bibnamefont {Chou}}, \bibinfo {author}
  {\bibfnamefont {A.~Y.}\ \bibnamefont {Cleland}}, \bibinfo {author}
  {\bibfnamefont {C.~R.}\ \bibnamefont {Conner}}, \bibinfo {author}
  {\bibfnamefont {{\'E}.}~\bibnamefont {Dumur}}, \bibinfo {author}
  {\bibfnamefont {J.}~\bibnamefont {Grebel}}, \bibinfo {author} {\bibfnamefont
  {I.}~\bibnamefont {Gutierrez}}, \bibinfo {author} {\bibfnamefont {B.~H.}\
  \bibnamefont {November}}, \bibinfo {author} {\bibfnamefont {R.~G.}\
  \bibnamefont {Povey}}, \bibinfo {author} {\bibfnamefont {S.~J.}\ \bibnamefont
  {Whiteley}}, \bibinfo {author} {\bibfnamefont {D.~D.}\ \bibnamefont
  {Awschalom}}, \bibinfo {author} {\bibfnamefont {D.~I.}\ \bibnamefont
  {Schuster}},\ and\ \bibinfo {author} {\bibfnamefont {A.~N.}\ \bibnamefont
  {Cleland}},\ }\bibfield  {title} {\bibinfo {title} {Quantum control of
  surface acoustic-wave phonons},\ }\href
  {https://doi.org/10.1038/s41586-018-0719-5} {\bibfield  {journal} {\bibinfo
  {journal} {Nature}\ }\textbf {\bibinfo {volume} {563}},\ \bibinfo {pages}
  {661} (\bibinfo {year} {2018})}\BibitemShut {NoStop}%
\bibitem [{\citenamefont {Bienfait}\ \emph {et~al.}(2019)\citenamefont
  {Bienfait}, \citenamefont {Satzinger}, \citenamefont {Zhong}, \citenamefont
  {Chang}, \citenamefont {Chou}, \citenamefont {Conner}, \citenamefont {Dumur},
  \citenamefont {Grebel}, \citenamefont {Peairs}, \citenamefont {Povey},\ and\
  \citenamefont {Cleland}}]{bienfait_phonon-mediated_2019}%
  \BibitemOpen
  \bibfield  {author} {\bibinfo {author} {\bibfnamefont {A.}~\bibnamefont
  {Bienfait}}, \bibinfo {author} {\bibfnamefont {K.~J.}\ \bibnamefont
  {Satzinger}}, \bibinfo {author} {\bibfnamefont {Y.~P.}\ \bibnamefont
  {Zhong}}, \bibinfo {author} {\bibfnamefont {H.-S.}\ \bibnamefont {Chang}},
  \bibinfo {author} {\bibfnamefont {M.-H.}\ \bibnamefont {Chou}}, \bibinfo
  {author} {\bibfnamefont {C.~R.}\ \bibnamefont {Conner}}, \bibinfo {author}
  {\bibfnamefont {{\'E}.}~\bibnamefont {Dumur}}, \bibinfo {author}
  {\bibfnamefont {J.}~\bibnamefont {Grebel}}, \bibinfo {author} {\bibfnamefont
  {G.~A.}\ \bibnamefont {Peairs}}, \bibinfo {author} {\bibfnamefont {R.~G.}\
  \bibnamefont {Povey}},\ and\ \bibinfo {author} {\bibfnamefont {A.~N.}\
  \bibnamefont {Cleland}},\ }\bibfield  {title} {\bibinfo {title}
  {Phonon-mediated quantum state transfer and remote qubit entanglement},\
  }\href {https://doi.org/10.1126/science.aaw8415} {\bibfield  {journal}
  {\bibinfo  {journal} {Science}\ }\textbf {\bibinfo {volume} {364}},\ \bibinfo
  {pages} {368} (\bibinfo {year} {2019})}\BibitemShut {NoStop}%
\bibitem [{\citenamefont {Bienfait}\ \emph {et~al.}(2020)\citenamefont
  {Bienfait}, \citenamefont {Zhong}, \citenamefont {Chang}, \citenamefont
  {Chou}, \citenamefont {Conner}, \citenamefont {Dumur}, \citenamefont
  {Grebel}, \citenamefont {Peairs}, \citenamefont {Povey}, \citenamefont
  {Satzinger},\ and\ \citenamefont {Cleland}}]{bienfait_quantum_2020}%
  \BibitemOpen
  \bibfield  {author} {\bibinfo {author} {\bibfnamefont {A.}~\bibnamefont
  {Bienfait}}, \bibinfo {author} {\bibfnamefont {Y.~P.}\ \bibnamefont {Zhong}},
  \bibinfo {author} {\bibfnamefont {H.-S.}\ \bibnamefont {Chang}}, \bibinfo
  {author} {\bibfnamefont {M.-H.}\ \bibnamefont {Chou}}, \bibinfo {author}
  {\bibfnamefont {C.~R.}\ \bibnamefont {Conner}}, \bibinfo {author}
  {\bibfnamefont {{\'E}.}~\bibnamefont {Dumur}}, \bibinfo {author}
  {\bibfnamefont {J.}~\bibnamefont {Grebel}}, \bibinfo {author} {\bibfnamefont
  {G.~A.}\ \bibnamefont {Peairs}}, \bibinfo {author} {\bibfnamefont {R.~G.}\
  \bibnamefont {Povey}}, \bibinfo {author} {\bibfnamefont {K.~J.}\ \bibnamefont
  {Satzinger}},\ and\ \bibinfo {author} {\bibfnamefont {A.~N.}\ \bibnamefont
  {Cleland}},\ }\bibfield  {title} {\bibinfo {title} {Quantum {{Erasure Using
  Entangled Surface Acoustic Phonons}}},\ }\href
  {https://doi.org/10.1103/PhysRevX.10.021055} {\bibfield  {journal} {\bibinfo
  {journal} {Phys. Rev. X}\ }\textbf {\bibinfo {volume} {10}},\ \bibinfo
  {pages} {021055} (\bibinfo {year} {2020})}\BibitemShut {NoStop}%
\bibitem [{\citenamefont {Degen}\ \emph {et~al.}(2017)\citenamefont {Degen},
  \citenamefont {Reinhard},\ and\ \citenamefont
  {Cappellaro}}]{degen_quantum_2017}%
  \BibitemOpen
  \bibfield  {author} {\bibinfo {author} {\bibfnamefont {C.~L.}\ \bibnamefont
  {Degen}}, \bibinfo {author} {\bibfnamefont {F.}~\bibnamefont {Reinhard}},\
  and\ \bibinfo {author} {\bibfnamefont {P.}~\bibnamefont {Cappellaro}},\
  }\bibfield  {title} {\bibinfo {title} {Quantum sensing},\ }\bibfield
  {journal} {\bibinfo  {journal} {Rev. Mod. Phys.}\ }\textbf {\bibinfo {volume}
  {89}},\ \href {https://doi.org/10.1103/RevModPhys.89.035002}
  {10.1103/RevModPhys.89.035002} (\bibinfo {year} {2017})\BibitemShut {NoStop}%
\bibitem [{\citenamefont {Urgell}\ \emph {et~al.}(2020)\citenamefont {Urgell},
  \citenamefont {Yang}, \citenamefont {De~Bonis}, \citenamefont {Samanta},
  \citenamefont {Esplandiu}, \citenamefont {Dong}, \citenamefont {Jin},\ and\
  \citenamefont {Bachtold}}]{urgell_cooling_2020}%
  \BibitemOpen
  \bibfield  {author} {\bibinfo {author} {\bibfnamefont {C.}~\bibnamefont
  {Urgell}}, \bibinfo {author} {\bibfnamefont {W.}~\bibnamefont {Yang}},
  \bibinfo {author} {\bibfnamefont {S.~L.}\ \bibnamefont {De~Bonis}}, \bibinfo
  {author} {\bibfnamefont {C.}~\bibnamefont {Samanta}}, \bibinfo {author}
  {\bibfnamefont {M.~J.}\ \bibnamefont {Esplandiu}}, \bibinfo {author}
  {\bibfnamefont {Q.}~\bibnamefont {Dong}}, \bibinfo {author} {\bibfnamefont
  {Y.}~\bibnamefont {Jin}},\ and\ \bibinfo {author} {\bibfnamefont
  {A.}~\bibnamefont {Bachtold}},\ }\bibfield  {title} {\bibinfo {title}
  {Cooling and self-oscillation in a nanotube electromechanical resonator},\
  }\href {https://doi.org/10.1038/s41567-019-0682-6} {\bibfield  {journal}
  {\bibinfo  {journal} {Nat. Phys.}\ }\textbf {\bibinfo {volume} {16}},\
  \bibinfo {pages} {32} (\bibinfo {year} {2020})}\BibitemShut {NoStop}%
\bibitem [{\citenamefont {MacCabe}\ \emph {et~al.}(2020)\citenamefont
  {MacCabe}, \citenamefont {Ren}, \citenamefont {Luo}, \citenamefont {Cohen},
  \citenamefont {Zhou}, \citenamefont {Sipahigil}, \citenamefont
  {Mirhosseini},\ and\ \citenamefont {Painter}}]{maccabe_nano-acoustic_2020}%
  \BibitemOpen
  \bibfield  {author} {\bibinfo {author} {\bibfnamefont {G.~S.}\ \bibnamefont
  {MacCabe}}, \bibinfo {author} {\bibfnamefont {H.}~\bibnamefont {Ren}},
  \bibinfo {author} {\bibfnamefont {J.}~\bibnamefont {Luo}}, \bibinfo {author}
  {\bibfnamefont {J.~D.}\ \bibnamefont {Cohen}}, \bibinfo {author}
  {\bibfnamefont {H.}~\bibnamefont {Zhou}}, \bibinfo {author} {\bibfnamefont
  {A.}~\bibnamefont {Sipahigil}}, \bibinfo {author} {\bibfnamefont
  {M.}~\bibnamefont {Mirhosseini}},\ and\ \bibinfo {author} {\bibfnamefont
  {O.}~\bibnamefont {Painter}},\ }\bibfield  {title} {\bibinfo {title}
  {Nano-acoustic resonator with ultralong phonon lifetime},\ }\href
  {https://doi.org/10.1126/science.abc7312} {\bibfield  {journal} {\bibinfo
  {journal} {Science}\ }\textbf {\bibinfo {volume} {370}},\ \bibinfo {pages}
  {840} (\bibinfo {year} {2020})}\BibitemShut {NoStop}%
\bibitem [{\citenamefont {Arute}\ \emph {et~al.}(2019)\citenamefont {Arute},
  \citenamefont {Arya}, \citenamefont {Babbush}, \citenamefont {Bacon},
  \citenamefont {Bardin}, \citenamefont {Barends}, \citenamefont {Biswas},
  \citenamefont {Boixo}, \citenamefont {Brandao}, \citenamefont {Buell},
  \citenamefont {Burkett}, \citenamefont {Chen}, \citenamefont {Chen},
  \citenamefont {Chiaro}, \citenamefont {Collins}, \citenamefont {Courtney},
  \citenamefont {Dunsworth}, \citenamefont {Farhi}, \citenamefont {Foxen},
  \citenamefont {Fowler}, \citenamefont {Gidney}, \citenamefont {Giustina},
  \citenamefont {Graff}, \citenamefont {Guerin}, \citenamefont {Habegger},
  \citenamefont {Harrigan}, \citenamefont {Hartmann}, \citenamefont {Ho},
  \citenamefont {Hoffmann}, \citenamefont {Huang}, \citenamefont {Humble},
  \citenamefont {Isakov}, \citenamefont {Jeffrey}, \citenamefont {Jiang},
  \citenamefont {Kafri}, \citenamefont {Kechedzhi}, \citenamefont {Kelly},
  \citenamefont {Klimov}, \citenamefont {Knysh}, \citenamefont {Korotkov},
  \citenamefont {Kostritsa}, \citenamefont {Landhuis}, \citenamefont
  {Lindmark}, \citenamefont {Lucero}, \citenamefont {Lyakh}, \citenamefont
  {Mandr{\`a}}, \citenamefont {McClean}, \citenamefont {McEwen}, \citenamefont
  {Megrant}, \citenamefont {Mi}, \citenamefont {Michielsen}, \citenamefont
  {Mohseni}, \citenamefont {Mutus}, \citenamefont {Naaman}, \citenamefont
  {Neeley}, \citenamefont {Neill}, \citenamefont {Niu}, \citenamefont {Ostby},
  \citenamefont {Petukhov}, \citenamefont {Platt}, \citenamefont {Quintana},
  \citenamefont {Rieffel}, \citenamefont {Roushan}, \citenamefont {Rubin},
  \citenamefont {Sank}, \citenamefont {Satzinger}, \citenamefont {Smelyanskiy},
  \citenamefont {Sung}, \citenamefont {Trevithick}, \citenamefont
  {Vainsencher}, \citenamefont {Villalonga}, \citenamefont {White},
  \citenamefont {Yao}, \citenamefont {Yeh}, \citenamefont {Zalcman},
  \citenamefont {Neven},\ and\ \citenamefont {Martinis}}]{arute_quantum_2019}%
  \BibitemOpen
  \bibfield  {author} {\bibinfo {author} {\bibfnamefont {F.}~\bibnamefont
  {Arute}}, \bibinfo {author} {\bibfnamefont {K.}~\bibnamefont {Arya}},
  \bibinfo {author} {\bibfnamefont {R.}~\bibnamefont {Babbush}}, \bibinfo
  {author} {\bibfnamefont {D.}~\bibnamefont {Bacon}}, \bibinfo {author}
  {\bibfnamefont {J.~C.}\ \bibnamefont {Bardin}}, \bibinfo {author}
  {\bibfnamefont {R.}~\bibnamefont {Barends}}, \bibinfo {author} {\bibfnamefont
  {R.}~\bibnamefont {Biswas}}, \bibinfo {author} {\bibfnamefont
  {S.}~\bibnamefont {Boixo}}, \bibinfo {author} {\bibfnamefont {F.~G. S.~L.}\
  \bibnamefont {Brandao}}, \bibinfo {author} {\bibfnamefont {D.~A.}\
  \bibnamefont {Buell}}, \bibinfo {author} {\bibfnamefont {B.}~\bibnamefont
  {Burkett}}, \bibinfo {author} {\bibfnamefont {Y.}~\bibnamefont {Chen}},
  \bibinfo {author} {\bibfnamefont {Z.}~\bibnamefont {Chen}}, \bibinfo {author}
  {\bibfnamefont {B.}~\bibnamefont {Chiaro}}, \bibinfo {author} {\bibfnamefont
  {R.}~\bibnamefont {Collins}}, \bibinfo {author} {\bibfnamefont
  {W.}~\bibnamefont {Courtney}}, \bibinfo {author} {\bibfnamefont
  {A.}~\bibnamefont {Dunsworth}}, \bibinfo {author} {\bibfnamefont
  {E.}~\bibnamefont {Farhi}}, \bibinfo {author} {\bibfnamefont
  {B.}~\bibnamefont {Foxen}}, \bibinfo {author} {\bibfnamefont
  {A.}~\bibnamefont {Fowler}}, \bibinfo {author} {\bibfnamefont
  {C.}~\bibnamefont {Gidney}}, \bibinfo {author} {\bibfnamefont
  {M.}~\bibnamefont {Giustina}}, \bibinfo {author} {\bibfnamefont
  {R.}~\bibnamefont {Graff}}, \bibinfo {author} {\bibfnamefont
  {K.}~\bibnamefont {Guerin}}, \bibinfo {author} {\bibfnamefont
  {S.}~\bibnamefont {Habegger}}, \bibinfo {author} {\bibfnamefont {M.~P.}\
  \bibnamefont {Harrigan}}, \bibinfo {author} {\bibfnamefont {M.~J.}\
  \bibnamefont {Hartmann}}, \bibinfo {author} {\bibfnamefont {A.}~\bibnamefont
  {Ho}}, \bibinfo {author} {\bibfnamefont {M.}~\bibnamefont {Hoffmann}},
  \bibinfo {author} {\bibfnamefont {T.}~\bibnamefont {Huang}}, \bibinfo
  {author} {\bibfnamefont {T.~S.}\ \bibnamefont {Humble}}, \bibinfo {author}
  {\bibfnamefont {S.~V.}\ \bibnamefont {Isakov}}, \bibinfo {author}
  {\bibfnamefont {E.}~\bibnamefont {Jeffrey}}, \bibinfo {author} {\bibfnamefont
  {Z.}~\bibnamefont {Jiang}}, \bibinfo {author} {\bibfnamefont
  {D.}~\bibnamefont {Kafri}}, \bibinfo {author} {\bibfnamefont
  {K.}~\bibnamefont {Kechedzhi}}, \bibinfo {author} {\bibfnamefont
  {J.}~\bibnamefont {Kelly}}, \bibinfo {author} {\bibfnamefont {P.~V.}\
  \bibnamefont {Klimov}}, \bibinfo {author} {\bibfnamefont {S.}~\bibnamefont
  {Knysh}}, \bibinfo {author} {\bibfnamefont {A.}~\bibnamefont {Korotkov}},
  \bibinfo {author} {\bibfnamefont {F.}~\bibnamefont {Kostritsa}}, \bibinfo
  {author} {\bibfnamefont {D.}~\bibnamefont {Landhuis}}, \bibinfo {author}
  {\bibfnamefont {M.}~\bibnamefont {Lindmark}}, \bibinfo {author}
  {\bibfnamefont {E.}~\bibnamefont {Lucero}}, \bibinfo {author} {\bibfnamefont
  {D.}~\bibnamefont {Lyakh}}, \bibinfo {author} {\bibfnamefont
  {S.}~\bibnamefont {Mandr{\`a}}}, \bibinfo {author} {\bibfnamefont {J.~R.}\
  \bibnamefont {McClean}}, \bibinfo {author} {\bibfnamefont {M.}~\bibnamefont
  {McEwen}}, \bibinfo {author} {\bibfnamefont {A.}~\bibnamefont {Megrant}},
  \bibinfo {author} {\bibfnamefont {X.}~\bibnamefont {Mi}}, \bibinfo {author}
  {\bibfnamefont {K.}~\bibnamefont {Michielsen}}, \bibinfo {author}
  {\bibfnamefont {M.}~\bibnamefont {Mohseni}}, \bibinfo {author} {\bibfnamefont
  {J.}~\bibnamefont {Mutus}}, \bibinfo {author} {\bibfnamefont
  {O.}~\bibnamefont {Naaman}}, \bibinfo {author} {\bibfnamefont
  {M.}~\bibnamefont {Neeley}}, \bibinfo {author} {\bibfnamefont
  {C.}~\bibnamefont {Neill}}, \bibinfo {author} {\bibfnamefont {M.~Y.}\
  \bibnamefont {Niu}}, \bibinfo {author} {\bibfnamefont {E.}~\bibnamefont
  {Ostby}}, \bibinfo {author} {\bibfnamefont {A.}~\bibnamefont {Petukhov}},
  \bibinfo {author} {\bibfnamefont {J.~C.}\ \bibnamefont {Platt}}, \bibinfo
  {author} {\bibfnamefont {C.}~\bibnamefont {Quintana}}, \bibinfo {author}
  {\bibfnamefont {E.~G.}\ \bibnamefont {Rieffel}}, \bibinfo {author}
  {\bibfnamefont {P.}~\bibnamefont {Roushan}}, \bibinfo {author} {\bibfnamefont
  {N.~C.}\ \bibnamefont {Rubin}}, \bibinfo {author} {\bibfnamefont
  {D.}~\bibnamefont {Sank}}, \bibinfo {author} {\bibfnamefont {K.~J.}\
  \bibnamefont {Satzinger}}, \bibinfo {author} {\bibfnamefont {V.}~\bibnamefont
  {Smelyanskiy}}, \bibinfo {author} {\bibfnamefont {K.~J.}\ \bibnamefont
  {Sung}}, \bibinfo {author} {\bibfnamefont {M.~D.}\ \bibnamefont
  {Trevithick}}, \bibinfo {author} {\bibfnamefont {A.}~\bibnamefont
  {Vainsencher}}, \bibinfo {author} {\bibfnamefont {B.}~\bibnamefont
  {Villalonga}}, \bibinfo {author} {\bibfnamefont {T.}~\bibnamefont {White}},
  \bibinfo {author} {\bibfnamefont {Z.~J.}\ \bibnamefont {Yao}}, \bibinfo
  {author} {\bibfnamefont {P.}~\bibnamefont {Yeh}}, \bibinfo {author}
  {\bibfnamefont {A.}~\bibnamefont {Zalcman}}, \bibinfo {author} {\bibfnamefont
  {H.}~\bibnamefont {Neven}},\ and\ \bibinfo {author} {\bibfnamefont {J.~M.}\
  \bibnamefont {Martinis}},\ }\bibfield  {title} {\bibinfo {title} {Quantum
  supremacy using a programmable superconducting processor},\ }\href
  {https://doi.org/10.1038/s41586-019-1666-5} {\bibfield  {journal} {\bibinfo
  {journal} {Nature}\ }\textbf {\bibinfo {volume} {574}},\ \bibinfo {pages}
  {505} (\bibinfo {year} {2019})}\BibitemShut {NoStop}%
\bibitem [{\citenamefont {Rigetti}\ \emph {et~al.}(2012)\citenamefont
  {Rigetti}, \citenamefont {Gambetta}, \citenamefont {Poletto}, \citenamefont
  {Plourde}, \citenamefont {Chow}, \citenamefont {C{\'o}rcoles}, \citenamefont
  {Smolin}, \citenamefont {Merkel}, \citenamefont {Rozen}, \citenamefont
  {Keefe}, \citenamefont {Rothwell}, \citenamefont {Ketchen},\ and\
  \citenamefont {Steffen}}]{rigetti_superconducting_2012}%
  \BibitemOpen
  \bibfield  {author} {\bibinfo {author} {\bibfnamefont {C.}~\bibnamefont
  {Rigetti}}, \bibinfo {author} {\bibfnamefont {J.~M.}\ \bibnamefont
  {Gambetta}}, \bibinfo {author} {\bibfnamefont {S.}~\bibnamefont {Poletto}},
  \bibinfo {author} {\bibfnamefont {B.~L.~T.}\ \bibnamefont {Plourde}},
  \bibinfo {author} {\bibfnamefont {J.~M.}\ \bibnamefont {Chow}}, \bibinfo
  {author} {\bibfnamefont {A.~D.}\ \bibnamefont {C{\'o}rcoles}}, \bibinfo
  {author} {\bibfnamefont {J.~A.}\ \bibnamefont {Smolin}}, \bibinfo {author}
  {\bibfnamefont {S.~T.}\ \bibnamefont {Merkel}}, \bibinfo {author}
  {\bibfnamefont {J.~R.}\ \bibnamefont {Rozen}}, \bibinfo {author}
  {\bibfnamefont {G.~A.}\ \bibnamefont {Keefe}}, \bibinfo {author}
  {\bibfnamefont {M.~B.}\ \bibnamefont {Rothwell}}, \bibinfo {author}
  {\bibfnamefont {M.~B.}\ \bibnamefont {Ketchen}},\ and\ \bibinfo {author}
  {\bibfnamefont {M.}~\bibnamefont {Steffen}},\ }\bibfield  {title} {\bibinfo
  {title} {Superconducting qubit in a waveguide cavity with a coherence time
  approaching 0.1 ms},\ }\href {https://doi.org/10.1103/PhysRevB.86.100506}
  {\bibfield  {journal} {\bibinfo  {journal} {Phys. Rev. B}\ }\textbf {\bibinfo
  {volume} {86}},\ \bibinfo {pages} {100506} (\bibinfo {year}
  {2012})}\BibitemShut {NoStop}%
\bibitem [{\citenamefont {Rips}\ and\ \citenamefont
  {Hartmann}(2013)}]{rips_quantum_2013}%
  \BibitemOpen
  \bibfield  {author} {\bibinfo {author} {\bibfnamefont {S.}~\bibnamefont
  {Rips}}\ and\ \bibinfo {author} {\bibfnamefont {M.~J.}\ \bibnamefont
  {Hartmann}},\ }\bibfield  {title} {\bibinfo {title} {Quantum {{Information
  Processing}} with {{Nanomechanical Qubits}}},\ }\href
  {https://doi.org/10.1103/PhysRevLett.110.120503} {\bibfield  {journal}
  {\bibinfo  {journal} {Phys. Rev. Lett.}\ }\textbf {\bibinfo {volume} {110}},\
  \bibinfo {pages} {120503} (\bibinfo {year} {2013})}\BibitemShut {NoStop}%
\bibitem [{\citenamefont {Rips}\ \emph {et~al.}(2014)\citenamefont {Rips},
  \citenamefont {{Wilson-Rae}},\ and\ \citenamefont
  {Hartmann}}]{rips_nonlinear_2014}%
  \BibitemOpen
  \bibfield  {author} {\bibinfo {author} {\bibfnamefont {S.}~\bibnamefont
  {Rips}}, \bibinfo {author} {\bibfnamefont {I.}~\bibnamefont {{Wilson-Rae}}},\
  and\ \bibinfo {author} {\bibfnamefont {M.~J.}\ \bibnamefont {Hartmann}},\
  }\bibfield  {title} {\bibinfo {title} {Nonlinear nanomechanical resonators
  for quantum optoelectromechanics},\ }\href
  {https://doi.org/10.1103/PhysRevA.89.013854} {\bibfield  {journal} {\bibinfo
  {journal} {Phys. Rev. A}\ }\textbf {\bibinfo {volume} {89}},\ \bibinfo
  {pages} {013854} (\bibinfo {year} {2014})}\BibitemShut {NoStop}%
\bibitem [{\citenamefont {Armour}\ \emph {et~al.}(2004)\citenamefont {Armour},
  \citenamefont {Blencowe},\ and\ \citenamefont
  {Zhang}}]{armour_classical_2004}%
  \BibitemOpen
  \bibfield  {author} {\bibinfo {author} {\bibfnamefont {A.~D.}\ \bibnamefont
  {Armour}}, \bibinfo {author} {\bibfnamefont {M.~P.}\ \bibnamefont
  {Blencowe}},\ and\ \bibinfo {author} {\bibfnamefont {Y.}~\bibnamefont
  {Zhang}},\ }\bibfield  {title} {\bibinfo {title} {Classical dynamics of a
  nanomechanical resonator coupled to a single-electron transistor},\ }\href
  {https://doi.org/10.1103/PhysRevB.69.125313} {\bibfield  {journal} {\bibinfo
  {journal} {Phys. Rev. B}\ }\textbf {\bibinfo {volume} {69}},\ \bibinfo
  {pages} {125313} (\bibinfo {year} {2004})}\BibitemShut {NoStop}%
\bibitem [{\citenamefont {Blanter}\ \emph {et~al.}(2004)\citenamefont
  {Blanter}, \citenamefont {Usmani},\ and\ \citenamefont
  {Nazarov}}]{blanter_single-electron_2004}%
  \BibitemOpen
  \bibfield  {author} {\bibinfo {author} {\bibfnamefont {Y.~M.}\ \bibnamefont
  {Blanter}}, \bibinfo {author} {\bibfnamefont {O.}~\bibnamefont {Usmani}},\
  and\ \bibinfo {author} {\bibfnamefont {a.~Y.~V.}\ \bibnamefont {Nazarov}},\
  }\bibfield  {title} {\bibinfo {title} {Single-{{Electron Tunneling}} with
  {{Strong Mechanical Feedback}}},\ }\href
  {https://doi.org/10.1103/PhysRevLett.93.136802} {\bibfield  {journal}
  {\bibinfo  {journal} {Phys. Rev. B.}\ }\textbf {\bibinfo {volume} {93}},\
  \bibinfo {pages} {136802} (\bibinfo {year} {2004})}\BibitemShut {NoStop}%
\bibitem [{\citenamefont {Chtchelkatchev}\ \emph {et~al.}(2004)\citenamefont
  {Chtchelkatchev}, \citenamefont {Belzig},\ and\ \citenamefont
  {Bruder}}]{chtchelkatchev_charge_2004}%
  \BibitemOpen
  \bibfield  {author} {\bibinfo {author} {\bibfnamefont {N.~M.}\ \bibnamefont
  {Chtchelkatchev}}, \bibinfo {author} {\bibfnamefont {W.}~\bibnamefont
  {Belzig}},\ and\ \bibinfo {author} {\bibfnamefont {C.}~\bibnamefont
  {Bruder}},\ }\bibfield  {title} {\bibinfo {title} {Charge transport through a
  single-electron transistor with a mechanically oscillating island},\ }\href
  {https://doi.org/10.1103/PhysRevB.70.193305} {\bibfield  {journal} {\bibinfo
  {journal} {Phys. Rev. B}\ }\textbf {\bibinfo {volume} {70}},\ \bibinfo
  {pages} {193305} (\bibinfo {year} {2004})}\BibitemShut {NoStop}%
\bibitem [{\citenamefont {Clerk}\ and\ \citenamefont
  {Bennett}(2005)}]{clerk_quantum_2005}%
  \BibitemOpen
  \bibfield  {author} {\bibinfo {author} {\bibfnamefont {A.~A.}\ \bibnamefont
  {Clerk}}\ and\ \bibinfo {author} {\bibfnamefont {S.}~\bibnamefont
  {Bennett}},\ }\bibfield  {title} {\bibinfo {title} {Quantum
  nanoelectromechanics with electrons, quasi-particles and {{Cooper}} pairs:
  Effective bath descriptions and strong feedback effects},\ }\href
  {https://doi.org/10.1088/1367-2630/7/1/238} {\bibfield  {journal} {\bibinfo
  {journal} {New J. of Phys.}\ }\textbf {\bibinfo {volume} {7}},\ \bibinfo
  {pages} {238} (\bibinfo {year} {2005})}\BibitemShut {NoStop}%
\bibitem [{\citenamefont {Koch}\ and\ \citenamefont {{von
  Oppen}}(2005)}]{koch_franck-condon_2005}%
  \BibitemOpen
  \bibfield  {author} {\bibinfo {author} {\bibfnamefont {J.}~\bibnamefont
  {Koch}}\ and\ \bibinfo {author} {\bibfnamefont {F.}~\bibnamefont {{von
  Oppen}}},\ }\bibfield  {title} {\bibinfo {title} {Franck-{{Condon Blockade}}
  and {{Giant Fano Factors}} in {{Transport}} through {{Single Molecules}}},\
  }\href {https://doi.org/10.1103/PhysRevLett.94.206804} {\bibfield  {journal}
  {\bibinfo  {journal} {Phys. Rev. Lett.}\ }\textbf {\bibinfo {volume} {94}},\
  \bibinfo {pages} {206804} (\bibinfo {year} {2005})}\BibitemShut {NoStop}%
\bibitem [{\citenamefont {Mozyrsky}\ \emph {et~al.}(2006)\citenamefont
  {Mozyrsky}, \citenamefont {Hastings},\ and\ \citenamefont
  {Martin}}]{mozyrsky_intermittent_2006}%
  \BibitemOpen
  \bibfield  {author} {\bibinfo {author} {\bibfnamefont {D.}~\bibnamefont
  {Mozyrsky}}, \bibinfo {author} {\bibfnamefont {M.~B.}\ \bibnamefont
  {Hastings}},\ and\ \bibinfo {author} {\bibfnamefont {I.}~\bibnamefont
  {Martin}},\ }\bibfield  {title} {\bibinfo {title} {Intermittent polaron
  dynamics: {{Born}}-{{Oppenheimer}} approximation out of equilibrium},\ }\href
  {https://doi.org/10.1103/PhysRevB.73.035104} {\bibfield  {journal} {\bibinfo
  {journal} {Phys. Rev. B}\ }\textbf {\bibinfo {volume} {73}},\ \bibinfo
  {pages} {035104} (\bibinfo {year} {2006})}\BibitemShut {NoStop}%
\bibitem [{\citenamefont {Doiron}\ \emph {et~al.}(2006)\citenamefont {Doiron},
  \citenamefont {Belzig},\ and\ \citenamefont
  {Bruder}}]{doiron_electrical_2006}%
  \BibitemOpen
  \bibfield  {author} {\bibinfo {author} {\bibfnamefont {C.~B.}\ \bibnamefont
  {Doiron}}, \bibinfo {author} {\bibfnamefont {W.}~\bibnamefont {Belzig}},\
  and\ \bibinfo {author} {\bibfnamefont {C.}~\bibnamefont {Bruder}},\
  }\bibfield  {title} {\bibinfo {title} {Electrical transport through a
  single-electron transistor strongly coupled to an oscillator},\ }\href
  {https://doi.org/10.1103/PhysRevB.74.205336} {\bibfield  {journal} {\bibinfo
  {journal} {Phys. Rev. B}\ }\textbf {\bibinfo {volume} {74}},\ \bibinfo
  {pages} {205336} (\bibinfo {year} {2006})}\BibitemShut {NoStop}%
\bibitem [{\citenamefont {Pistolesi}\ and\ \citenamefont
  {Labarthe}(2007)}]{pistolesi_current_2007}%
  \BibitemOpen
  \bibfield  {author} {\bibinfo {author} {\bibfnamefont {F.}~\bibnamefont
  {Pistolesi}}\ and\ \bibinfo {author} {\bibfnamefont {S.}~\bibnamefont
  {Labarthe}},\ }\bibfield  {title} {\bibinfo {title} {Current blockade in
  classical single-electron nanomechanical resonator},\ }\href
  {https://doi.org/10.1103/PhysRevB.76.165317} {\bibfield  {journal} {\bibinfo
  {journal} {Phys. Rev. B}\ }\textbf {\bibinfo {volume} {76}},\ \bibinfo
  {pages} {165317} (\bibinfo {year} {2007})}\BibitemShut {NoStop}%
\bibitem [{\citenamefont {{de Bonis}}\ \emph {et~al.}(2018)\citenamefont {{de
  Bonis}}, \citenamefont {Urgell}, \citenamefont {Yang}, \citenamefont
  {Samanta}, \citenamefont {Noury}, \citenamefont {{Vergara-Cruz}},
  \citenamefont {Dong}, \citenamefont {Jin},\ and\ \citenamefont
  {Bachtold}}]{de_bonis_ultrasensitive_2018}%
  \BibitemOpen
  \bibfield  {author} {\bibinfo {author} {\bibfnamefont {S.~L.}\ \bibnamefont
  {{de Bonis}}}, \bibinfo {author} {\bibfnamefont {C.}~\bibnamefont {Urgell}},
  \bibinfo {author} {\bibfnamefont {W.}~\bibnamefont {Yang}}, \bibinfo {author}
  {\bibfnamefont {C.}~\bibnamefont {Samanta}}, \bibinfo {author} {\bibfnamefont
  {A.}~\bibnamefont {Noury}}, \bibinfo {author} {\bibfnamefont
  {J.}~\bibnamefont {{Vergara-Cruz}}}, \bibinfo {author} {\bibfnamefont
  {Q.}~\bibnamefont {Dong}}, \bibinfo {author} {\bibfnamefont {Y.}~\bibnamefont
  {Jin}},\ and\ \bibinfo {author} {\bibfnamefont {A.}~\bibnamefont
  {Bachtold}},\ }\bibfield  {title} {\bibinfo {title} {Ultrasensitive
  {{Displacement Noise Measurement}} of {{Carbon Nanotube Mechanical
  Resonators}}},\ }\href {https://doi.org/10.1021/acs.nanolett.8b02437}
  {\bibfield  {journal} {\bibinfo  {journal} {Nano Lett.}\ }\textbf {\bibinfo
  {volume} {18}},\ \bibinfo {pages} {5324} (\bibinfo {year}
  {2018})}\BibitemShut {NoStop}%
\bibitem [{\citenamefont {Khivrich}\ \emph {et~al.}(2019)\citenamefont
  {Khivrich}, \citenamefont {Clerk},\ and\ \citenamefont
  {Ilani}}]{khivrich_nanomechanical_2019}%
  \BibitemOpen
  \bibfield  {author} {\bibinfo {author} {\bibfnamefont {I.}~\bibnamefont
  {Khivrich}}, \bibinfo {author} {\bibfnamefont {A.~A.}\ \bibnamefont
  {Clerk}},\ and\ \bibinfo {author} {\bibfnamefont {S.}~\bibnamefont {Ilani}},\
  }\bibfield  {title} {\bibinfo {title} {Nanomechanical pump\textendash probe
  measurements of insulating electronic states in a carbon nanotube},\ }\href
  {https://doi.org/10.1038/s41565-018-0341-6} {\bibfield  {journal} {\bibinfo
  {journal} {Nat. Nanotechnol.}\ }\textbf {\bibinfo {volume} {14}},\ \bibinfo
  {pages} {161} (\bibinfo {year} {2019})}\BibitemShut {NoStop}%
\bibitem [{\citenamefont {Blien}\ \emph {et~al.}(2020)\citenamefont {Blien},
  \citenamefont {Steger}, \citenamefont {H{\"u}ttner}, \citenamefont {Graaf},\
  and\ \citenamefont {H{\"u}ttel}}]{blien_quantum_2020}%
  \BibitemOpen
  \bibfield  {author} {\bibinfo {author} {\bibfnamefont {S.}~\bibnamefont
  {Blien}}, \bibinfo {author} {\bibfnamefont {P.}~\bibnamefont {Steger}},
  \bibinfo {author} {\bibfnamefont {N.}~\bibnamefont {H{\"u}ttner}}, \bibinfo
  {author} {\bibfnamefont {R.}~\bibnamefont {Graaf}},\ and\ \bibinfo {author}
  {\bibfnamefont {A.~K.}\ \bibnamefont {H{\"u}ttel}},\ }\bibfield  {title}
  {\bibinfo {title} {Quantum capacitance mediated carbon nanotube
  optomechanics},\ }\href {https://doi.org/10.1038/s41467-020-15433-3}
  {\bibfield  {journal} {\bibinfo  {journal} {Nat. Comm.}\ }\textbf {\bibinfo
  {volume} {11}},\ \bibinfo {pages} {1636} (\bibinfo {year}
  {2020})}\BibitemShut {NoStop}%
\bibitem [{\citenamefont {Wen}\ \emph {et~al.}(2020)\citenamefont {Wen},
  \citenamefont {Ares}, \citenamefont {Schupp}, \citenamefont {Pei},
  \citenamefont {Briggs},\ and\ \citenamefont {Laird}}]{wen_coherent_2020}%
  \BibitemOpen
  \bibfield  {author} {\bibinfo {author} {\bibfnamefont {Y.}~\bibnamefont
  {Wen}}, \bibinfo {author} {\bibfnamefont {N.}~\bibnamefont {Ares}}, \bibinfo
  {author} {\bibfnamefont {F.~J.}\ \bibnamefont {Schupp}}, \bibinfo {author}
  {\bibfnamefont {T.}~\bibnamefont {Pei}}, \bibinfo {author} {\bibfnamefont
  {G.~a.~D.}\ \bibnamefont {Briggs}},\ and\ \bibinfo {author} {\bibfnamefont
  {E.~A.}\ \bibnamefont {Laird}},\ }\bibfield  {title} {\bibinfo {title} {A
  coherent nanomechanical oscillator driven by single-electron tunnelling},\
  }\href {https://doi.org/10.1038/s41567-019-0683-5} {\bibfield  {journal}
  {\bibinfo  {journal} {Nat. Phys.}\ }\textbf {\bibinfo {volume} {16}},\
  \bibinfo {pages} {75} (\bibinfo {year} {2020})}\BibitemShut {NoStop}%
\bibitem [{\citenamefont {Benyamini}\ \emph {et~al.}(2014)\citenamefont
  {Benyamini}, \citenamefont {Hamo}, \citenamefont {Kusminskiy}, \citenamefont
  {{von Oppen}},\ and\ \citenamefont {Ilani}}]{benyamini_real-space_2014}%
  \BibitemOpen
  \bibfield  {author} {\bibinfo {author} {\bibfnamefont {A.}~\bibnamefont
  {Benyamini}}, \bibinfo {author} {\bibfnamefont {A.}~\bibnamefont {Hamo}},
  \bibinfo {author} {\bibfnamefont {S.~V.}\ \bibnamefont {Kusminskiy}},
  \bibinfo {author} {\bibfnamefont {F.}~\bibnamefont {{von Oppen}}},\ and\
  \bibinfo {author} {\bibfnamefont {S.}~\bibnamefont {Ilani}},\ }\bibfield
  {title} {\bibinfo {title} {Real-space tailoring of the electron\textendash
  phonon coupling in ultraclean nanotube mechanical resonators},\ }\href
  {https://doi.org/10.1038/nphys2842} {\bibfield  {journal} {\bibinfo
  {journal} {Nat. Phys.}\ }\textbf {\bibinfo {volume} {10}},\ \bibinfo {pages}
  {151} (\bibinfo {year} {2014})}\BibitemShut {NoStop}%
\bibitem [{\citenamefont {Hamo}\ \emph {et~al.}(2016)\citenamefont {Hamo},
  \citenamefont {Benyamini}, \citenamefont {Shapir}, \citenamefont {Khivrich},
  \citenamefont {Waissman}, \citenamefont {Kaasbjerg}, \citenamefont {Oreg},
  \citenamefont {{von Oppen}},\ and\ \citenamefont
  {Ilani}}]{hamo_electron_2016}%
  \BibitemOpen
  \bibfield  {author} {\bibinfo {author} {\bibfnamefont {A.}~\bibnamefont
  {Hamo}}, \bibinfo {author} {\bibfnamefont {A.}~\bibnamefont {Benyamini}},
  \bibinfo {author} {\bibfnamefont {I.}~\bibnamefont {Shapir}}, \bibinfo
  {author} {\bibfnamefont {I.}~\bibnamefont {Khivrich}}, \bibinfo {author}
  {\bibfnamefont {J.}~\bibnamefont {Waissman}}, \bibinfo {author}
  {\bibfnamefont {K.}~\bibnamefont {Kaasbjerg}}, \bibinfo {author}
  {\bibfnamefont {Y.}~\bibnamefont {Oreg}}, \bibinfo {author} {\bibfnamefont
  {F.}~\bibnamefont {{von Oppen}}},\ and\ \bibinfo {author} {\bibfnamefont
  {S.}~\bibnamefont {Ilani}},\ }\bibfield  {title} {\bibinfo {title} {Electron
  attraction mediated by {{Coulomb}} repulsion},\ }\href
  {https://doi.org/10.1038/nature18639} {\bibfield  {journal} {\bibinfo
  {journal} {Nature}\ }\textbf {\bibinfo {volume} {535}},\ \bibinfo {pages}
  {395} (\bibinfo {year} {2016})}\BibitemShut {NoStop}%
\bibitem [{\citenamefont {{van der Wiel}}\ \emph {et~al.}(2002)\citenamefont
  {{van der Wiel}}, \citenamefont {De~Franceschi}, \citenamefont {Elzerman},
  \citenamefont {Fujisawa}, \citenamefont {Tarucha},\ and\ \citenamefont
  {Kouwenhoven}}]{van_der_wiel_electron_2002}%
  \BibitemOpen
  \bibfield  {author} {\bibinfo {author} {\bibfnamefont {W.~G.}\ \bibnamefont
  {{van der Wiel}}}, \bibinfo {author} {\bibfnamefont {S.}~\bibnamefont
  {De~Franceschi}}, \bibinfo {author} {\bibfnamefont {J.~M.}\ \bibnamefont
  {Elzerman}}, \bibinfo {author} {\bibfnamefont {T.}~\bibnamefont {Fujisawa}},
  \bibinfo {author} {\bibfnamefont {S.}~\bibnamefont {Tarucha}},\ and\ \bibinfo
  {author} {\bibfnamefont {L.~P.}\ \bibnamefont {Kouwenhoven}},\ }\bibfield
  {title} {\bibinfo {title} {Electron transport through double quantum dots},\
  }\href {https://doi.org/10.1103/RevModPhys.75.1} {\bibfield  {journal}
  {\bibinfo  {journal} {Rev. Mod. Phys.}\ }\textbf {\bibinfo {volume} {75}},\
  \bibinfo {pages} {1} (\bibinfo {year} {2002})}\BibitemShut {NoStop}%
\bibitem [{\citenamefont {Galperin}\ \emph {et~al.}(2005)\citenamefont
  {Galperin}, \citenamefont {Ratner},\ and\ \citenamefont
  {Nitzan}}]{galperin_hysteresis_2005}%
  \BibitemOpen
  \bibfield  {author} {\bibinfo {author} {\bibfnamefont {M.}~\bibnamefont
  {Galperin}}, \bibinfo {author} {\bibfnamefont {M.~A.}\ \bibnamefont
  {Ratner}},\ and\ \bibinfo {author} {\bibfnamefont {A.}~\bibnamefont
  {Nitzan}},\ }\bibfield  {title} {\bibinfo {title} {Hysteresis, {{Switching}},
  and {{Negative Differential Resistance}} in {{Molecular Junctions}}: {{A
  Polaron Model}}},\ }\href {https://doi.org/10.1021/nl048216c} {\bibfield
  {journal} {\bibinfo  {journal} {Nano Lett.}\ }\textbf {\bibinfo {volume}
  {5}},\ \bibinfo {pages} {125} (\bibinfo {year} {2005})}\BibitemShut {NoStop}%
\bibitem [{\citenamefont {Micchi}\ \emph {et~al.}(2015)\citenamefont {Micchi},
  \citenamefont {Avriller},\ and\ \citenamefont
  {Pistolesi}}]{micchi_mechanical_2015}%
  \BibitemOpen
  \bibfield  {author} {\bibinfo {author} {\bibfnamefont {G.}~\bibnamefont
  {Micchi}}, \bibinfo {author} {\bibfnamefont {R.}~\bibnamefont {Avriller}},\
  and\ \bibinfo {author} {\bibfnamefont {F.}~\bibnamefont {Pistolesi}},\
  }\bibfield  {title} {\bibinfo {title} {Mechanical {{Signatures}} of the
  {{Current Blockade Instability}} in {{Suspended Carbon Nanotubes}}},\ }\href
  {https://doi.org/10.1103/PhysRevLett.115.206802} {\bibfield  {journal}
  {\bibinfo  {journal} {Phys. Rev. Lett.}\ }\textbf {\bibinfo {volume} {115}},\
  \bibinfo {pages} {206802} (\bibinfo {year} {2015})}\BibitemShut {NoStop}%
\bibitem [{\citenamefont {Avriller}\ \emph {et~al.}(2018)\citenamefont
  {Avriller}, \citenamefont {Murr},\ and\ \citenamefont
  {Pistolesi}}]{avriller_bistability_2018}%
  \BibitemOpen
  \bibfield  {author} {\bibinfo {author} {\bibfnamefont {R.}~\bibnamefont
  {Avriller}}, \bibinfo {author} {\bibfnamefont {B.}~\bibnamefont {Murr}},\
  and\ \bibinfo {author} {\bibfnamefont {F.}~\bibnamefont {Pistolesi}},\
  }\bibfield  {title} {\bibinfo {title} {Bistability and displacement
  fluctuations in a quantum nanomechanical oscillator},\ }\href
  {https://doi.org/10.1103/PhysRevB.97.155414} {\bibfield  {journal} {\bibinfo
  {journal} {Phys. Rev. B}\ }\textbf {\bibinfo {volume} {97}},\ \bibinfo
  {pages} {155414} (\bibinfo {year} {2018})}\BibitemShut {NoStop}%
\bibitem [{\citenamefont {Hioe}\ and\ \citenamefont
  {Montroll}(1975)}]{hioe_quantum_1975}%
  \BibitemOpen
  \bibfield  {author} {\bibinfo {author} {\bibfnamefont {F.~T.}\ \bibnamefont
  {Hioe}}\ and\ \bibinfo {author} {\bibfnamefont {E.~W.}\ \bibnamefont
  {Montroll}},\ }\bibfield  {title} {\bibinfo {title} {Quantum theory of
  anharmonic oscillators. {{I}}. {{Energy}} levels of oscillators with positive
  quartic anharmonicity},\ }\href {https://doi.org/10.1063/1.522747} {\bibfield
   {journal} {\bibinfo  {journal} {J. Math. Phys.}\ }\textbf {\bibinfo {volume}
  {16}},\ \bibinfo {pages} {1945} (\bibinfo {year} {1975})}\BibitemShut
  {NoStop}%
\bibitem [{\citenamefont {Schreier}\ \emph {et~al.}(2008)\citenamefont
  {Schreier}, \citenamefont {Houck}, \citenamefont {Koch}, \citenamefont
  {Schuster}, \citenamefont {Johnson}, \citenamefont {Chow}, \citenamefont
  {Gambetta}, \citenamefont {Majer}, \citenamefont {Frunzio}, \citenamefont
  {Devoret}, \citenamefont {Girvin},\ and\ \citenamefont
  {Schoelkopf}}]{schreier_suppressing_2008}%
  \BibitemOpen
  \bibfield  {author} {\bibinfo {author} {\bibfnamefont {J.~A.}\ \bibnamefont
  {Schreier}}, \bibinfo {author} {\bibfnamefont {A.~A.}\ \bibnamefont {Houck}},
  \bibinfo {author} {\bibfnamefont {J.}~\bibnamefont {Koch}}, \bibinfo {author}
  {\bibfnamefont {D.~I.}\ \bibnamefont {Schuster}}, \bibinfo {author}
  {\bibfnamefont {B.~R.}\ \bibnamefont {Johnson}}, \bibinfo {author}
  {\bibfnamefont {J.~M.}\ \bibnamefont {Chow}}, \bibinfo {author}
  {\bibfnamefont {J.~M.}\ \bibnamefont {Gambetta}}, \bibinfo {author}
  {\bibfnamefont {J.}~\bibnamefont {Majer}}, \bibinfo {author} {\bibfnamefont
  {L.}~\bibnamefont {Frunzio}}, \bibinfo {author} {\bibfnamefont {M.~H.}\
  \bibnamefont {Devoret}}, \bibinfo {author} {\bibfnamefont {S.~M.}\
  \bibnamefont {Girvin}},\ and\ \bibinfo {author} {\bibfnamefont {R.~J.}\
  \bibnamefont {Schoelkopf}},\ }\bibfield  {title} {\bibinfo {title}
  {Suppressing charge noise decoherence in superconducting charge qubits},\
  }\href {https://doi.org/10.1103/PhysRevB.77.180502} {\bibfield  {journal}
  {\bibinfo  {journal} {Phys. Rev. B}\ }\textbf {\bibinfo {volume} {77}},\
  \bibinfo {pages} {180502} (\bibinfo {year} {2008})}\BibitemShut {NoStop}%
\bibitem [{\citenamefont {Hioe}\ \emph {et~al.}(1978)\citenamefont {Hioe},
  \citenamefont {MacMillen},\ and\ \citenamefont
  {Montroll}}]{hioe_quantum_1978}%
  \BibitemOpen
  \bibfield  {author} {\bibinfo {author} {\bibfnamefont {F.~T.}\ \bibnamefont
  {Hioe}}, \bibinfo {author} {\bibfnamefont {D.}~\bibnamefont {MacMillen}},\
  and\ \bibinfo {author} {\bibfnamefont {E.~W.}\ \bibnamefont {Montroll}},\
  }\bibfield  {title} {\bibinfo {title} {Quantum theory of anharmonic
  oscillators: Energy levels of a single and a pair of coupled oscillators with
  quartic coupling},\ }\href@noop {} {\bibfield  {journal} {\bibinfo  {journal}
  {Phys. Rep.}\ }\textbf {\bibinfo {volume} {43}},\ \bibinfo {pages} {305}
  (\bibinfo {year} {1978})}\BibitemShut {NoStop}%
\bibitem [{\citenamefont {Collin}\ \emph {et~al.}(2004)\citenamefont {Collin},
  \citenamefont {Ithier}, \citenamefont {Aassime}, \citenamefont {Joyez},
  \citenamefont {Vion},\ and\ \citenamefont {Esteve}}]{collin_nmr-like_2004}%
  \BibitemOpen
  \bibfield  {author} {\bibinfo {author} {\bibfnamefont {E.}~\bibnamefont
  {Collin}}, \bibinfo {author} {\bibfnamefont {G.}~\bibnamefont {Ithier}},
  \bibinfo {author} {\bibfnamefont {A.}~\bibnamefont {Aassime}}, \bibinfo
  {author} {\bibfnamefont {P.}~\bibnamefont {Joyez}}, \bibinfo {author}
  {\bibfnamefont {D.}~\bibnamefont {Vion}},\ and\ \bibinfo {author}
  {\bibfnamefont {D.}~\bibnamefont {Esteve}},\ }\bibfield  {title} {\bibinfo
  {title} {{{NMR}}-like {{Control}} of a {{Quantum Bit Superconducting
  Circuit}}},\ }\href {https://doi.org/10.1103/PhysRevLett.93.157005}
  {\bibfield  {journal} {\bibinfo  {journal} {Phys. Rev. Lett.}\ }\textbf
  {\bibinfo {volume} {93}},\ \bibinfo {pages} {157005} (\bibinfo {year}
  {2004})}\BibitemShut {NoStop}%
\bibitem [{\citenamefont {Majer}\ \emph {et~al.}(2007)\citenamefont {Majer},
  \citenamefont {Chow}, \citenamefont {Gambetta}, \citenamefont {Koch},
  \citenamefont {Johnson}, \citenamefont {Schreier}, \citenamefont {Frunzio},
  \citenamefont {Schuster}, \citenamefont {Houck}, \citenamefont {Wallraff},
  \citenamefont {Blais}, \citenamefont {Devoret}, \citenamefont {Girvin},\ and\
  \citenamefont {Schoelkopf}}]{majer_coupling_2007}%
  \BibitemOpen
  \bibfield  {author} {\bibinfo {author} {\bibfnamefont {J.}~\bibnamefont
  {Majer}}, \bibinfo {author} {\bibfnamefont {J.~M.}\ \bibnamefont {Chow}},
  \bibinfo {author} {\bibfnamefont {J.~M.}\ \bibnamefont {Gambetta}}, \bibinfo
  {author} {\bibfnamefont {J.}~\bibnamefont {Koch}}, \bibinfo {author}
  {\bibfnamefont {B.~R.}\ \bibnamefont {Johnson}}, \bibinfo {author}
  {\bibfnamefont {J.~A.}\ \bibnamefont {Schreier}}, \bibinfo {author}
  {\bibfnamefont {L.}~\bibnamefont {Frunzio}}, \bibinfo {author} {\bibfnamefont
  {D.~I.}\ \bibnamefont {Schuster}}, \bibinfo {author} {\bibfnamefont {A.~A.}\
  \bibnamefont {Houck}}, \bibinfo {author} {\bibfnamefont {A.}~\bibnamefont
  {Wallraff}}, \bibinfo {author} {\bibfnamefont {A.}~\bibnamefont {Blais}},
  \bibinfo {author} {\bibfnamefont {M.~H.}\ \bibnamefont {Devoret}}, \bibinfo
  {author} {\bibfnamefont {S.~M.}\ \bibnamefont {Girvin}},\ and\ \bibinfo
  {author} {\bibfnamefont {R.~J.}\ \bibnamefont {Schoelkopf}},\ }\bibfield
  {title} {\bibinfo {title} {Coupling superconducting qubits via a cavity
  bus},\ }\href {https://doi.org/10.1038/nature06184} {\bibfield  {journal}
  {\bibinfo  {journal} {Nature}\ }\textbf {\bibinfo {volume} {449}},\ \bibinfo
  {pages} {443} (\bibinfo {year} {2007})}\BibitemShut {NoStop}%
\bibitem [{\citenamefont {Houck}\ \emph {et~al.}(2008)\citenamefont {Houck},
  \citenamefont {Schreier}, \citenamefont {Johnson}, \citenamefont {Chow},
  \citenamefont {Koch}, \citenamefont {Gambetta}, \citenamefont {Schuster},
  \citenamefont {Frunzio}, \citenamefont {Devoret}, \citenamefont {Girvin},\
  and\ \citenamefont {Schoelkopf}}]{houck_controlling_2008}%
  \BibitemOpen
  \bibfield  {author} {\bibinfo {author} {\bibfnamefont {A.~A.}\ \bibnamefont
  {Houck}}, \bibinfo {author} {\bibfnamefont {J.~A.}\ \bibnamefont {Schreier}},
  \bibinfo {author} {\bibfnamefont {B.~R.}\ \bibnamefont {Johnson}}, \bibinfo
  {author} {\bibfnamefont {J.~M.}\ \bibnamefont {Chow}}, \bibinfo {author}
  {\bibfnamefont {J.}~\bibnamefont {Koch}}, \bibinfo {author} {\bibfnamefont
  {J.~M.}\ \bibnamefont {Gambetta}}, \bibinfo {author} {\bibfnamefont {D.~I.}\
  \bibnamefont {Schuster}}, \bibinfo {author} {\bibfnamefont {L.}~\bibnamefont
  {Frunzio}}, \bibinfo {author} {\bibfnamefont {M.~H.}\ \bibnamefont
  {Devoret}}, \bibinfo {author} {\bibfnamefont {S.~M.}\ \bibnamefont
  {Girvin}},\ and\ \bibinfo {author} {\bibfnamefont {R.~J.}\ \bibnamefont
  {Schoelkopf}},\ }\bibfield  {title} {\bibinfo {title} {Controlling the
  {{Spontaneous Emission}} of a {{Superconducting Transmon Qubit}}},\ }\href
  {https://doi.org/10.1103/PhysRevLett.101.080502} {\bibfield  {journal}
  {\bibinfo  {journal} {Phys. Rev. Lett.}\ }\textbf {\bibinfo {volume} {101}},\
  \bibinfo {pages} {080502} (\bibinfo {year} {2008})}\BibitemShut {NoStop}%
\bibitem [{\citenamefont {Blais}\ \emph {et~al.}(2004)\citenamefont {Blais},
  \citenamefont {Huang}, \citenamefont {Wallraff}, \citenamefont {Girvin},\
  and\ \citenamefont {Schoelkopf}}]{blais_cavity_2004}%
  \BibitemOpen
  \bibfield  {author} {\bibinfo {author} {\bibfnamefont {A.}~\bibnamefont
  {Blais}}, \bibinfo {author} {\bibfnamefont {R.-S.}\ \bibnamefont {Huang}},
  \bibinfo {author} {\bibfnamefont {A.}~\bibnamefont {Wallraff}}, \bibinfo
  {author} {\bibfnamefont {S.~M.}\ \bibnamefont {Girvin}},\ and\ \bibinfo
  {author} {\bibfnamefont {R.~J.}\ \bibnamefont {Schoelkopf}},\ }\bibfield
  {title} {\bibinfo {title} {Cavity quantum electrodynamics for superconducting
  electrical circuits: {{An}} architecture for quantum computation},\ }\href
  {https://doi.org/10.1103/PhysRevA.69.062320} {\bibfield  {journal} {\bibinfo
  {journal} {Phys. Rev. A}\ }\textbf {\bibinfo {volume} {69}},\ \bibinfo
  {pages} {062320} (\bibinfo {year} {2004})}\BibitemShut {NoStop}%
\bibitem [{\citenamefont {Koch}\ \emph {et~al.}(2007)\citenamefont {Koch},
  \citenamefont {Yu}, \citenamefont {Gambetta}, \citenamefont {Houck},
  \citenamefont {Schuster}, \citenamefont {Majer}, \citenamefont {Blais},
  \citenamefont {Devoret}, \citenamefont {Girvin},\ and\ \citenamefont
  {Schoelkopf}}]{koch_charge-insensitive_2007}%
  \BibitemOpen
  \bibfield  {author} {\bibinfo {author} {\bibfnamefont {J.}~\bibnamefont
  {Koch}}, \bibinfo {author} {\bibfnamefont {T.~M.}\ \bibnamefont {Yu}},
  \bibinfo {author} {\bibfnamefont {J.}~\bibnamefont {Gambetta}}, \bibinfo
  {author} {\bibfnamefont {A.~A.}\ \bibnamefont {Houck}}, \bibinfo {author}
  {\bibfnamefont {D.~I.}\ \bibnamefont {Schuster}}, \bibinfo {author}
  {\bibfnamefont {J.}~\bibnamefont {Majer}}, \bibinfo {author} {\bibfnamefont
  {A.}~\bibnamefont {Blais}}, \bibinfo {author} {\bibfnamefont {M.~H.}\
  \bibnamefont {Devoret}}, \bibinfo {author} {\bibfnamefont {S.~M.}\
  \bibnamefont {Girvin}},\ and\ \bibinfo {author} {\bibfnamefont {R.~J.}\
  \bibnamefont {Schoelkopf}},\ }\bibfield  {title} {\bibinfo {title}
  {Charge-insensitive qubit design derived from the {{Cooper}} pair box},\
  }\href {https://doi.org/10.1103/PhysRevA.76.042319} {\bibfield  {journal}
  {\bibinfo  {journal} {Phys. Rev. A}\ }\textbf {\bibinfo {volume} {76}},\
  \bibinfo {pages} {042319} (\bibinfo {year} {2007})}\BibitemShut {NoStop}%
\bibitem [{\citenamefont {Scarlino}\ \emph {et~al.}(2019)\citenamefont
  {Scarlino}, \citenamefont {{van Woerkom}}, \citenamefont {Stockklauser},
  \citenamefont {Koski}, \citenamefont {Collodo}, \citenamefont {Gasparinetti},
  \citenamefont {Reichl}, \citenamefont {Wegscheider}, \citenamefont {Ihn},
  \citenamefont {Ensslin},\ and\ \citenamefont
  {Wallraff}}]{scarlino_all-microwave_2019}%
  \BibitemOpen
  \bibfield  {author} {\bibinfo {author} {\bibfnamefont {P.}~\bibnamefont
  {Scarlino}}, \bibinfo {author} {\bibfnamefont {D.~J.}\ \bibnamefont {{van
  Woerkom}}}, \bibinfo {author} {\bibfnamefont {A.}~\bibnamefont
  {Stockklauser}}, \bibinfo {author} {\bibfnamefont {J.~V.}\ \bibnamefont
  {Koski}}, \bibinfo {author} {\bibfnamefont {M.~C.}\ \bibnamefont {Collodo}},
  \bibinfo {author} {\bibfnamefont {S.}~\bibnamefont {Gasparinetti}}, \bibinfo
  {author} {\bibfnamefont {C.}~\bibnamefont {Reichl}}, \bibinfo {author}
  {\bibfnamefont {W.}~\bibnamefont {Wegscheider}}, \bibinfo {author}
  {\bibfnamefont {T.}~\bibnamefont {Ihn}}, \bibinfo {author} {\bibfnamefont
  {K.}~\bibnamefont {Ensslin}},\ and\ \bibinfo {author} {\bibfnamefont
  {A.}~\bibnamefont {Wallraff}},\ }\bibfield  {title} {\bibinfo {title}
  {All-{{Microwave Control}} and {{Dispersive Readout}} of {{Gate}}-{{Defined
  Quantum Dot Qubits}} in {{Circuit Quantum Electrodynamics}}},\ }\href
  {https://doi.org/10.1103/PhysRevLett.122.206802} {\bibfield  {journal}
  {\bibinfo  {journal} {Phys. Rev. Lett.}\ }\textbf {\bibinfo {volume} {122}},\
  \bibinfo {pages} {206802} (\bibinfo {year} {2019})}\BibitemShut {NoStop}%
\bibitem [{\citenamefont {Hauss}\ \emph {et~al.}(2008)\citenamefont {Hauss},
  \citenamefont {Fedorov}, \citenamefont {Andr{\'e}}, \citenamefont {Brosco},
  \citenamefont {Hutter}, \citenamefont {Kothari}, \citenamefont {Yeshwanth},
  \citenamefont {Shnirman},\ and\ \citenamefont
  {Sch{\"o}n}}]{hauss_dissipation_2008}%
  \BibitemOpen
  \bibfield  {author} {\bibinfo {author} {\bibfnamefont {J.}~\bibnamefont
  {Hauss}}, \bibinfo {author} {\bibfnamefont {A.}~\bibnamefont {Fedorov}},
  \bibinfo {author} {\bibfnamefont {S.}~\bibnamefont {Andr{\'e}}}, \bibinfo
  {author} {\bibfnamefont {V.}~\bibnamefont {Brosco}}, \bibinfo {author}
  {\bibfnamefont {C.}~\bibnamefont {Hutter}}, \bibinfo {author} {\bibfnamefont
  {R.}~\bibnamefont {Kothari}}, \bibinfo {author} {\bibfnamefont
  {S.}~\bibnamefont {Yeshwanth}}, \bibinfo {author} {\bibfnamefont
  {A.}~\bibnamefont {Shnirman}},\ and\ \bibinfo {author} {\bibfnamefont
  {G.}~\bibnamefont {Sch{\"o}n}},\ }\bibfield  {title} {\bibinfo {title}
  {Dissipation in circuit quantum electrodynamics: Lasing and cooling of a
  low-frequency oscillator},\ }\href
  {https://doi.org/10.1088/1367-2630/10/9/095018} {\bibfield  {journal}
  {\bibinfo  {journal} {New J. Phys.}\ }\textbf {\bibinfo {volume} {10}},\
  \bibinfo {pages} {095018} (\bibinfo {year} {2008})}\BibitemShut {NoStop}%
\bibitem [{\citenamefont {{Cohen-Tannoudji}}\ \emph {et~al.}(1992)\citenamefont
  {{Cohen-Tannoudji}}, \citenamefont {{Dupont-Roc}},\ and\ \citenamefont
  {Grynberg}}]{cohen-tannoudji_atom-photon_1992}%
  \BibitemOpen
  \bibfield  {author} {\bibinfo {author} {\bibfnamefont {C.}~\bibnamefont
  {{Cohen-Tannoudji}}}, \bibinfo {author} {\bibfnamefont {J.}~\bibnamefont
  {{Dupont-Roc}}},\ and\ \bibinfo {author} {\bibfnamefont {G.}~\bibnamefont
  {Grynberg}},\ }\href@noop {} {\emph {\bibinfo {title} {Atom-Photon
  Interactions: Basic Processes and Applications}}}\ (\bibinfo  {publisher}
  {{Wiley}},\ \bibinfo {address} {{New York}},\ \bibinfo {year}
  {1992})\BibitemShut {NoStop}%
\bibitem [{\citenamefont {Rigetti}\ and\ \citenamefont
  {Devoret}(2010)}]{rigetti_fully_2010}%
  \BibitemOpen
  \bibfield  {author} {\bibinfo {author} {\bibfnamefont {C.}~\bibnamefont
  {Rigetti}}\ and\ \bibinfo {author} {\bibfnamefont {M.}~\bibnamefont
  {Devoret}},\ }\bibfield  {title} {\bibinfo {title} {Fully microwave-tunable
  universal gates in superconducting qubits with linear couplings and fixed
  transition frequencies},\ }\href {https://doi.org/10.1103/PhysRevB.81.134507}
  {\bibfield  {journal} {\bibinfo  {journal} {Phys. Rev. B}\ }\textbf {\bibinfo
  {volume} {81}},\ \bibinfo {pages} {134507} (\bibinfo {year}
  {2010})}\BibitemShut {NoStop}%
\bibitem [{\citenamefont {Chow}\ \emph {et~al.}(2011)\citenamefont {Chow},
  \citenamefont {C{\'o}rcoles}, \citenamefont {Gambetta}, \citenamefont
  {Rigetti}, \citenamefont {Johnson}, \citenamefont {Smolin}, \citenamefont
  {Rozen}, \citenamefont {Keefe}, \citenamefont {Rothwell}, \citenamefont
  {Ketchen},\ and\ \citenamefont {Steffen}}]{chow_simple_2011}%
  \BibitemOpen
  \bibfield  {author} {\bibinfo {author} {\bibfnamefont {J.~M.}\ \bibnamefont
  {Chow}}, \bibinfo {author} {\bibfnamefont {A.~D.}\ \bibnamefont
  {C{\'o}rcoles}}, \bibinfo {author} {\bibfnamefont {J.~M.}\ \bibnamefont
  {Gambetta}}, \bibinfo {author} {\bibfnamefont {C.}~\bibnamefont {Rigetti}},
  \bibinfo {author} {\bibfnamefont {B.~R.}\ \bibnamefont {Johnson}}, \bibinfo
  {author} {\bibfnamefont {J.~A.}\ \bibnamefont {Smolin}}, \bibinfo {author}
  {\bibfnamefont {J.~R.}\ \bibnamefont {Rozen}}, \bibinfo {author}
  {\bibfnamefont {G.~A.}\ \bibnamefont {Keefe}}, \bibinfo {author}
  {\bibfnamefont {M.~B.}\ \bibnamefont {Rothwell}}, \bibinfo {author}
  {\bibfnamefont {M.~B.}\ \bibnamefont {Ketchen}},\ and\ \bibinfo {author}
  {\bibfnamefont {M.}~\bibnamefont {Steffen}},\ }\bibfield  {title} {\bibinfo
  {title} {Simple {{All}}-{{Microwave Entangling Gate}} for
  {{Fixed}}-{{Frequency Superconducting Qubits}}},\ }\href
  {https://doi.org/10.1103/PhysRevLett.107.080502} {\bibfield  {journal}
  {\bibinfo  {journal} {Phys. Rev. Lett.}\ }\textbf {\bibinfo {volume} {107}},\
  \bibinfo {pages} {080502} (\bibinfo {year} {2011})}\BibitemShut {NoStop}%
\bibitem [{\citenamefont {Yang}\ \emph {et~al.}(2020)\citenamefont {Yang},
  \citenamefont {Urgell}, \citenamefont {De~Bonis}, \citenamefont {Marganska},
  \citenamefont {Grifoni},\ and\ \citenamefont
  {Bachtold}}]{yang_fabry-perot_2020}%
  \BibitemOpen
  \bibfield  {author} {\bibinfo {author} {\bibfnamefont {W.}~\bibnamefont
  {Yang}}, \bibinfo {author} {\bibfnamefont {C.}~\bibnamefont {Urgell}},
  \bibinfo {author} {\bibfnamefont {S.~L.}\ \bibnamefont {De~Bonis}}, \bibinfo
  {author} {\bibfnamefont {M.}~\bibnamefont {Marganska}}, \bibinfo {author}
  {\bibfnamefont {M.}~\bibnamefont {Grifoni}},\ and\ \bibinfo {author}
  {\bibfnamefont {A.}~\bibnamefont {Bachtold}},\ }\bibfield  {title} {\bibinfo
  {title} {Fabry-{{P\'erot}} oscillations in correlated carbon nanotubes},\
  }\href@noop {} {\bibfield  {journal} {\bibinfo  {journal} {arXiv:2003.08226
  [cond-mat]}\ } (\bibinfo {year} {2020})},\ \Eprint
  {https://arxiv.org/abs/2003.08226} {arXiv:2003.08226 [cond-mat]} \BibitemShut
  {NoStop}%
\bibitem [{\citenamefont {Viennot}\ \emph {et~al.}(2015)\citenamefont
  {Viennot}, \citenamefont {Dartiailh}, \citenamefont {Cottet},\ and\
  \citenamefont {Kontos}}]{viennot_coherent_2015}%
  \BibitemOpen
  \bibfield  {author} {\bibinfo {author} {\bibfnamefont {J.~J.}\ \bibnamefont
  {Viennot}}, \bibinfo {author} {\bibfnamefont {M.~C.}\ \bibnamefont
  {Dartiailh}}, \bibinfo {author} {\bibfnamefont {A.}~\bibnamefont {Cottet}},\
  and\ \bibinfo {author} {\bibfnamefont {T.}~\bibnamefont {Kontos}},\
  }\bibfield  {title} {\bibinfo {title} {Coherent coupling of a single spin to
  microwave cavity photons},\ }\href {https://doi.org/10.1126/science.aaa3786}
  {\bibfield  {journal} {\bibinfo  {journal} {Science}\ }\textbf {\bibinfo
  {volume} {349}},\ \bibinfo {pages} {408} (\bibinfo {year}
  {2015})}\BibitemShut {NoStop}%
\bibitem [{\citenamefont {Cubaynes}\ \emph {et~al.}(2019)\citenamefont
  {Cubaynes}, \citenamefont {Delbecq}, \citenamefont {Dartiailh}, \citenamefont
  {Assouly}, \citenamefont {Desjardins}, \citenamefont {Contamin},
  \citenamefont {Bruhat}, \citenamefont {Leghtas}, \citenamefont {Mallet},
  \citenamefont {Cottet},\ and\ \citenamefont {Kontos}}]{cubaynes_highly_2019}%
  \BibitemOpen
  \bibfield  {author} {\bibinfo {author} {\bibfnamefont {T.}~\bibnamefont
  {Cubaynes}}, \bibinfo {author} {\bibfnamefont {M.~R.}\ \bibnamefont
  {Delbecq}}, \bibinfo {author} {\bibfnamefont {M.~C.}\ \bibnamefont
  {Dartiailh}}, \bibinfo {author} {\bibfnamefont {R.}~\bibnamefont {Assouly}},
  \bibinfo {author} {\bibfnamefont {M.~M.}\ \bibnamefont {Desjardins}},
  \bibinfo {author} {\bibfnamefont {L.~C.}\ \bibnamefont {Contamin}}, \bibinfo
  {author} {\bibfnamefont {L.~E.}\ \bibnamefont {Bruhat}}, \bibinfo {author}
  {\bibfnamefont {Z.}~\bibnamefont {Leghtas}}, \bibinfo {author} {\bibfnamefont
  {F.}~\bibnamefont {Mallet}}, \bibinfo {author} {\bibfnamefont
  {A.}~\bibnamefont {Cottet}},\ and\ \bibinfo {author} {\bibfnamefont
  {T.}~\bibnamefont {Kontos}},\ }\bibfield  {title} {\bibinfo {title} {Highly
  coherent spin states in carbon nanotubes coupled to cavity photons},\ }\href
  {https://doi.org/10.1038/s41534-019-0169-4} {\bibfield  {journal} {\bibinfo
  {journal} {npj Quantum Information}\ }\textbf {\bibinfo {volume} {5}},\
  \bibinfo {pages} {1} (\bibinfo {year} {2019})}\BibitemShut {NoStop}%
\bibitem [{\citenamefont {Hebestreit}\ \emph {et~al.}(2018)\citenamefont
  {Hebestreit}, \citenamefont {Frimmer}, \citenamefont {Reimann},\ and\
  \citenamefont {Novotny}}]{hebestreit_sensing_2018}%
  \BibitemOpen
  \bibfield  {author} {\bibinfo {author} {\bibfnamefont {E.}~\bibnamefont
  {Hebestreit}}, \bibinfo {author} {\bibfnamefont {M.}~\bibnamefont {Frimmer}},
  \bibinfo {author} {\bibfnamefont {R.}~\bibnamefont {Reimann}},\ and\ \bibinfo
  {author} {\bibfnamefont {L.}~\bibnamefont {Novotny}},\ }\bibfield  {title}
  {\bibinfo {title} {Sensing {{Static Forces}} with {{Free}}-{{Falling
  Nanoparticles}}},\ }\href {https://doi.org/10.1103/PhysRevLett.121.063602}
  {\bibfield  {journal} {\bibinfo  {journal} {Phys. Rev. Lett.}\ }\textbf
  {\bibinfo {volume} {121}},\ \bibinfo {pages} {063602} (\bibinfo {year}
  {2018})}\BibitemShut {NoStop}%
\bibitem [{\citenamefont {Hug}\ \emph {et~al.}(1999)\citenamefont {Hug},
  \citenamefont {Stiefel}, \citenamefont {{van Schendel}}, \citenamefont
  {Moser}, \citenamefont {Martin},\ and\ \citenamefont
  {G{\"u}ntherodt}}]{hug_low_1999}%
  \BibitemOpen
  \bibfield  {author} {\bibinfo {author} {\bibfnamefont {H.~J.}\ \bibnamefont
  {Hug}}, \bibinfo {author} {\bibfnamefont {B.}~\bibnamefont {Stiefel}},
  \bibinfo {author} {\bibfnamefont {P.~J.~A.}\ \bibnamefont {{van Schendel}}},
  \bibinfo {author} {\bibfnamefont {A.}~\bibnamefont {Moser}}, \bibinfo
  {author} {\bibfnamefont {S.}~\bibnamefont {Martin}},\ and\ \bibinfo {author}
  {\bibfnamefont {H.-J.}\ \bibnamefont {G{\"u}ntherodt}},\ }\bibfield  {title}
  {\bibinfo {title} {A low temperature ultrahigh vaccum scanning force
  microscope},\ }\href {https://doi.org/10.1063/1.1149970} {\bibfield
  {journal} {\bibinfo  {journal} {Review of Scientific Instruments}\ }\textbf
  {\bibinfo {volume} {70}},\ \bibinfo {pages} {3625} (\bibinfo {year}
  {1999})}\BibitemShut {NoStop}%
\bibitem [{\citenamefont {{Ribezzi-Crivellari}}\ \emph
  {et~al.}(2013)\citenamefont {{Ribezzi-Crivellari}}, \citenamefont {Huguet},\
  and\ \citenamefont {Ritort}}]{ribezzi-crivellari_counter-propagating_2013}%
  \BibitemOpen
  \bibfield  {author} {\bibinfo {author} {\bibfnamefont {M.}~\bibnamefont
  {{Ribezzi-Crivellari}}}, \bibinfo {author} {\bibfnamefont {J.~M.}\
  \bibnamefont {Huguet}},\ and\ \bibinfo {author} {\bibfnamefont
  {F.}~\bibnamefont {Ritort}},\ }\bibfield  {title} {\bibinfo {title}
  {Counter-propagating dual-trap optical tweezers based on linear momentum
  conservation},\ }\href {https://doi.org/10.1063/1.4799289} {\bibfield
  {journal} {\bibinfo  {journal} {Review of Scientific Instruments}\ }\textbf
  {\bibinfo {volume} {84}},\ \bibinfo {pages} {043104} (\bibinfo {year}
  {2013})}\BibitemShut {NoStop}%
\bibitem [{\citenamefont {Clerk}\ \emph {et~al.}(2010)\citenamefont {Clerk},
  \citenamefont {Devoret}, \citenamefont {Girvin}, \citenamefont {Marquardt},\
  and\ \citenamefont {Schoelkopf}}]{clerk_introduction_2010}%
  \BibitemOpen
  \bibfield  {author} {\bibinfo {author} {\bibfnamefont {A.~A.}\ \bibnamefont
  {Clerk}}, \bibinfo {author} {\bibfnamefont {M.~H.}\ \bibnamefont {Devoret}},
  \bibinfo {author} {\bibfnamefont {S.~M.}\ \bibnamefont {Girvin}}, \bibinfo
  {author} {\bibfnamefont {F.}~\bibnamefont {Marquardt}},\ and\ \bibinfo
  {author} {\bibfnamefont {R.~J.}\ \bibnamefont {Schoelkopf}},\ }\bibfield
  {title} {\bibinfo {title} {Introduction to quantum noise, measurement, and
  amplification},\ }\href {https://doi.org/10.1103/RevModPhys.82.1155}
  {\bibfield  {journal} {\bibinfo  {journal} {Reviews of Modern Physics}\
  }\textbf {\bibinfo {volume} {82}},\ \bibinfo {pages} {1155} (\bibinfo {year}
  {2010})}\BibitemShut {NoStop}%
\bibitem [{\citenamefont {Grabert}\ and\ \citenamefont
  {Devoret}(2013)}]{grabert_single_2013}%
  \BibitemOpen
  \bibfield  {author} {\bibinfo {author} {\bibfnamefont {H.}~\bibnamefont
  {Grabert}}\ and\ \bibinfo {author} {\bibfnamefont {M.~H.}\ \bibnamefont
  {Devoret}},\ }\href@noop {} {\emph {\bibinfo {title} {Single {{Charge
  Tunneling}}: {{Coulomb Blockade Phenomena In Nanostructures}}}}}\ (\bibinfo
  {publisher} {{Springer Science \& Business Media}},\ \bibinfo {year}
  {2013})\BibitemShut {NoStop}%
\bibitem [{\citenamefont {Braak}(2011)}]{braak_integrability_2011}%
  \BibitemOpen
  \bibfield  {author} {\bibinfo {author} {\bibfnamefont {D.}~\bibnamefont
  {Braak}},\ }\bibfield  {title} {\bibinfo {title} {Integrability of the {{Rabi
  Model}}},\ }\href {https://doi.org/10.1103/PhysRevLett.107.100401} {\bibfield
   {journal} {\bibinfo  {journal} {Phys. Rev. Lett.}\ }\textbf {\bibinfo
  {volume} {107}},\ \bibinfo {pages} {100401} (\bibinfo {year}
  {2011})}\BibitemShut {NoStop}%
\bibitem [{\citenamefont {Zueco}\ \emph {et~al.}(2009)\citenamefont {Zueco},
  \citenamefont {Reuther}, \citenamefont {Kohler},\ and\ \citenamefont
  {H{\"a}nggi}}]{zueco_qubit-oscillator_2009}%
  \BibitemOpen
  \bibfield  {author} {\bibinfo {author} {\bibfnamefont {D.}~\bibnamefont
  {Zueco}}, \bibinfo {author} {\bibfnamefont {G.~M.}\ \bibnamefont {Reuther}},
  \bibinfo {author} {\bibfnamefont {S.}~\bibnamefont {Kohler}},\ and\ \bibinfo
  {author} {\bibfnamefont {P.}~\bibnamefont {H{\"a}nggi}},\ }\bibfield  {title}
  {\bibinfo {title} {Qubit-oscillator dynamics in the dispersive regime:
  {{Analytical}} theory beyond the rotating-wave approximation},\ }\href
  {https://doi.org/10.1103/PhysRevA.80.033846} {\bibfield  {journal} {\bibinfo
  {journal} {Phys. Rev. A}\ }\textbf {\bibinfo {volume} {80}},\ \bibinfo
  {pages} {033846} (\bibinfo {year} {2009})}\BibitemShut {NoStop}%
\end{thebibliography}%
\end{document}